\documentclass[reprint,amsmath,amssymb,aps,superscriptaddress,prl]{revtex4-2}
\usepackage{graphicx}
\usepackage{hyperref}
\usepackage{footmisc}
\usepackage{xcolor}
\newcommand{\tc}[1]{{#1}}
\begin{document}

\title{Ideal Quantum Tele-amplification up to a Selected Energy Cut-off using Linear Optics}

\author{Joshua J. Guanzon}
\email{joshua.guanzon@uq.net.au}
\affiliation{Centre for Quantum Computation and Communication Technology, School of Mathematics and Physics, University of Queensland, St Lucia, Queensland 4072, Australia}

\author{Matthew S. Winnel}
\affiliation{Centre for Quantum Computation and Communication Technology, School of Mathematics and Physics, University of Queensland, St Lucia, Queensland 4072, Australia}

\author{Austin P. Lund}
\affiliation{Dahlem Center for Complex Quantum Systems, Freie Universit\"at Berlin, 14195 Berlin, Germany}
\affiliation{Centre for Quantum Computation and Communication Technology, School of Mathematics and Physics, University of Queensland, St Lucia, Queensland 4072, Australia}

\author{Timothy C. Ralph}
\affiliation{Centre for Quantum Computation and Communication Technology, School of Mathematics and Physics, University of Queensland, St Lucia, Queensland 4072, Australia}

\date{\today}

\begin{abstract}

We introduce a linear optical technique that can implement \textit{ideal} quantum tele-amplification up to the $n^\mathrm{th}$ Fock state, where $n$ can be any positive integer. Here tele-amplification consists of both quantum teleportation and noiseless linear amplification (NLA). This simple protocol consists of a beam-splitter and an $(n+1)$-splitter, with $n$ ancillary photons and detection of $n$ photons. For a given target fidelity, our technique improves success probability and physical resource costs by orders of magnitude over current alternative teleportation and NLA schemes. We show how this protocol can also be used as a loss-tolerant quantum relay for entanglement distribution and distillation. 

\end{abstract}

\maketitle

The ability to amplify an arbitrary state in a linear, or phase insensitive, manner is useful for a wide variety of quantum protocols. Unfortunately, the uncertainty principle means \textit{deterministic} linear amplification will always introduce noise, which diminishes the output state's quantum characteristics~\cite{heffner1962fundamental,caves1981quantum}. However, \textit{noiseless} linear amplification (NLA) is possible for non-deterministic amplifiers, which work with some success probability $\mathbb{P}<1$~\cite{ralph2009nondeterministic,xiang2010heralded,ferreyrol2010implementation}. Applications of NLA include quantum secure communication~\cite{gisin2010proposal,blandino2012improving,xu2013improving,ghalaii2020long,zhou2020device,li2021improving}, quantum repeaters~\cite{dias2017quantum,dias2020quantum,seshadreesan2020continuous}, entanglement distillation~\cite{zhang2012protecting,seshadreesan2019continuous}, quantum sensing~\cite{usuga2010noise,zhao2017quantum,xia2019repeater}, and quantum error correction~\cite{ralph2011quantum,dias2018quantum}.

Quantum tele-amplification protocols implement quantum teleportation~\cite{bennett1993teleporting} and NLA simultaneously. In this regard, Pegg \textit{et al.} proposed a non-deterministic teleporter for low-energy states called the one-photon quantum scissor ($1$-QS), named for its ability to cut or truncate an arbitrary state up to its one-photon Fock state $|\psi\rangle \equiv \sum_{j=0}^{\infty} c_j|j\rangle \rightarrow \sum_{j=0}^{1} c_j|j\rangle $~\cite{pegg1998optical}. Ralph and Lund later realised adjusting a beam-splitter in the $1$-QS modified the output state's amplitudes $|\psi\rangle\xrightarrow{1\text{-QS}} \sum_{j=0}^{1} g^jc_j|j\rangle $~\cite{ralph2009nondeterministic}. Hence, for low-energy states, the $1$-QS can also perform an ideal NLA operation $g^{a^\dagger a}$ up to the one-photon Fock state with $g\in(0,\infty)$ gain; this was subsequently experimentally verified~\cite{xiang2010heralded,ferreyrol2010implementation}. To overcome the low-energy limitation, it was proposed to split up the input state, before applying multiple $1$-QS in parallel~\cite{ralph2009nondeterministic,xiang2010heralded}. However, for a finite number of $1$-QS, this protocol introduces extra undesirable factors to the Fock states, distorting the output state away from the ideal. Other NLA proposals are similarly non-ideal~\cite{fiuravsek2009engineering,zavatta2011high,jeffers2010nondeterministic}.

Rather than multiple $1$-QS in parallel, here we propose to generalise the $1$-QS to the $n$-photons quantum scissor ($n$-QS), for any $n\in\mathbb{N}^+$. Previous generalisations of the QS were only for specific sizes $n\in\{1,3,7\}$~\cite{winnel2020generalized}, and our fully generalised $n$-QS protocol contains these previous results~\footnote{See Supplemental Material at [URL will be inserted by publisher] for the details about the technical proofs for the $n$-QS operations, probability of success with improvements, fidelity of output state, comparisons with other NLA protocols, \tc{loss-tolerant entanglement distillation analysis, and experimental imperfections analysis,} which includes Ref.~\cite{scheel2004permanents,gard2015introduction,lund2017quantum,brualdi1991combinatorial,percus2012combinatorial,ralph2009nondeterministic,winnel2020generalized,tserkis2019quantifying,garcia2009reverse,winnel2021overcoming,reddy2019exceeding,lita2008counting,marsili2013detecting,miller2003demonstration,slussarenko2019photonic,su2019conversion,xiang2010heralded,andersen2013high}}. Our $n$-QS protocol is a fully scalable linear optical scheme, which can perform tele-amplification on an arbitrary state perfectly up to the $n^\mathrm{th}$ Fock state. Other tele-amplification proposals are restricted to specific types of input states~\cite{neergaard2013quantum}. The $2$-QS case is of particular experimental interest, as it should be immediately accessible with current technology. 

In this Letter, we first describe our $n$-QS protocol, including its operational mechanism and probability of success. \tc{We show that as an NLA it can produce amplified states with fidelities that are unreachable by previous linear-optical NLA protocols.} Next, we explain how the $n$-QS is also useful as a high-fidelity continuous-variable teleporter, with orders of magnitude advantages over current alternatives. \tc{We then show that the $n$-QS can be used as a loss-tolerant relay for entanglement distillation. Finally, we discuss how our scheme is tolerant to standard resource and detector imperfections, and hence remains advantageous under practical conditions.}

\begin{figure*}[htbp]
    \begin{center}
        \includegraphics[width=\linewidth]{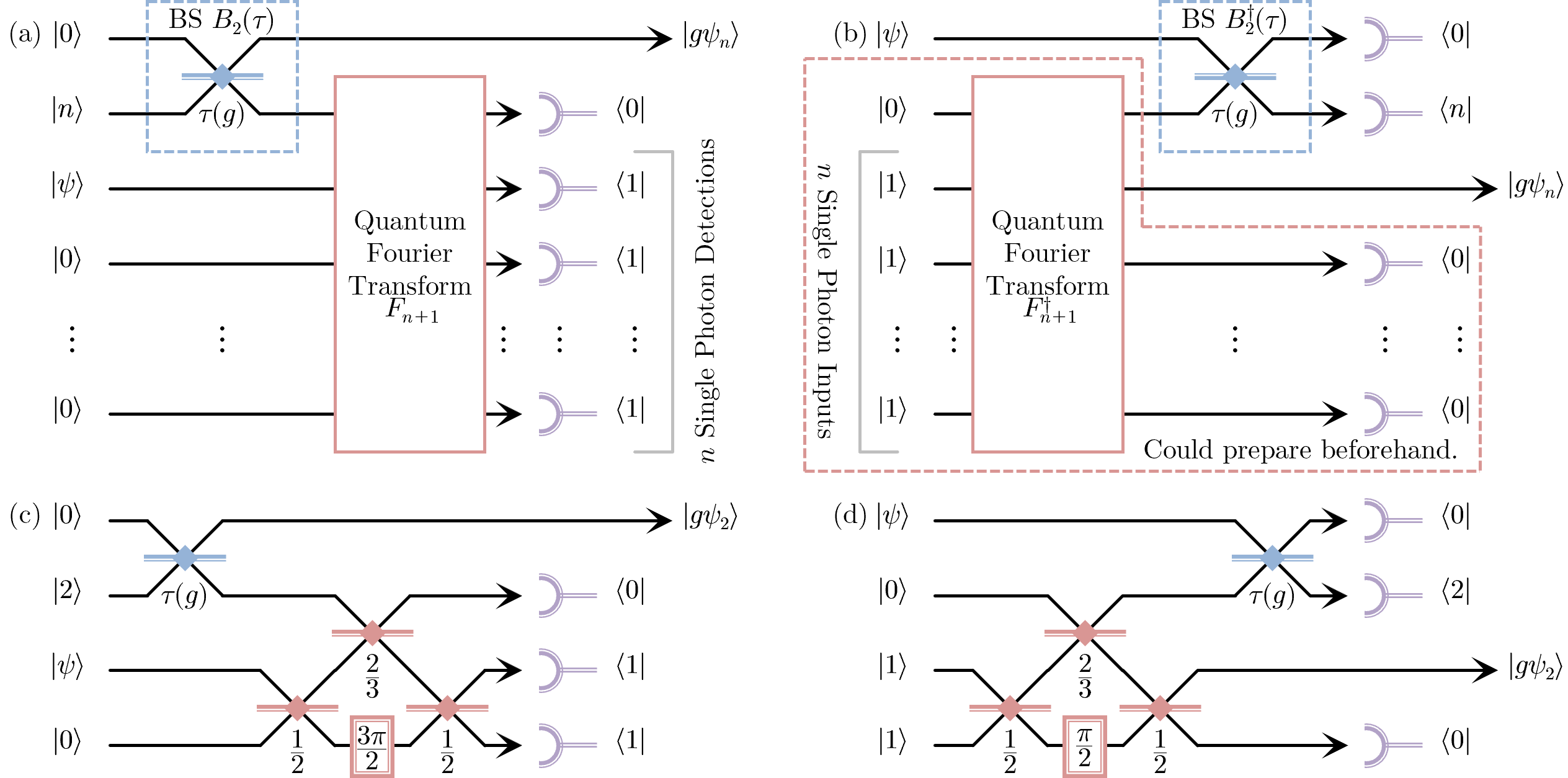}
        \caption{\label{fig:nQS} 
            Schematic of our scalable $n$-photons quantum scissors ($n$-QS) protocol, which implements noiseless linear amplification (NLA) or de-amplification of an arbitrary state $|\psi\rangle \rightarrow g^{a^\dagger a}|\psi_n\rangle = |g\psi_n\rangle$, up to the $n^\mathrm{th}$ Fock state with perfect fidelity. The gain $g\in(0,\infty)$ is chosen by modifying the transmissivity $\tau = g^2/(1+g^2)\in(0,1)$ of the beam-splitter (BS). The Quantum Fourier Transform (QFT) is a coherent $(n+1)$-splitter. This $n$-QS protocol requires either (a) $n$ bunched photons (BP) or (b) $n$ single photons (SP) as a resource. The linear optical networks for the $2$-QS is shown for (c) BP or (d) SP resources.}
    \end{center}
\end{figure*}

\textit{Noiseless linear amplifier.---}The $n$-QS operation on an arbitrary bosonic state $|\psi\rangle$ truncates the Fock components after $n$ and performs NLA $g^{a^\dagger a}$ as follows
\begin{align}
    |\psi\rangle &\equiv \sum_{j=0}^{\infty} c_j|j\rangle \xrightarrow{n\text{-QS}} |g\psi_n\rangle = N \sum_{j=0}^{n} g^j c_j|j\rangle. \label{eq:nQS}
\end{align}
This is implemented via Fig.~\ref{fig:nQS} using a beam-splitter (BS) and a fixed coherent $(n+1)$-splitter called the Quantum Fourier Transform (QFT), with $n$ extra resource photons and $n$ photon detections. The amount of amplification or de-amplification gain $g\in(0,\infty)$ can be freely chosen by setting the BS transmissivity to $\tau=g^2/(1+g^2)$. The $n$-QS operation only occurs if the correct outcomes are measured. Two different architectures are shown in Fig.~\ref{fig:nQS}, with (a) requiring $|n\rangle$ bunched photons ($n$-QSBP or BP), while (b) requiring $\otimes^n|1\rangle$ single photons ($n$-QSSP or SP); we will differentiate these devices by their state resources. Due to recent experimental advances, such as in boson sampling, $n$-QSSP may be easier to implement; for example, Ref.~\cite{su2017multiphoton} experimentally implements the QFT up to fourth order with single-photon inputs. 

The action of the BS $B_2(\tau)$ is $a_1^\dagger \rightarrow \sqrt{1-\tau}a_2^\dagger - \sqrt{\tau} a_1^\dagger$, which describes how the photons are scattered for a given transmissivity $\tau$. Similarly, the action of an $m$ mode linear optical network $\vec{a}^\dagger \rightarrow U_m \vec{a}^\dagger$ is captured by an $m\times m$ unitary scattering matrix $U_m$. The QFT optical device has the scattering matrix $(F_{n+1})_{j,k} \equiv e^{-\frac{2i\pi(j-1)(k-1)}{n+1}}/\sqrt{n+1}$. This definition justifies the interpretation of the QFT as a coherent $(n+1)$-splitter, as it scatters photons equally amongst its $n+1$ output ports with fixed phases. An arbitrary unitary $U_m$ can always be decomposed into a network of at most $m(m-1)/2$ beam-splitters and phase shifts~\cite{reck1994experimental,clements2016optimal}; however, only around half of the QFT network is needed since only the first two ports are used. As an example, we use Ref.~\cite{clements2016optimal} to decompose $2$-QS into a four BS network, as shown in Fig.~\ref{fig:nQS} for either (c) BP $|2\rangle$, or (d) SP $|1\rangle|1\rangle$ resources. The $2$-QS is the smallest network whose useful tele-amplification properties were previously not known. 

Here we will highlight the key elements which prove Fig.~\ref{fig:nQS} implements the $n$-QS transformation in Eq.~\eqref{eq:nQS} $\forall n\in\mathbb{N}^+$. Firstly, one can show that $B_2(g)|0\rangle|n\rangle = \frac{1}{(g^2+1)^{n/2}} \sum_{j=0}^n g^j (-1)^j \sqrt{\binom{n}{j}} |j\rangle|n-j\rangle$, which already has the gain $g^j$ coefficients, though with unwanted $(-1)^j \sqrt{\binom{n}{j}}$ factors. The red dashed box in (b) produces the two-mode output $|R_n\rangle = \otimes^{n-1} \langle 0| F_{n+1}^\dagger |0\rangle \otimes^n |1\rangle = \frac{\sqrt{n!}}{(n+1)^{n/2}} \sum_{j=0}^n (-1)^j\sqrt{\binom{n}{j}^{-1}}|n-j\rangle|j\rangle$. In other words, the state $|R_n\rangle$ distorts the Fock states in an inverse manner to $B_2|0\rangle|n\rangle$. Hence, by combining one mode from each of these states, the overall action of the $n$-QSSP is
\begin{align}
    \langle 0|\langle n| B_2^\dagger |R_n\rangle = \frac{\sqrt{n!}}{(n+1)^{n/2}} \frac{1}{(g^2+1)^{n/2}} \sum_{j=0}^n g^j |j \rangle\langle j|.
\end{align}
The $n$-QSBP is described by the same operator, since it is the conjugate transpose of this expression. The $n$-QS therefore applies ideal $g^{a^\dagger a}$ up to the $n^\mathrm{th}$ Fock state 
\begin{align}
    |g\psi_n\rangle &= \frac{\sqrt{n!}}{(n+1)^{n/2}} \frac{1}{(g^2+1)^{n/2}} \sum_{j=0}^n g^j c_j |j\rangle. \label{eq:gpsi}
\end{align}
The Supplemental Material contains the full proof~\cite{Note1}. 

\begin{figure}[htbp]
    \begin{center}
        \includegraphics[width=\linewidth]{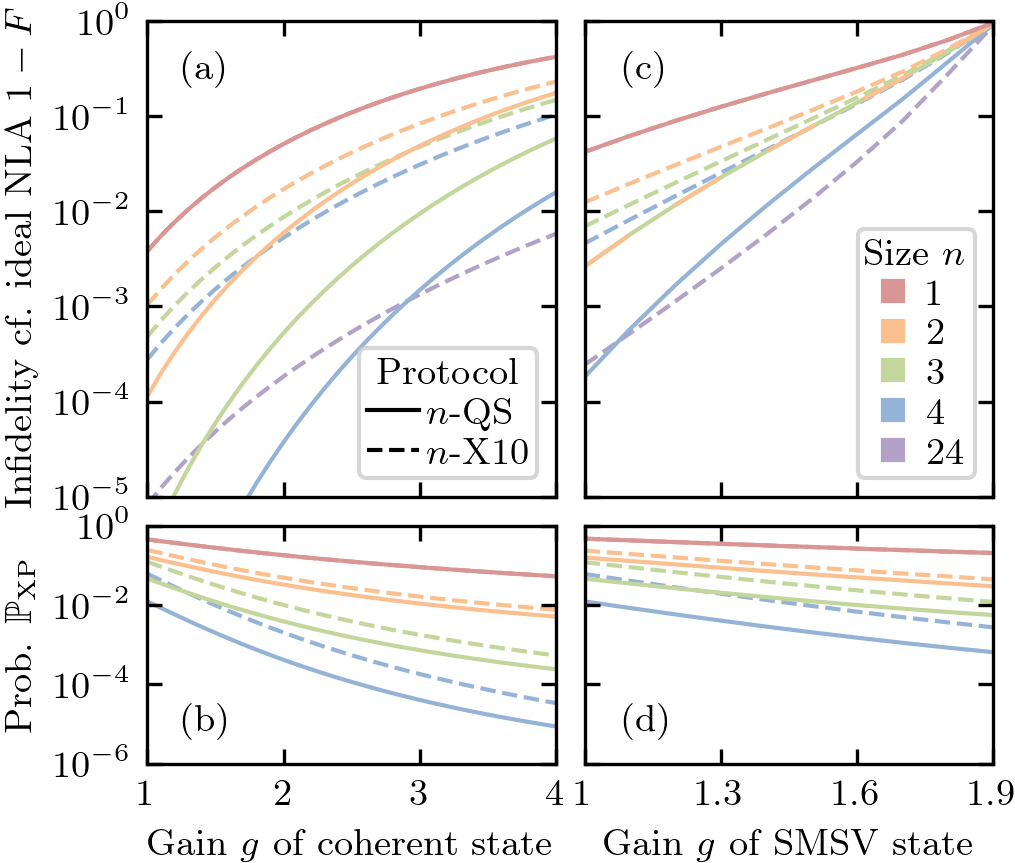}
        \caption{\label{fig:PFg} 
            A comparison of our $n$-QS NLA protocol, as per Fig.~\ref{fig:nQS}, against the $n$-X10 NLA protocol~\cite{xiang2010heralded}. The left plots considers a coherent state input with $\alpha=0.3$ amplitude, and the right plots considers a single-mode squeezed vacuum (SMSV) state input with $s\approx 0.29$ squeezing (such that these states have the same average photon number). (a) and (c) shows the infidelity $1-F$ relative to a perfect NLA output state, while (b) and (d) shows the probability of success $\mathbb{P}_\text{XP}$.
            }
    \end{center}
\end{figure}

The $n$-QS has a success probability of $\mathbb{P} = \langle g \psi_n | g\psi_n \rangle$, which can be improved depending on whether we are considering the BP or SP configuration. For $n$-QSBP, it is not required that the vacuum state $\langle0|$ be detected at the first output port of $F_{n+1}$, since the QFT is highly symmetric. If $\langle0|$ was instead detected in the $(m_0+1)^\mathrm{th}$ output port $\otimes^{m_0} \langle1| \otimes \langle0| \otimes^{n-m_0} \langle1|$, $m_0\in\lbrace0,\cdots,n\rbrace$, the output state will be $|g\psi_n\rangle$ with an extra phase shift that can be corrected by $C_1(m_0) = e^{\frac{2i\pi m_0}{n+1} a^\dagger a}$~\cite{Note1}. Utilizing all $n+1$ heralding events enhances the success probability by $\mathbb{P}_\text{BP}=(n+1)\mathbb{P}$. For $n$-QSSP, the $|R_n\rangle$ resource state from the red dashed box in Fig.~\ref{fig:nQS}(b) could be prepared and stored beforehand; assuming $|R_n\rangle$ is deterministically available increases the success probability $\mathbb{P}_\text{SP}$ to at least $\frac{(n+1)^n}{(n+1)!}\mathbb{P}$~\cite{Note1}. Note that $\mathbb{P}_\text{SP}<\mathbb{P}_\text{BP}$ for $n\in\lbrace1,2\rbrace$, $\mathbb{P}_\text{SP}=\mathbb{P}_\text{BP}$ for $n=3$, and $\mathbb{P}_\text{SP}>\mathbb{P}_\text{BP}$ for $n>3$~\cite{Note1}. Since there is no difference between the output states of these configurations, we will use $\mathbb{P}_\text{XP}=\max(\mathbb{P}_\text{SP},\mathbb{P}_\text{BP})$ depending on the size $n$ under consideration. 

We will now contrast our protocol with Xiang, \textit{et al.} 2010 (X10) linear optical NLA protocol~\cite{xiang2010heralded}. An $n$ sized X10 network has $n$ copies of $1$-QS in parallel between two $n$-splitters, hence requires approximately the same amount of physical resources as an $n$-QS. One advantage of the simplified $n$-QS structure is that setting a particular gain requires changing just one BS, while $n$-X10 requires changing $n$ BS concurrently. The output state from $n$-X10 has both the cut-off and distorted coefficients
\begin{align}
    |\psi\rangle \xrightarrow{n\text{-X}10} |g\phi_n\rangle = N' \sum_{j=0}^{n} \frac{1}{(n-j)!n^j} g^j c_j|j\rangle, 
\end{align}
hence the NLA is not ideal in general~\cite{ralph2009nondeterministic}. The fidelity $F$ can quantify how far away these output states are from the ideal NLA output state $g^{a^\dagger a}|\psi\rangle=|g\psi\rangle$~\cite{Note1}. It is clear an $n_\text{max}$-QS can amplify any arbitrary state with a $n_\text{max}$ upper energy limit with perfect fidelity.  This feat cannot be replicated by any finite sized $n$-X10, or by any previous linear-optical NLA protocol~\cite{fiuravsek2009engineering,zavatta2011high}.   

In Fig.~\ref{fig:PFg} we consider amplifying a coherent state and a single-mode squeezed vacuum (SMSV) state. Our $n$-QS has superior fidelity scaling, and hence for a required target fidelity needs much less resources with better success probability than the $n$-X10. For example, Fig.~\ref{fig:PFg}(a) and (b) shows amplifying the coherent state by $g\approx3$ with 99.9\% fidelity requires only an $4$-QS with $10^{-5}$ success probability, as opposed to a much larger $24$-X10 with $10^{-24}$ success probability. Fig.~\ref{fig:PFg}(c) and (d) emphasize the flexibility of our $n$-QS protocol, in that we can choose the best $n$ size for a given input; since SMSV states contain only even photon numbers, it is best to use even sized $n$-QS (odd sizes will give the same fidelity as one size down). These graphs also show the $n$-QS has fidelity advantages even with amplifying SMSV states near maximum squeezing given by \tc{$g_\text{max}^2\ s =1$ (here $g_\text{max}\approx1.9$).}

\begin{figure}[htbp]
    \begin{center}
        \includegraphics[width=\linewidth]{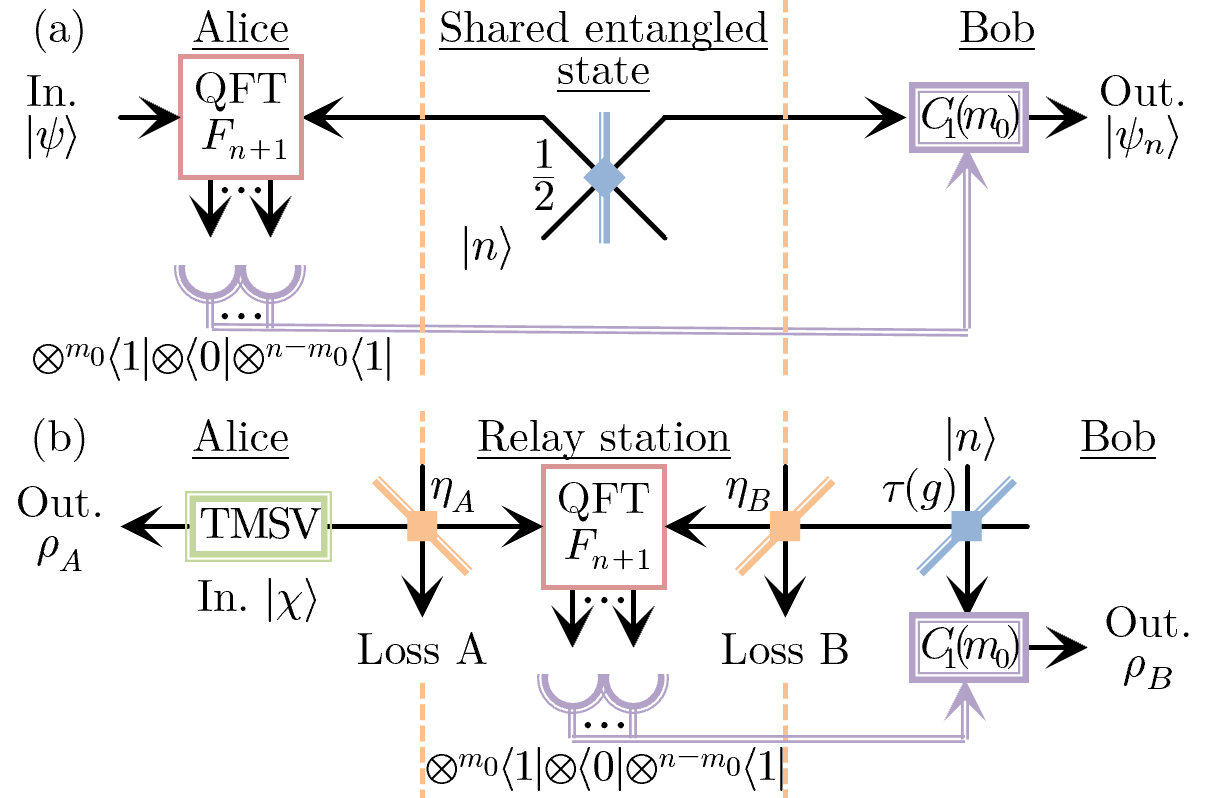}
        \caption{\label{fig:nprotocols} 
            Our scalable $n$-QS structure can be applied to many situations besides NLA, with significant improvements over existing protocols. We investigate applications for (a) quantum teleportation and (b) entanglement distillation as a loss-tolerant quantum relay. Shown is the BP variants.}
    \end{center}
\end{figure}

\textit{Quantum teleporter.---}Quantum teleportation is a key primitive in quantum protocols~\cite{kimble2008quantum,gottesman1999demonstrating,chen2021directly}, since it allows for the transfer and manipulation of quantum information through a shared entangled state; this is possible in both discrete variable~\cite{bennett1993teleporting} and continuous variable (CV)~\cite{braunstein1998teleportation} regimes. Andersen and Ralph in 2013 (AR13) proposed a CV teleportation scheme~\cite{andersen2013high}, which could in principle reach high fidelities with lower energy requirements than standard CV teleportation~\cite{braunstein1998teleportation}. However, in a similar manner as X10, a finite sized AR13 protocol distorts the output state. We will demonstrate our $n$-QS with $g=1$, as in Fig.~\ref{fig:nprotocols}(a), is a better protocol for high-fidelity teleportation. We restrict ourselves to linear-optical systems, hence both $n$-AR13 and $n$-QS are non-deterministic, and require a comparable amount of physical resources. 

\begin{figure}[htbp]
    \begin{center}
        \includegraphics[width=\linewidth]{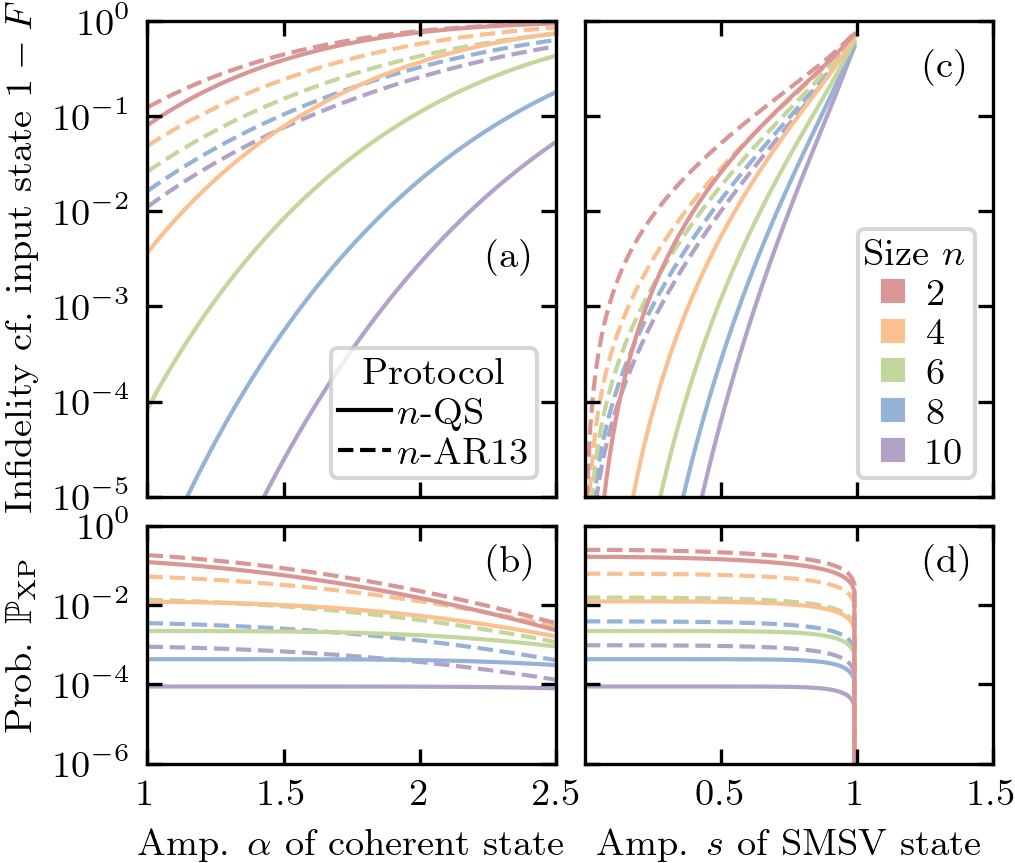}
        \caption{\label{fig:PFamp} 
            A comparison of our $n$-QS teleportation protocol, as per Fig.~\ref{fig:nprotocols}(a), against the $n$-AR13 high-fidelity teleportation protocol~\cite{andersen2013high}. We consider teleporting an $\alpha$ amplitude coherent state on the left and an $s$ squeezed SMSV state on the right. (a) and (c) is the infidelity $1-F$ of the teleported output state relative to the input state, while (b) and (d) is the protocol's probability of success $\mathbb{P}_\text{XP}$.}
    \end{center}
\end{figure}

We consider teleporting coherent and SMSV states with various amplitudes in Fig.~\ref{fig:PFamp}; we chose higher valued energy states to show the advantage of our scheme for larger $n$. It is clear our $n$-QS scales with many orders of magnitude better fidelity in comparison to $n$-AR13, while the probability of success scales comparatively. For example, teleporting an SMSV using a $4$-QS results in superior fidelity and success probability, while requiring less resources compared to a $10$-AR13. 

The AR13 paper illustrated the effectiveness of their protocol by analysing the teleportation of a coherent state superposition $|\alpha\rangle + |-\alpha\rangle$ with $\alpha = 2$. The authors note that to achieve just over 99\% fidelity, the standard teleportation approach requires 500 average photons (30 dB of squeezing)~\cite{braunstein1998teleportation}, while $n$-AR13 requires an $n=100$ photon entangled state~\cite{andersen2013high}. To reach the same fidelity, our $n$-QS protocol requires just $n=10$ photons.

\begin{figure}[htbp]
    \begin{center}
        \includegraphics[width=\linewidth]{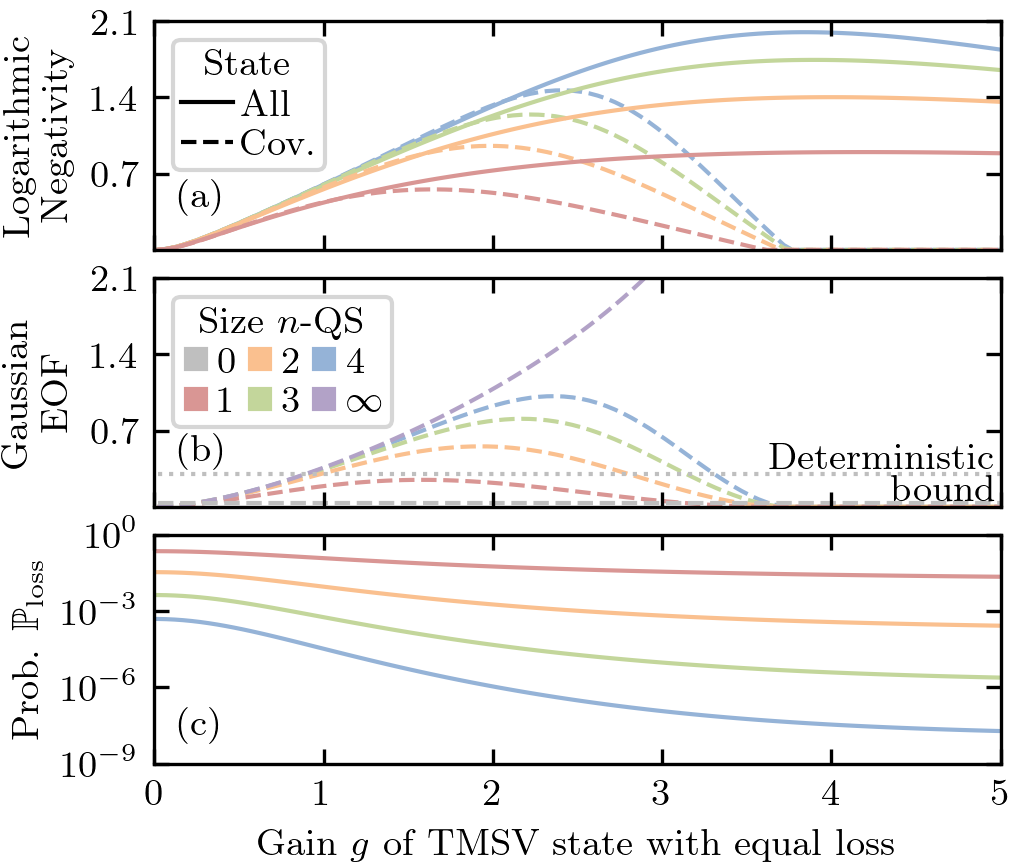}
        \caption{\label{fig:ent} 
            The amount of entanglement which can be recovered, using an equally $\eta_A=\eta_B=\sqrt{\eta}$ distributed $n$-QS as a quantum relay, as per Fig.~\ref{fig:nprotocols}(b). We consider a $\chi=0.25$ amplitude two-mode squeezed vacuum (TMSV) state into a lossy channel with $\eta=0.05$ total transmission. The entanglement is measured using (a) log negativity and (b) Gaussian entanglement of formation (EOF). The solid line considers all correlations (i.e. the entire state), while the dashed lines only considers second moments (i.e. the covariance matrix). (c) is the protocol's probability of success $\mathbb{P}_\text{loss}$.}
    \end{center}
\end{figure}

\textit{Loss-tolerant quantum relay.---}Here we consider distilling entanglement through a loss channel with $\eta\in[0,1]$ total transmissivity. The $n$-QS can function as a quantum relay by distributing the QFT measurement component over the channel, as shown in Fig.~\ref{fig:nprotocols}(b), such that $\eta_A\eta_B=\eta$. The distributed $1$-QS has previously been shown to be uniquely loss tolerant, in that it can overcome the repeaterless PLOB bound without quantum memories~\cite{winnel2021overcoming}; the only other known scheme that can also do this feat is the twin-field QKD protocol and its variants~\cite{lucamarini2018overcoming,liu2019experimental,curty2019simple,chen2020sending}. We confirm that the complete set of distributed $n$-QS are also loss tolerant with improved usage rates. In other words, instead of having the entire NLA at Bob's side ($\eta_A=\eta,\eta_B=1$), by placing the QFT measurement half way ($\eta_A=\sqrt{\eta},\eta_B=\sqrt{\eta}$), we improve success probability scaling from $\eta^n$ to $\eta^{n/2}$~\cite{Note1}. Note here we consider distilling a two-mode squeezed vacuum (TMSV) or EPR state with $\chi=0.25$ squeezing. 

The logarithmic negativity (LN) is an entanglement monotone~\cite{audenaert2003entanglement,plenio2005logarithmic}, and an upper bound for distillable entanglement~\cite{vidal2002computable}. The LN is shown by the solid lines in Fig.~\ref{fig:ent}(a), which increases with larger $n$ sizes. Maximum LN occurs with gain approximately $g_\text{max}\chi\sqrt{\eta_A/\eta_B} \approx 1$ (here $g_\text{max}\approx4$), which corresponds to an output state that is uniformly distributed in the Fock basis~\cite{Note1}. The dashed lines in these graphs only considers the second moment covariance correlations, which are more relevant for Gaussian and CV protocols~\cite{adesso2004extremal}. 

The entanglement of formation (EOF) is an entanglement metric~\cite{bennett1996mixed}, whose properties for multi-mode Gaussian states are known~\cite{wolf2004gaussian,marian2008entanglement,tserkis2017quantifying,tserkis2019quantifying,onoe2020multipartite}. Fig~\ref{fig:ent}(b) is the Gaussian EOF, which closely resembles the covariance-based LN, as expected. The gray horizontal lines are pure loss channels with no QS, where the dashed line has the same initial squeezing $\chi=0.25$, and the dotted line is the deterministic bound with infinite squeezing $\chi=1$; it's clear this bound can be beaten by small sized $n$-QS. Increasing loss doesn't significantly change the maximum amount of distillable entanglement, which is another experimental appealing feature of this loss tolerant protocol~\cite{Note1}.

\tc{\textit{Experimental imperfections.---}Finally, we examined the effect of noisy, inefficient photon detectors and sources. We showed that our $n$-QSBP protocol is tolerant to experimental imperfections, in the same sense as the already experimentally verified $1$-QS~\cite{xiang2010heralded}. In other words, an imperfect $n$-QSBP as an amplifier, teleporter or relay results in relative improvements with increased $n$, in a similar fashion as the ideal graphs in this Letter. Unfortunately, the $n$-QSSP is not tolerant to experimental imperfections. This is because of how the entanglement resource is prepared, and a different preparation scheme could help to improve an imperfect $n$-QSSP. See the Supplemental Material for more details~\cite{Note1}.}

\textit{Conclusion.---}We introduced the generalised $n\in\mathbb{N}^+$ quantum scissors, which can perform perfect fidelity tele-amplification up to the $n^\mathrm{th}$ Fock state. We proved that this operation can be implemented using two simple scalable linear-optical networks, with either $n$ single or $n$ bunched ancillary photons. As a consequence, our $n$-QS protocol is shown to have substantial advantages over existing NLA and teleportation schemes, in terms of fidelity scaling, success probability and physical resources. Finally, we showed that a distributed $n$-QS quantum relay is loss-tolerant with fast rates, hence is useful as building blocks for quantum repeater networks. \\

\begin{acknowledgments}
APL acknowledges support from BMBF (QPIC) and the Einstein Research Unit on Quantum Devices. This research was supported by the Australian Research Council (ARC) under the Centre of Excellence for Quantum Computation and Communication Technology (Project No. CE170100012). \\
\end{acknowledgments}

\textit{Note added.---}The authors recently became aware of a new related work which investigated noiseless quantum tele-amplifiers from a different angle~\cite{fiuravsek2021teleportation}, based on the continuous-variable teleportation protocol~\cite{braunstein1998teleportation}.

\bibliography{main}

\begin{thebibliography}{65}%
\makeatletter
\providecommand \@ifxundefined [1]{%
 \@ifx{#1\undefined}
}%
\providecommand \@ifnum [1]{%
 \ifnum #1\expandafter \@firstoftwo
 \else \expandafter \@secondoftwo
 \fi
}%
\providecommand \@ifx [1]{%
 \ifx #1\expandafter \@firstoftwo
 \else \expandafter \@secondoftwo
 \fi
}%
\providecommand \natexlab [1]{#1}%
\providecommand \enquote  [1]{``#1''}%
\providecommand \bibnamefont  [1]{#1}%
\providecommand \bibfnamefont [1]{#1}%
\providecommand \citenamefont [1]{#1}%
\providecommand \href@noop [0]{\@secondoftwo}%
\providecommand \href [0]{\begingroup \@sanitize@url \@href}%
\providecommand \@href[1]{\@@startlink{#1}\@@href}%
\providecommand \@@href[1]{\endgroup#1\@@endlink}%
\providecommand \@sanitize@url [0]{\catcode `\\12\catcode `\$12\catcode
  `\&12\catcode `\#12\catcode `\^12\catcode `\_12\catcode `\%12\relax}%
\providecommand \@@startlink[1]{}%
\providecommand \@@endlink[0]{}%
\providecommand \url  [0]{\begingroup\@sanitize@url \@url }%
\providecommand \@url [1]{\endgroup\@href {#1}{\urlprefix }}%
\providecommand \urlprefix  [0]{URL }%
\providecommand \Eprint [0]{\href }%
\providecommand \doibase [0]{https://doi.org/}%
\providecommand \selectlanguage [0]{\@gobble}%
\providecommand \bibinfo  [0]{\@secondoftwo}%
\providecommand \bibfield  [0]{\@secondoftwo}%
\providecommand \translation [1]{[#1]}%
\providecommand \BibitemOpen [0]{}%
\providecommand \bibitemStop [0]{}%
\providecommand \bibitemNoStop [0]{.\EOS\space}%
\providecommand \EOS [0]{\spacefactor3000\relax}%
\providecommand \BibitemShut  [1]{\csname bibitem#1\endcsname}%
\let\auto@bib@innerbib\@empty
\bibitem [{\citenamefont {Heffner}(1962)}]{heffner1962fundamental}%
  \BibitemOpen
  \bibfield  {author} {\bibinfo {author} {\bibfnamefont {H.}~\bibnamefont
  {Heffner}},\ }\href@noop {} {\bibfield  {journal} {\bibinfo  {journal}
  {Proceedings of the IRE}\ }\textbf {\bibinfo {volume} {50}},\ \bibinfo
  {pages} {1604} (\bibinfo {year} {1962})}\BibitemShut {NoStop}%
\bibitem [{\citenamefont {Caves}(1981)}]{caves1981quantum}%
  \BibitemOpen
  \bibfield  {author} {\bibinfo {author} {\bibfnamefont {C.~M.}\ \bibnamefont
  {Caves}},\ }\href@noop {} {\bibfield  {journal} {\bibinfo  {journal}
  {Physical Review D}\ }\textbf {\bibinfo {volume} {23}},\ \bibinfo {pages}
  {1693} (\bibinfo {year} {1981})}\BibitemShut {NoStop}%
\bibitem [{\citenamefont {Ralph}\ and\ \citenamefont
  {Lund}(2009)}]{ralph2009nondeterministic}%
  \BibitemOpen
  \bibfield  {author} {\bibinfo {author} {\bibfnamefont {T.~C.}\ \bibnamefont
  {Ralph}}\ and\ \bibinfo {author} {\bibfnamefont {A.~P.}\ \bibnamefont
  {Lund}},\ }in\ \href@noop {} {\emph {\bibinfo {booktitle} {AIP Conference
  Proceedings}}},\ Vol.\ \bibinfo {volume} {1110}\ (\bibinfo {organization}
  {American Institute of Physics},\ \bibinfo {year} {2009})\ pp.\ \bibinfo
  {pages} {155--160}\BibitemShut {NoStop}%
\bibitem [{\citenamefont {Xiang}\ \emph {et~al.}(2010)\citenamefont {Xiang},
  \citenamefont {Ralph}, \citenamefont {Lund}, \citenamefont {Walk},\ and\
  \citenamefont {Pryde}}]{xiang2010heralded}%
  \BibitemOpen
  \bibfield  {author} {\bibinfo {author} {\bibfnamefont {G.-Y.}\ \bibnamefont
  {Xiang}}, \bibinfo {author} {\bibfnamefont {T.~C.}\ \bibnamefont {Ralph}},
  \bibinfo {author} {\bibfnamefont {A.~P.}\ \bibnamefont {Lund}}, \bibinfo
  {author} {\bibfnamefont {N.}~\bibnamefont {Walk}},\ and\ \bibinfo {author}
  {\bibfnamefont {G.~J.}\ \bibnamefont {Pryde}},\ }\href@noop {} {\bibfield
  {journal} {\bibinfo  {journal} {Nature Photonics}\ }\textbf {\bibinfo
  {volume} {4}},\ \bibinfo {pages} {316} (\bibinfo {year} {2010})}\BibitemShut
  {NoStop}%
\bibitem [{\citenamefont {Ferreyrol}\ \emph {et~al.}(2010)\citenamefont
  {Ferreyrol}, \citenamefont {Barbieri}, \citenamefont {Blandino},
  \citenamefont {Fossier}, \citenamefont {Tualle-Brouri},\ and\ \citenamefont
  {Grangier}}]{ferreyrol2010implementation}%
  \BibitemOpen
  \bibfield  {author} {\bibinfo {author} {\bibfnamefont {F.}~\bibnamefont
  {Ferreyrol}}, \bibinfo {author} {\bibfnamefont {M.}~\bibnamefont {Barbieri}},
  \bibinfo {author} {\bibfnamefont {R.}~\bibnamefont {Blandino}}, \bibinfo
  {author} {\bibfnamefont {S.}~\bibnamefont {Fossier}}, \bibinfo {author}
  {\bibfnamefont {R.}~\bibnamefont {Tualle-Brouri}},\ and\ \bibinfo {author}
  {\bibfnamefont {P.}~\bibnamefont {Grangier}},\ }\href@noop {} {\bibfield
  {journal} {\bibinfo  {journal} {Physical review letters}\ }\textbf {\bibinfo
  {volume} {104}},\ \bibinfo {pages} {123603} (\bibinfo {year}
  {2010})}\BibitemShut {NoStop}%
\bibitem [{\citenamefont {Gisin}\ \emph {et~al.}(2010)\citenamefont {Gisin},
  \citenamefont {Pironio},\ and\ \citenamefont
  {Sangouard}}]{gisin2010proposal}%
  \BibitemOpen
  \bibfield  {author} {\bibinfo {author} {\bibfnamefont {N.}~\bibnamefont
  {Gisin}}, \bibinfo {author} {\bibfnamefont {S.}~\bibnamefont {Pironio}},\
  and\ \bibinfo {author} {\bibfnamefont {N.}~\bibnamefont {Sangouard}},\
  }\href@noop {} {\bibfield  {journal} {\bibinfo  {journal} {Physical review
  letters}\ }\textbf {\bibinfo {volume} {105}},\ \bibinfo {pages} {070501}
  (\bibinfo {year} {2010})}\BibitemShut {NoStop}%
\bibitem [{\citenamefont {Blandino}\ \emph {et~al.}(2012)\citenamefont
  {Blandino}, \citenamefont {Leverrier}, \citenamefont {Barbieri},
  \citenamefont {Etesse}, \citenamefont {Grangier},\ and\ \citenamefont
  {Tualle-Brouri}}]{blandino2012improving}%
  \BibitemOpen
  \bibfield  {author} {\bibinfo {author} {\bibfnamefont {R.}~\bibnamefont
  {Blandino}}, \bibinfo {author} {\bibfnamefont {A.}~\bibnamefont {Leverrier}},
  \bibinfo {author} {\bibfnamefont {M.}~\bibnamefont {Barbieri}}, \bibinfo
  {author} {\bibfnamefont {J.}~\bibnamefont {Etesse}}, \bibinfo {author}
  {\bibfnamefont {P.}~\bibnamefont {Grangier}},\ and\ \bibinfo {author}
  {\bibfnamefont {R.}~\bibnamefont {Tualle-Brouri}},\ }\href@noop {} {\bibfield
   {journal} {\bibinfo  {journal} {Physical Review A}\ }\textbf {\bibinfo
  {volume} {86}},\ \bibinfo {pages} {012327} (\bibinfo {year}
  {2012})}\BibitemShut {NoStop}%
\bibitem [{\citenamefont {Xu}\ \emph {et~al.}(2013)\citenamefont {Xu},
  \citenamefont {Tang}, \citenamefont {Chen}, \citenamefont {Zhang},\ and\
  \citenamefont {Zhu}}]{xu2013improving}%
  \BibitemOpen
  \bibfield  {author} {\bibinfo {author} {\bibfnamefont {B.}~\bibnamefont
  {Xu}}, \bibinfo {author} {\bibfnamefont {C.}~\bibnamefont {Tang}}, \bibinfo
  {author} {\bibfnamefont {H.}~\bibnamefont {Chen}}, \bibinfo {author}
  {\bibfnamefont {W.}~\bibnamefont {Zhang}},\ and\ \bibinfo {author}
  {\bibfnamefont {F.}~\bibnamefont {Zhu}},\ }\href@noop {} {\bibfield
  {journal} {\bibinfo  {journal} {Physical Review A}\ }\textbf {\bibinfo
  {volume} {87}},\ \bibinfo {pages} {062311} (\bibinfo {year}
  {2013})}\BibitemShut {NoStop}%
\bibitem [{\citenamefont {Ghalaii}\ \emph {et~al.}(2020)\citenamefont
  {Ghalaii}, \citenamefont {Ottaviani}, \citenamefont {Kumar}, \citenamefont
  {Pirandola},\ and\ \citenamefont {Razavi}}]{ghalaii2020long}%
  \BibitemOpen
  \bibfield  {author} {\bibinfo {author} {\bibfnamefont {M.}~\bibnamefont
  {Ghalaii}}, \bibinfo {author} {\bibfnamefont {C.}~\bibnamefont {Ottaviani}},
  \bibinfo {author} {\bibfnamefont {R.}~\bibnamefont {Kumar}}, \bibinfo
  {author} {\bibfnamefont {S.}~\bibnamefont {Pirandola}},\ and\ \bibinfo
  {author} {\bibfnamefont {M.}~\bibnamefont {Razavi}},\ }\href@noop {}
  {\bibfield  {journal} {\bibinfo  {journal} {IEEE Journal of Selected Topics
  in Quantum Electronics}\ }\textbf {\bibinfo {volume} {26}},\ \bibinfo {pages}
  {1} (\bibinfo {year} {2020})}\BibitemShut {NoStop}%
\bibitem [{\citenamefont {Zhou}\ \emph {et~al.}(2020)\citenamefont {Zhou},
  \citenamefont {Sheng},\ and\ \citenamefont {Long}}]{zhou2020device}%
  \BibitemOpen
  \bibfield  {author} {\bibinfo {author} {\bibfnamefont {L.}~\bibnamefont
  {Zhou}}, \bibinfo {author} {\bibfnamefont {Y.-B.}\ \bibnamefont {Sheng}},\
  and\ \bibinfo {author} {\bibfnamefont {G.-L.}\ \bibnamefont {Long}},\
  }\href@noop {} {\bibfield  {journal} {\bibinfo  {journal} {Science Bulletin}\
  }\textbf {\bibinfo {volume} {65}},\ \bibinfo {pages} {12} (\bibinfo {year}
  {2020})}\BibitemShut {NoStop}%
\bibitem [{\citenamefont {Li}\ \emph {et~al.}(2021)\citenamefont {Li},
  \citenamefont {Guo}, \citenamefont {Ruan},\ and\ \citenamefont
  {Zhao}}]{li2021improving}%
  \BibitemOpen
  \bibfield  {author} {\bibinfo {author} {\bibfnamefont {Y.}~\bibnamefont
  {Li}}, \bibinfo {author} {\bibfnamefont {Y.}~\bibnamefont {Guo}}, \bibinfo
  {author} {\bibfnamefont {X.}~\bibnamefont {Ruan}},\ and\ \bibinfo {author}
  {\bibfnamefont {W.}~\bibnamefont {Zhao}},\ }\href@noop {} {\bibfield
  {journal} {\bibinfo  {journal} {International Journal of Theoretical
  Physics}\ }\textbf {\bibinfo {volume} {60}},\ \bibinfo {pages} {1949}
  (\bibinfo {year} {2021})}\BibitemShut {NoStop}%
\bibitem [{\citenamefont {Dias}\ and\ \citenamefont
  {Ralph}(2017)}]{dias2017quantum}%
  \BibitemOpen
  \bibfield  {author} {\bibinfo {author} {\bibfnamefont {J.}~\bibnamefont
  {Dias}}\ and\ \bibinfo {author} {\bibfnamefont {T.~C.}\ \bibnamefont
  {Ralph}},\ }\href@noop {} {\bibfield  {journal} {\bibinfo  {journal}
  {Physical Review A}\ }\textbf {\bibinfo {volume} {95}},\ \bibinfo {pages}
  {022312} (\bibinfo {year} {2017})}\BibitemShut {NoStop}%
\bibitem [{\citenamefont {Dias}\ \emph {et~al.}(2020)\citenamefont {Dias},
  \citenamefont {Winnel}, \citenamefont {Hosseinidehaj},\ and\ \citenamefont
  {Ralph}}]{dias2020quantum}%
  \BibitemOpen
  \bibfield  {author} {\bibinfo {author} {\bibfnamefont {J.}~\bibnamefont
  {Dias}}, \bibinfo {author} {\bibfnamefont {M.~S.}\ \bibnamefont {Winnel}},
  \bibinfo {author} {\bibfnamefont {N.}~\bibnamefont {Hosseinidehaj}},\ and\
  \bibinfo {author} {\bibfnamefont {T.~C.}\ \bibnamefont {Ralph}},\ }\href@noop
  {} {\bibfield  {journal} {\bibinfo  {journal} {Physical Review A}\ }\textbf
  {\bibinfo {volume} {102}},\ \bibinfo {pages} {052425} (\bibinfo {year}
  {2020})}\BibitemShut {NoStop}%
\bibitem [{\citenamefont {Seshadreesan}\ \emph {et~al.}(2020)\citenamefont
  {Seshadreesan}, \citenamefont {Krovi},\ and\ \citenamefont
  {Guha}}]{seshadreesan2020continuous}%
  \BibitemOpen
  \bibfield  {author} {\bibinfo {author} {\bibfnamefont {K.~P.}\ \bibnamefont
  {Seshadreesan}}, \bibinfo {author} {\bibfnamefont {H.}~\bibnamefont
  {Krovi}},\ and\ \bibinfo {author} {\bibfnamefont {S.}~\bibnamefont {Guha}},\
  }\href@noop {} {\bibfield  {journal} {\bibinfo  {journal} {Physical Review
  Research}\ }\textbf {\bibinfo {volume} {2}},\ \bibinfo {pages} {013310}
  (\bibinfo {year} {2020})}\BibitemShut {NoStop}%
\bibitem [{\citenamefont {Zhang}\ \emph {et~al.}(2012)\citenamefont {Zhang},
  \citenamefont {Yang}, \citenamefont {Zou}, \citenamefont {Shi},\ and\
  \citenamefont {Guo}}]{zhang2012protecting}%
  \BibitemOpen
  \bibfield  {author} {\bibinfo {author} {\bibfnamefont {S.~L.}\ \bibnamefont
  {Zhang}}, \bibinfo {author} {\bibfnamefont {S.}~\bibnamefont {Yang}},
  \bibinfo {author} {\bibfnamefont {X.~B.}\ \bibnamefont {Zou}}, \bibinfo
  {author} {\bibfnamefont {B.~S.}\ \bibnamefont {Shi}},\ and\ \bibinfo {author}
  {\bibfnamefont {G.~C.}\ \bibnamefont {Guo}},\ }\href@noop {} {\bibfield
  {journal} {\bibinfo  {journal} {Physical Review A}\ }\textbf {\bibinfo
  {volume} {86}},\ \bibinfo {pages} {034302} (\bibinfo {year}
  {2012})}\BibitemShut {NoStop}%
\bibitem [{\citenamefont {Seshadreesan}\ \emph {et~al.}(2019)\citenamefont
  {Seshadreesan}, \citenamefont {Krovi},\ and\ \citenamefont
  {Guha}}]{seshadreesan2019continuous}%
  \BibitemOpen
  \bibfield  {author} {\bibinfo {author} {\bibfnamefont {K.~P.}\ \bibnamefont
  {Seshadreesan}}, \bibinfo {author} {\bibfnamefont {H.}~\bibnamefont
  {Krovi}},\ and\ \bibinfo {author} {\bibfnamefont {S.}~\bibnamefont {Guha}},\
  }\href@noop {} {\bibfield  {journal} {\bibinfo  {journal} {Physical Review
  A}\ }\textbf {\bibinfo {volume} {100}},\ \bibinfo {pages} {022315} (\bibinfo
  {year} {2019})}\BibitemShut {NoStop}%
\bibitem [{\citenamefont {Usuga}\ \emph {et~al.}(2010)\citenamefont {Usuga},
  \citenamefont {M{\"u}ller}, \citenamefont {Wittmann}, \citenamefont {Marek},
  \citenamefont {Filip}, \citenamefont {Marquardt}, \citenamefont {Leuchs},\
  and\ \citenamefont {Andersen}}]{usuga2010noise}%
  \BibitemOpen
  \bibfield  {author} {\bibinfo {author} {\bibfnamefont {M.~A.}\ \bibnamefont
  {Usuga}}, \bibinfo {author} {\bibfnamefont {C.~R.}\ \bibnamefont
  {M{\"u}ller}}, \bibinfo {author} {\bibfnamefont {C.}~\bibnamefont
  {Wittmann}}, \bibinfo {author} {\bibfnamefont {P.}~\bibnamefont {Marek}},
  \bibinfo {author} {\bibfnamefont {R.}~\bibnamefont {Filip}}, \bibinfo
  {author} {\bibfnamefont {C.}~\bibnamefont {Marquardt}}, \bibinfo {author}
  {\bibfnamefont {G.}~\bibnamefont {Leuchs}},\ and\ \bibinfo {author}
  {\bibfnamefont {U.~L.}\ \bibnamefont {Andersen}},\ }\href@noop {} {\bibfield
  {journal} {\bibinfo  {journal} {Nature Physics}\ }\textbf {\bibinfo {volume}
  {6}},\ \bibinfo {pages} {767} (\bibinfo {year} {2010})}\BibitemShut {NoStop}%
\bibitem [{\citenamefont {Zhao}\ \emph {et~al.}(2017)\citenamefont {Zhao},
  \citenamefont {Dias}, \citenamefont {Haw}, \citenamefont {Symul},
  \citenamefont {Bradshaw}, \citenamefont {Blandino}, \citenamefont {Ralph},
  \citenamefont {Assad},\ and\ \citenamefont {Lam}}]{zhao2017quantum}%
  \BibitemOpen
  \bibfield  {author} {\bibinfo {author} {\bibfnamefont {J.}~\bibnamefont
  {Zhao}}, \bibinfo {author} {\bibfnamefont {J.}~\bibnamefont {Dias}}, \bibinfo
  {author} {\bibfnamefont {J.~Y.}\ \bibnamefont {Haw}}, \bibinfo {author}
  {\bibfnamefont {T.}~\bibnamefont {Symul}}, \bibinfo {author} {\bibfnamefont
  {M.}~\bibnamefont {Bradshaw}}, \bibinfo {author} {\bibfnamefont
  {R.}~\bibnamefont {Blandino}}, \bibinfo {author} {\bibfnamefont
  {T.}~\bibnamefont {Ralph}}, \bibinfo {author} {\bibfnamefont {S.~M.}\
  \bibnamefont {Assad}},\ and\ \bibinfo {author} {\bibfnamefont {P.~K.}\
  \bibnamefont {Lam}},\ }\href@noop {} {\bibfield  {journal} {\bibinfo
  {journal} {Optica}\ }\textbf {\bibinfo {volume} {4}},\ \bibinfo {pages}
  {1421} (\bibinfo {year} {2017})}\BibitemShut {NoStop}%
\bibitem [{\citenamefont {Xia}\ \emph {et~al.}(2019)\citenamefont {Xia},
  \citenamefont {Zhuang}, \citenamefont {Clark},\ and\ \citenamefont
  {Zhang}}]{xia2019repeater}%
  \BibitemOpen
  \bibfield  {author} {\bibinfo {author} {\bibfnamefont {Y.}~\bibnamefont
  {Xia}}, \bibinfo {author} {\bibfnamefont {Q.}~\bibnamefont {Zhuang}},
  \bibinfo {author} {\bibfnamefont {W.}~\bibnamefont {Clark}},\ and\ \bibinfo
  {author} {\bibfnamefont {Z.}~\bibnamefont {Zhang}},\ }\href@noop {}
  {\bibfield  {journal} {\bibinfo  {journal} {Physical Review A}\ }\textbf
  {\bibinfo {volume} {99}},\ \bibinfo {pages} {012328} (\bibinfo {year}
  {2019})}\BibitemShut {NoStop}%
\bibitem [{\citenamefont {Ralph}(2011)}]{ralph2011quantum}%
  \BibitemOpen
  \bibfield  {author} {\bibinfo {author} {\bibfnamefont {T.~C.}\ \bibnamefont
  {Ralph}},\ }\href@noop {} {\bibfield  {journal} {\bibinfo  {journal}
  {Physical Review A}\ }\textbf {\bibinfo {volume} {84}},\ \bibinfo {pages}
  {022339} (\bibinfo {year} {2011})}\BibitemShut {NoStop}%
\bibitem [{\citenamefont {Dias}\ and\ \citenamefont
  {Ralph}(2018)}]{dias2018quantum}%
  \BibitemOpen
  \bibfield  {author} {\bibinfo {author} {\bibfnamefont {J.}~\bibnamefont
  {Dias}}\ and\ \bibinfo {author} {\bibfnamefont {T.~C.}\ \bibnamefont
  {Ralph}},\ }\href@noop {} {\bibfield  {journal} {\bibinfo  {journal}
  {Physical Review A}\ }\textbf {\bibinfo {volume} {97}},\ \bibinfo {pages}
  {032335} (\bibinfo {year} {2018})}\BibitemShut {NoStop}%
\bibitem [{\citenamefont {Bennett}\ \emph {et~al.}(1993)\citenamefont
  {Bennett}, \citenamefont {Brassard}, \citenamefont {Cr{\'e}peau},
  \citenamefont {Jozsa}, \citenamefont {Peres},\ and\ \citenamefont
  {Wootters}}]{bennett1993teleporting}%
  \BibitemOpen
  \bibfield  {author} {\bibinfo {author} {\bibfnamefont {C.~H.}\ \bibnamefont
  {Bennett}}, \bibinfo {author} {\bibfnamefont {G.}~\bibnamefont {Brassard}},
  \bibinfo {author} {\bibfnamefont {C.}~\bibnamefont {Cr{\'e}peau}}, \bibinfo
  {author} {\bibfnamefont {R.}~\bibnamefont {Jozsa}}, \bibinfo {author}
  {\bibfnamefont {A.}~\bibnamefont {Peres}},\ and\ \bibinfo {author}
  {\bibfnamefont {W.~K.}\ \bibnamefont {Wootters}},\ }\href@noop {} {\bibfield
  {journal} {\bibinfo  {journal} {Physical review letters}\ }\textbf {\bibinfo
  {volume} {70}},\ \bibinfo {pages} {1895} (\bibinfo {year}
  {1993})}\BibitemShut {NoStop}%
\bibitem [{\citenamefont {Pegg}\ \emph {et~al.}(1998)\citenamefont {Pegg},
  \citenamefont {Phillips},\ and\ \citenamefont {Barnett}}]{pegg1998optical}%
  \BibitemOpen
  \bibfield  {author} {\bibinfo {author} {\bibfnamefont {D.~T.}\ \bibnamefont
  {Pegg}}, \bibinfo {author} {\bibfnamefont {L.~S.}\ \bibnamefont {Phillips}},\
  and\ \bibinfo {author} {\bibfnamefont {S.~M.}\ \bibnamefont {Barnett}},\
  }\href@noop {} {\bibfield  {journal} {\bibinfo  {journal} {Physical review
  letters}\ }\textbf {\bibinfo {volume} {81}},\ \bibinfo {pages} {1604}
  (\bibinfo {year} {1998})}\BibitemShut {NoStop}%
\bibitem [{\citenamefont
  {Fiur{\'a}{\v{s}}ek}(2009)}]{fiuravsek2009engineering}%
  \BibitemOpen
  \bibfield  {author} {\bibinfo {author} {\bibfnamefont {J.}~\bibnamefont
  {Fiur{\'a}{\v{s}}ek}},\ }\href@noop {} {\bibfield  {journal} {\bibinfo
  {journal} {Physical Review A}\ }\textbf {\bibinfo {volume} {80}},\ \bibinfo
  {pages} {053822} (\bibinfo {year} {2009})}\BibitemShut {NoStop}%
\bibitem [{\citenamefont {Zavatta}\ \emph {et~al.}(2011)\citenamefont
  {Zavatta}, \citenamefont {Fiur{\'a}{\v{s}}ek},\ and\ \citenamefont
  {Bellini}}]{zavatta2011high}%
  \BibitemOpen
  \bibfield  {author} {\bibinfo {author} {\bibfnamefont {A.}~\bibnamefont
  {Zavatta}}, \bibinfo {author} {\bibfnamefont {J.}~\bibnamefont
  {Fiur{\'a}{\v{s}}ek}},\ and\ \bibinfo {author} {\bibfnamefont
  {M.}~\bibnamefont {Bellini}},\ }\href@noop {} {\bibfield  {journal} {\bibinfo
   {journal} {Nature photonics}\ }\textbf {\bibinfo {volume} {5}},\ \bibinfo
  {pages} {52} (\bibinfo {year} {2011})}\BibitemShut {NoStop}%
\bibitem [{\citenamefont {Jeffers}(2010)}]{jeffers2010nondeterministic}%
  \BibitemOpen
  \bibfield  {author} {\bibinfo {author} {\bibfnamefont {J.}~\bibnamefont
  {Jeffers}},\ }\href@noop {} {\bibfield  {journal} {\bibinfo  {journal}
  {Physical Review A}\ }\textbf {\bibinfo {volume} {82}},\ \bibinfo {pages}
  {063828} (\bibinfo {year} {2010})}\BibitemShut {NoStop}%
\bibitem [{\citenamefont {Winnel}\ \emph {et~al.}(2020)\citenamefont {Winnel},
  \citenamefont {Hosseinidehaj},\ and\ \citenamefont
  {Ralph}}]{winnel2020generalized}%
  \BibitemOpen
  \bibfield  {author} {\bibinfo {author} {\bibfnamefont {M.~S.}\ \bibnamefont
  {Winnel}}, \bibinfo {author} {\bibfnamefont {N.}~\bibnamefont
  {Hosseinidehaj}},\ and\ \bibinfo {author} {\bibfnamefont {T.~C.}\
  \bibnamefont {Ralph}},\ }\href@noop {} {\bibfield  {journal} {\bibinfo
  {journal} {Physical Review A}\ }\textbf {\bibinfo {volume} {102}},\ \bibinfo
  {pages} {063715} (\bibinfo {year} {2020})}\BibitemShut {NoStop}%
\bibitem [{Note1()}]{Note1}%
  \BibitemOpen
  \bibinfo {note} {See Supplemental Material at [URL will be inserted by
  publisher] for the details about the technical proofs for the $n$-QS
  operations, probability of success with improvements, fidelity of output
  state, comparisons with other NLA protocols, {loss-tolerant entanglement
  distillation analysis, and experimental imperfections analysis,} which
  includes Ref.~\cite
  {scheel2004permanents,gard2015introduction,lund2017quantum,brualdi1991combinatorial,percus2012combinatorial,ralph2009nondeterministic,winnel2020generalized,tserkis2019quantifying,garcia2009reverse,winnel2021overcoming,reddy2019exceeding,lita2008counting,marsili2013detecting,miller2003demonstration,slussarenko2019photonic,su2019conversion,xiang2010heralded,andersen2013high}}\BibitemShut
  {NoStop}%
\bibitem [{\citenamefont {Neergaard-Nielsen}\ \emph {et~al.}(2013)\citenamefont
  {Neergaard-Nielsen}, \citenamefont {Eto}, \citenamefont {Lee}, \citenamefont
  {Jeong},\ and\ \citenamefont {Sasaki}}]{neergaard2013quantum}%
  \BibitemOpen
  \bibfield  {author} {\bibinfo {author} {\bibfnamefont {J.~S.}\ \bibnamefont
  {Neergaard-Nielsen}}, \bibinfo {author} {\bibfnamefont {Y.}~\bibnamefont
  {Eto}}, \bibinfo {author} {\bibfnamefont {C.-W.}\ \bibnamefont {Lee}},
  \bibinfo {author} {\bibfnamefont {H.}~\bibnamefont {Jeong}},\ and\ \bibinfo
  {author} {\bibfnamefont {M.}~\bibnamefont {Sasaki}},\ }\href@noop {}
  {\bibfield  {journal} {\bibinfo  {journal} {Nature Photonics}\ }\textbf
  {\bibinfo {volume} {7}},\ \bibinfo {pages} {439} (\bibinfo {year}
  {2013})}\BibitemShut {NoStop}%
\bibitem [{\citenamefont {Su}\ \emph {et~al.}(2017)\citenamefont {Su},
  \citenamefont {Li}, \citenamefont {Rohde}, \citenamefont {Huang},
  \citenamefont {Wang}, \citenamefont {Li}, \citenamefont {Liu}, \citenamefont
  {Dowling}, \citenamefont {Lu},\ and\ \citenamefont
  {Pan}}]{su2017multiphoton}%
  \BibitemOpen
  \bibfield  {author} {\bibinfo {author} {\bibfnamefont {Z.-E.}\ \bibnamefont
  {Su}}, \bibinfo {author} {\bibfnamefont {Y.}~\bibnamefont {Li}}, \bibinfo
  {author} {\bibfnamefont {P.~P.}\ \bibnamefont {Rohde}}, \bibinfo {author}
  {\bibfnamefont {H.-L.}\ \bibnamefont {Huang}}, \bibinfo {author}
  {\bibfnamefont {X.-L.}\ \bibnamefont {Wang}}, \bibinfo {author}
  {\bibfnamefont {L.}~\bibnamefont {Li}}, \bibinfo {author} {\bibfnamefont
  {N.-L.}\ \bibnamefont {Liu}}, \bibinfo {author} {\bibfnamefont {J.~P.}\
  \bibnamefont {Dowling}}, \bibinfo {author} {\bibfnamefont {C.-Y.}\
  \bibnamefont {Lu}},\ and\ \bibinfo {author} {\bibfnamefont {J.-W.}\
  \bibnamefont {Pan}},\ }\href@noop {} {\bibfield  {journal} {\bibinfo
  {journal} {Physical review letters}\ }\textbf {\bibinfo {volume} {119}},\
  \bibinfo {pages} {080502} (\bibinfo {year} {2017})}\BibitemShut {NoStop}%
\bibitem [{\citenamefont {Reck}\ \emph {et~al.}(1994)\citenamefont {Reck},
  \citenamefont {Zeilinger}, \citenamefont {Bernstein},\ and\ \citenamefont
  {Bertani}}]{reck1994experimental}%
  \BibitemOpen
  \bibfield  {author} {\bibinfo {author} {\bibfnamefont {M.}~\bibnamefont
  {Reck}}, \bibinfo {author} {\bibfnamefont {A.}~\bibnamefont {Zeilinger}},
  \bibinfo {author} {\bibfnamefont {H.~J.}\ \bibnamefont {Bernstein}},\ and\
  \bibinfo {author} {\bibfnamefont {P.}~\bibnamefont {Bertani}},\ }\href@noop
  {} {\bibfield  {journal} {\bibinfo  {journal} {Physical Review Letters}\
  }\textbf {\bibinfo {volume} {73}},\ \bibinfo {pages} {58} (\bibinfo {year}
  {1994})}\BibitemShut {NoStop}%
\bibitem [{\citenamefont {Clements}\ \emph {et~al.}(2016)\citenamefont
  {Clements}, \citenamefont {Humphreys}, \citenamefont {Metcalf}, \citenamefont
  {Kolthammer},\ and\ \citenamefont {Walmsley}}]{clements2016optimal}%
  \BibitemOpen
  \bibfield  {author} {\bibinfo {author} {\bibfnamefont {W.~R.}\ \bibnamefont
  {Clements}}, \bibinfo {author} {\bibfnamefont {P.~C.}\ \bibnamefont
  {Humphreys}}, \bibinfo {author} {\bibfnamefont {B.~J.}\ \bibnamefont
  {Metcalf}}, \bibinfo {author} {\bibfnamefont {W.~S.}\ \bibnamefont
  {Kolthammer}},\ and\ \bibinfo {author} {\bibfnamefont {I.~A.}\ \bibnamefont
  {Walmsley}},\ }\href@noop {} {\bibfield  {journal} {\bibinfo  {journal}
  {Optica}\ }\textbf {\bibinfo {volume} {3}},\ \bibinfo {pages} {1460}
  (\bibinfo {year} {2016})}\BibitemShut {NoStop}%
\bibitem [{\citenamefont {Kimble}(2008)}]{kimble2008quantum}%
  \BibitemOpen
  \bibfield  {author} {\bibinfo {author} {\bibfnamefont {H.~J.}\ \bibnamefont
  {Kimble}},\ }\href@noop {} {\bibfield  {journal} {\bibinfo  {journal}
  {Nature}\ }\textbf {\bibinfo {volume} {453}},\ \bibinfo {pages} {1023}
  (\bibinfo {year} {2008})}\BibitemShut {NoStop}%
\bibitem [{\citenamefont {Gottesman}\ and\ \citenamefont
  {Chuang}(1999)}]{gottesman1999demonstrating}%
  \BibitemOpen
  \bibfield  {author} {\bibinfo {author} {\bibfnamefont {D.}~\bibnamefont
  {Gottesman}}\ and\ \bibinfo {author} {\bibfnamefont {I.~L.}\ \bibnamefont
  {Chuang}},\ }\href@noop {} {\bibfield  {journal} {\bibinfo  {journal}
  {Nature}\ }\textbf {\bibinfo {volume} {402}},\ \bibinfo {pages} {390}
  (\bibinfo {year} {1999})}\BibitemShut {NoStop}%
\bibitem [{\citenamefont {Chen}\ \emph {et~al.}(2021)\citenamefont {Chen},
  \citenamefont {Li}, \citenamefont {Liu}, \citenamefont {Wu}, \citenamefont
  {Su}, \citenamefont {Wang}, \citenamefont {Li}, \citenamefont {Liu},
  \citenamefont {Lu},\ and\ \citenamefont {Pan}}]{chen2021directly}%
  \BibitemOpen
  \bibfield  {author} {\bibinfo {author} {\bibfnamefont {M.-C.}\ \bibnamefont
  {Chen}}, \bibinfo {author} {\bibfnamefont {Y.}~\bibnamefont {Li}}, \bibinfo
  {author} {\bibfnamefont {R.-Z.}\ \bibnamefont {Liu}}, \bibinfo {author}
  {\bibfnamefont {D.}~\bibnamefont {Wu}}, \bibinfo {author} {\bibfnamefont
  {Z.-E.}\ \bibnamefont {Su}}, \bibinfo {author} {\bibfnamefont {X.-L.}\
  \bibnamefont {Wang}}, \bibinfo {author} {\bibfnamefont {L.}~\bibnamefont
  {Li}}, \bibinfo {author} {\bibfnamefont {N.-L.}\ \bibnamefont {Liu}},
  \bibinfo {author} {\bibfnamefont {C.-Y.}\ \bibnamefont {Lu}},\ and\ \bibinfo
  {author} {\bibfnamefont {J.-W.}\ \bibnamefont {Pan}},\ }\href@noop {}
  {\bibfield  {journal} {\bibinfo  {journal} {Physical Review Letters}\
  }\textbf {\bibinfo {volume} {127}},\ \bibinfo {pages} {030402} (\bibinfo
  {year} {2021})}\BibitemShut {NoStop}%
\bibitem [{\citenamefont {Braunstein}\ and\ \citenamefont
  {Kimble}(1998)}]{braunstein1998teleportation}%
  \BibitemOpen
  \bibfield  {author} {\bibinfo {author} {\bibfnamefont {S.~L.}\ \bibnamefont
  {Braunstein}}\ and\ \bibinfo {author} {\bibfnamefont {H.~J.}\ \bibnamefont
  {Kimble}},\ }\href@noop {} {\bibfield  {journal} {\bibinfo  {journal}
  {Physical Review Letters}\ }\textbf {\bibinfo {volume} {80}},\ \bibinfo
  {pages} {869} (\bibinfo {year} {1998})}\BibitemShut {NoStop}%
\bibitem [{\citenamefont {Andersen}\ and\ \citenamefont
  {Ralph}(2013)}]{andersen2013high}%
  \BibitemOpen
  \bibfield  {author} {\bibinfo {author} {\bibfnamefont {U.~L.}\ \bibnamefont
  {Andersen}}\ and\ \bibinfo {author} {\bibfnamefont {T.~C.}\ \bibnamefont
  {Ralph}},\ }\href@noop {} {\bibfield  {journal} {\bibinfo  {journal}
  {Physical review letters}\ }\textbf {\bibinfo {volume} {111}},\ \bibinfo
  {pages} {050504} (\bibinfo {year} {2013})}\BibitemShut {NoStop}%
\bibitem [{\citenamefont {Winnel}\ \emph {et~al.}(2021)\citenamefont {Winnel},
  \citenamefont {Guanzon}, \citenamefont {Hosseinidehaj},\ and\ \citenamefont
  {Ralph}}]{winnel2021overcoming}%
  \BibitemOpen
  \bibfield  {author} {\bibinfo {author} {\bibfnamefont {M.~S.}\ \bibnamefont
  {Winnel}}, \bibinfo {author} {\bibfnamefont {J.~J.}\ \bibnamefont {Guanzon}},
  \bibinfo {author} {\bibfnamefont {N.}~\bibnamefont {Hosseinidehaj}},\ and\
  \bibinfo {author} {\bibfnamefont {T.~C.}\ \bibnamefont {Ralph}},\ }\href@noop
  {} {\bibfield  {journal} {\bibinfo  {journal} {arXiv preprint
  arXiv:2105.03586}\ } (\bibinfo {year} {2021})}\BibitemShut {NoStop}%
\bibitem [{\citenamefont {Lucamarini}\ \emph {et~al.}(2018)\citenamefont
  {Lucamarini}, \citenamefont {Yuan}, \citenamefont {Dynes},\ and\
  \citenamefont {Shields}}]{lucamarini2018overcoming}%
  \BibitemOpen
  \bibfield  {author} {\bibinfo {author} {\bibfnamefont {M.}~\bibnamefont
  {Lucamarini}}, \bibinfo {author} {\bibfnamefont {Z.~L.}\ \bibnamefont
  {Yuan}}, \bibinfo {author} {\bibfnamefont {J.~F.}\ \bibnamefont {Dynes}},\
  and\ \bibinfo {author} {\bibfnamefont {A.~J.}\ \bibnamefont {Shields}},\
  }\href@noop {} {\bibfield  {journal} {\bibinfo  {journal} {Nature}\ }\textbf
  {\bibinfo {volume} {557}},\ \bibinfo {pages} {400} (\bibinfo {year}
  {2018})}\BibitemShut {NoStop}%
\bibitem [{\citenamefont {Liu}\ \emph {et~al.}(2019)\citenamefont {Liu},
  \citenamefont {Yu}, \citenamefont {Zhang}, \citenamefont {Guan},
  \citenamefont {Chen}, \citenamefont {Zhang}, \citenamefont {Hu},
  \citenamefont {Li}, \citenamefont {Jiang}, \citenamefont {Lin}, \citenamefont
  {Chen}, \citenamefont {You}, \citenamefont {Wang}, \citenamefont {Wang},
  \citenamefont {Zhang},\ and\ \citenamefont {Pan}}]{liu2019experimental}%
  \BibitemOpen
  \bibfield  {author} {\bibinfo {author} {\bibfnamefont {Y.}~\bibnamefont
  {Liu}}, \bibinfo {author} {\bibfnamefont {Z.-W.}\ \bibnamefont {Yu}},
  \bibinfo {author} {\bibfnamefont {W.}~\bibnamefont {Zhang}}, \bibinfo
  {author} {\bibfnamefont {J.-Y.}\ \bibnamefont {Guan}}, \bibinfo {author}
  {\bibfnamefont {J.-P.}\ \bibnamefont {Chen}}, \bibinfo {author}
  {\bibfnamefont {C.}~\bibnamefont {Zhang}}, \bibinfo {author} {\bibfnamefont
  {X.-L.}\ \bibnamefont {Hu}}, \bibinfo {author} {\bibfnamefont
  {H.}~\bibnamefont {Li}}, \bibinfo {author} {\bibfnamefont {C.}~\bibnamefont
  {Jiang}}, \bibinfo {author} {\bibfnamefont {J.}~\bibnamefont {Lin}}, \bibinfo
  {author} {\bibfnamefont {T.-Y.}\ \bibnamefont {Chen}}, \bibinfo {author}
  {\bibfnamefont {L.}~\bibnamefont {You}}, \bibinfo {author} {\bibfnamefont
  {Z.}~\bibnamefont {Wang}}, \bibinfo {author} {\bibfnamefont {X.-B.}\
  \bibnamefont {Wang}}, \bibinfo {author} {\bibfnamefont {Q.}~\bibnamefont
  {Zhang}},\ and\ \bibinfo {author} {\bibfnamefont {J.-W.}\ \bibnamefont
  {Pan}},\ }\href@noop {} {\bibfield  {journal} {\bibinfo  {journal} {Physical
  Review Letters}\ }\textbf {\bibinfo {volume} {123}},\ \bibinfo {pages}
  {100505} (\bibinfo {year} {2019})}\BibitemShut {NoStop}%
\bibitem [{\citenamefont {Curty}\ \emph {et~al.}(2019)\citenamefont {Curty},
  \citenamefont {Azuma},\ and\ \citenamefont {Lo}}]{curty2019simple}%
  \BibitemOpen
  \bibfield  {author} {\bibinfo {author} {\bibfnamefont {M.}~\bibnamefont
  {Curty}}, \bibinfo {author} {\bibfnamefont {K.}~\bibnamefont {Azuma}},\ and\
  \bibinfo {author} {\bibfnamefont {H.-K.}\ \bibnamefont {Lo}},\ }\href@noop {}
  {\bibfield  {journal} {\bibinfo  {journal} {npj Quantum Information}\
  }\textbf {\bibinfo {volume} {5}},\ \bibinfo {pages} {1} (\bibinfo {year}
  {2019})}\BibitemShut {NoStop}%
\bibitem [{\citenamefont {Chen}\ \emph {et~al.}(2020)\citenamefont {Chen},
  \citenamefont {Zhang}, \citenamefont {Liu}, \citenamefont {Jiang},
  \citenamefont {Zhang}, \citenamefont {Hu}, \citenamefont {Guan},
  \citenamefont {Yu}, \citenamefont {Xu}, \citenamefont {Lin}, \citenamefont
  {Li}, \citenamefont {Chen}, \citenamefont {Li}, \citenamefont {You},
  \citenamefont {Wang}, \citenamefont {Wang}, \citenamefont {Zhang},\ and\
  \citenamefont {Pan}}]{chen2020sending}%
  \BibitemOpen
  \bibfield  {author} {\bibinfo {author} {\bibfnamefont {J.-P.}\ \bibnamefont
  {Chen}}, \bibinfo {author} {\bibfnamefont {C.}~\bibnamefont {Zhang}},
  \bibinfo {author} {\bibfnamefont {Y.}~\bibnamefont {Liu}}, \bibinfo {author}
  {\bibfnamefont {C.}~\bibnamefont {Jiang}}, \bibinfo {author} {\bibfnamefont
  {W.}~\bibnamefont {Zhang}}, \bibinfo {author} {\bibfnamefont {X.-L.}\
  \bibnamefont {Hu}}, \bibinfo {author} {\bibfnamefont {J.-Y.}\ \bibnamefont
  {Guan}}, \bibinfo {author} {\bibfnamefont {Z.-W.}\ \bibnamefont {Yu}},
  \bibinfo {author} {\bibfnamefont {H.}~\bibnamefont {Xu}}, \bibinfo {author}
  {\bibfnamefont {J.}~\bibnamefont {Lin}}, \bibinfo {author} {\bibfnamefont
  {M.-J.}\ \bibnamefont {Li}}, \bibinfo {author} {\bibfnamefont
  {H.}~\bibnamefont {Chen}}, \bibinfo {author} {\bibfnamefont {H.}~\bibnamefont
  {Li}}, \bibinfo {author} {\bibfnamefont {L.}~\bibnamefont {You}}, \bibinfo
  {author} {\bibfnamefont {Z.}~\bibnamefont {Wang}}, \bibinfo {author}
  {\bibfnamefont {X.-B.}\ \bibnamefont {Wang}}, \bibinfo {author}
  {\bibfnamefont {Q.}~\bibnamefont {Zhang}},\ and\ \bibinfo {author}
  {\bibfnamefont {J.-W.}\ \bibnamefont {Pan}},\ }\href@noop {} {\bibfield
  {journal} {\bibinfo  {journal} {Physical review letters}\ }\textbf {\bibinfo
  {volume} {124}},\ \bibinfo {pages} {070501} (\bibinfo {year}
  {2020})}\BibitemShut {NoStop}%
\bibitem [{\citenamefont {Audenaert}\ \emph {et~al.}(2003)\citenamefont
  {Audenaert}, \citenamefont {Plenio},\ and\ \citenamefont
  {Eisert}}]{audenaert2003entanglement}%
  \BibitemOpen
  \bibfield  {author} {\bibinfo {author} {\bibfnamefont {K.}~\bibnamefont
  {Audenaert}}, \bibinfo {author} {\bibfnamefont {M.~B.}\ \bibnamefont
  {Plenio}},\ and\ \bibinfo {author} {\bibfnamefont {J.}~\bibnamefont
  {Eisert}},\ }\href@noop {} {\bibfield  {journal} {\bibinfo  {journal}
  {Physical review letters}\ }\textbf {\bibinfo {volume} {90}},\ \bibinfo
  {pages} {027901} (\bibinfo {year} {2003})}\BibitemShut {NoStop}%
\bibitem [{\citenamefont {Plenio}(2005)}]{plenio2005logarithmic}%
  \BibitemOpen
  \bibfield  {author} {\bibinfo {author} {\bibfnamefont {M.~B.}\ \bibnamefont
  {Plenio}},\ }\href@noop {} {\bibfield  {journal} {\bibinfo  {journal}
  {Physical review letters}\ }\textbf {\bibinfo {volume} {95}},\ \bibinfo
  {pages} {090503} (\bibinfo {year} {2005})}\BibitemShut {NoStop}%
\bibitem [{\citenamefont {Vidal}\ and\ \citenamefont
  {Werner}(2002)}]{vidal2002computable}%
  \BibitemOpen
  \bibfield  {author} {\bibinfo {author} {\bibfnamefont {G.}~\bibnamefont
  {Vidal}}\ and\ \bibinfo {author} {\bibfnamefont {R.~F.}\ \bibnamefont
  {Werner}},\ }\href@noop {} {\bibfield  {journal} {\bibinfo  {journal}
  {Physical Review A}\ }\textbf {\bibinfo {volume} {65}},\ \bibinfo {pages}
  {032314} (\bibinfo {year} {2002})}\BibitemShut {NoStop}%
\bibitem [{\citenamefont {Adesso}\ \emph {et~al.}(2004)\citenamefont {Adesso},
  \citenamefont {Serafini},\ and\ \citenamefont
  {Illuminati}}]{adesso2004extremal}%
  \BibitemOpen
  \bibfield  {author} {\bibinfo {author} {\bibfnamefont {G.}~\bibnamefont
  {Adesso}}, \bibinfo {author} {\bibfnamefont {A.}~\bibnamefont {Serafini}},\
  and\ \bibinfo {author} {\bibfnamefont {F.}~\bibnamefont {Illuminati}},\
  }\href@noop {} {\bibfield  {journal} {\bibinfo  {journal} {Physical Review
  A}\ }\textbf {\bibinfo {volume} {70}},\ \bibinfo {pages} {022318} (\bibinfo
  {year} {2004})}\BibitemShut {NoStop}%
\bibitem [{\citenamefont {Bennett}\ \emph {et~al.}(1996)\citenamefont
  {Bennett}, \citenamefont {DiVincenzo}, \citenamefont {Smolin},\ and\
  \citenamefont {Wootters}}]{bennett1996mixed}%
  \BibitemOpen
  \bibfield  {author} {\bibinfo {author} {\bibfnamefont {C.~H.}\ \bibnamefont
  {Bennett}}, \bibinfo {author} {\bibfnamefont {D.~P.}\ \bibnamefont
  {DiVincenzo}}, \bibinfo {author} {\bibfnamefont {J.~A.}\ \bibnamefont
  {Smolin}},\ and\ \bibinfo {author} {\bibfnamefont {W.~K.}\ \bibnamefont
  {Wootters}},\ }\href@noop {} {\bibfield  {journal} {\bibinfo  {journal}
  {Physical Review A}\ }\textbf {\bibinfo {volume} {54}},\ \bibinfo {pages}
  {3824} (\bibinfo {year} {1996})}\BibitemShut {NoStop}%
\bibitem [{\citenamefont {Wolf}\ \emph {et~al.}(2004)\citenamefont {Wolf},
  \citenamefont {Giedke}, \citenamefont {Kr{\"u}ger}, \citenamefont {Werner},\
  and\ \citenamefont {Cirac}}]{wolf2004gaussian}%
  \BibitemOpen
  \bibfield  {author} {\bibinfo {author} {\bibfnamefont {M.~M.}\ \bibnamefont
  {Wolf}}, \bibinfo {author} {\bibfnamefont {G.}~\bibnamefont {Giedke}},
  \bibinfo {author} {\bibfnamefont {O.}~\bibnamefont {Kr{\"u}ger}}, \bibinfo
  {author} {\bibfnamefont {R.~F.}\ \bibnamefont {Werner}},\ and\ \bibinfo
  {author} {\bibfnamefont {J.~I.}\ \bibnamefont {Cirac}},\ }\href@noop {}
  {\bibfield  {journal} {\bibinfo  {journal} {Physical Review A}\ }\textbf
  {\bibinfo {volume} {69}},\ \bibinfo {pages} {052320} (\bibinfo {year}
  {2004})}\BibitemShut {NoStop}%
\bibitem [{\citenamefont {Marian}\ and\ \citenamefont
  {Marian}(2008)}]{marian2008entanglement}%
  \BibitemOpen
  \bibfield  {author} {\bibinfo {author} {\bibfnamefont {P.}~\bibnamefont
  {Marian}}\ and\ \bibinfo {author} {\bibfnamefont {T.~A.}\ \bibnamefont
  {Marian}},\ }\href@noop {} {\bibfield  {journal} {\bibinfo  {journal}
  {Physical review letters}\ }\textbf {\bibinfo {volume} {101}},\ \bibinfo
  {pages} {220403} (\bibinfo {year} {2008})}\BibitemShut {NoStop}%
\bibitem [{\citenamefont {Tserkis}\ and\ \citenamefont
  {Ralph}(2017)}]{tserkis2017quantifying}%
  \BibitemOpen
  \bibfield  {author} {\bibinfo {author} {\bibfnamefont {S.}~\bibnamefont
  {Tserkis}}\ and\ \bibinfo {author} {\bibfnamefont {T.~C.}\ \bibnamefont
  {Ralph}},\ }\href@noop {} {\bibfield  {journal} {\bibinfo  {journal}
  {Physical Review A}\ }\textbf {\bibinfo {volume} {96}},\ \bibinfo {pages}
  {062338} (\bibinfo {year} {2017})}\BibitemShut {NoStop}%
\bibitem [{\citenamefont {Tserkis}\ \emph {et~al.}(2019)\citenamefont
  {Tserkis}, \citenamefont {Onoe},\ and\ \citenamefont
  {Ralph}}]{tserkis2019quantifying}%
  \BibitemOpen
  \bibfield  {author} {\bibinfo {author} {\bibfnamefont {S.}~\bibnamefont
  {Tserkis}}, \bibinfo {author} {\bibfnamefont {S.}~\bibnamefont {Onoe}},\ and\
  \bibinfo {author} {\bibfnamefont {T.~C.}\ \bibnamefont {Ralph}},\ }\href@noop
  {} {\bibfield  {journal} {\bibinfo  {journal} {Physical Review A}\ }\textbf
  {\bibinfo {volume} {99}},\ \bibinfo {pages} {052337} (\bibinfo {year}
  {2019})}\BibitemShut {NoStop}%
\bibitem [{\citenamefont {Onoe}\ \emph {et~al.}(2020)\citenamefont {Onoe},
  \citenamefont {Tserkis}, \citenamefont {Lund},\ and\ \citenamefont
  {Ralph}}]{onoe2020multipartite}%
  \BibitemOpen
  \bibfield  {author} {\bibinfo {author} {\bibfnamefont {S.}~\bibnamefont
  {Onoe}}, \bibinfo {author} {\bibfnamefont {S.}~\bibnamefont {Tserkis}},
  \bibinfo {author} {\bibfnamefont {A.~P.}\ \bibnamefont {Lund}},\ and\
  \bibinfo {author} {\bibfnamefont {T.~C.}\ \bibnamefont {Ralph}},\ }\href@noop
  {} {\bibfield  {journal} {\bibinfo  {journal} {Physical Review A}\ }\textbf
  {\bibinfo {volume} {102}},\ \bibinfo {pages} {042408} (\bibinfo {year}
  {2020})}\BibitemShut {NoStop}%
\bibitem [{\citenamefont
  {Fiur{\'a}{\v{s}}ek}(2021)}]{fiuravsek2021teleportation}%
  \BibitemOpen
  \bibfield  {author} {\bibinfo {author} {\bibfnamefont {J.}~\bibnamefont
  {Fiur{\'a}{\v{s}}ek}},\ }\href@noop {} {\bibfield  {journal} {\bibinfo
  {journal} {arXiv preprint arXiv:2110.06040}\ } (\bibinfo {year}
  {2021})}\BibitemShut {NoStop}%
\bibitem [{\citenamefont {Scheel}(2004)}]{scheel2004permanents}%
  \BibitemOpen
  \bibfield  {author} {\bibinfo {author} {\bibfnamefont {S.}~\bibnamefont
  {Scheel}},\ }\href@noop {} {\bibfield  {journal} {\bibinfo  {journal} {arXiv
  preprint quant-ph/0406127}\ } (\bibinfo {year} {2004})}\BibitemShut {NoStop}%
\bibitem [{\citenamefont {Gard}\ \emph {et~al.}(2015)\citenamefont {Gard},
  \citenamefont {Motes}, \citenamefont {Olson}, \citenamefont {Rohde},\ and\
  \citenamefont {Dowling}}]{gard2015introduction}%
  \BibitemOpen
  \bibfield  {author} {\bibinfo {author} {\bibfnamefont {B.~T.}\ \bibnamefont
  {Gard}}, \bibinfo {author} {\bibfnamefont {K.~R.}\ \bibnamefont {Motes}},
  \bibinfo {author} {\bibfnamefont {J.~P.}\ \bibnamefont {Olson}}, \bibinfo
  {author} {\bibfnamefont {P.~P.}\ \bibnamefont {Rohde}},\ and\ \bibinfo
  {author} {\bibfnamefont {J.~P.}\ \bibnamefont {Dowling}},\ }in\ \href@noop {}
  {\emph {\bibinfo {booktitle} {From atomic to mesoscale: The role of quantum
  coherence in systems of various complexities}}}\ (\bibinfo  {publisher}
  {World Scientific},\ \bibinfo {year} {2015})\ pp.\ \bibinfo {pages}
  {167--192}\BibitemShut {NoStop}%
\bibitem [{\citenamefont {Lund}\ \emph {et~al.}(2017)\citenamefont {Lund},
  \citenamefont {Bremner},\ and\ \citenamefont {Ralph}}]{lund2017quantum}%
  \BibitemOpen
  \bibfield  {author} {\bibinfo {author} {\bibfnamefont {A.~P.}\ \bibnamefont
  {Lund}}, \bibinfo {author} {\bibfnamefont {M.~J.}\ \bibnamefont {Bremner}},\
  and\ \bibinfo {author} {\bibfnamefont {T.~C.}\ \bibnamefont {Ralph}},\
  }\href@noop {} {\bibfield  {journal} {\bibinfo  {journal} {npj Quantum
  Information}\ }\textbf {\bibinfo {volume} {3}},\ \bibinfo {pages} {1}
  (\bibinfo {year} {2017})}\BibitemShut {NoStop}%
\bibitem [{\citenamefont {Brualdi}\ \emph {et~al.}(1991)\citenamefont
  {Brualdi}, \citenamefont {Ryser} \emph {et~al.}}]{brualdi1991combinatorial}%
  \BibitemOpen
  \bibfield  {author} {\bibinfo {author} {\bibfnamefont {R.~A.}\ \bibnamefont
  {Brualdi}}, \bibinfo {author} {\bibfnamefont {H.~J.}\ \bibnamefont {Ryser}},
  \emph {et~al.},\ }\href@noop {} {\emph {\bibinfo {title} {Combinatorial
  matrix theory}}},\ Vol.~\bibinfo {volume} {39}\ (\bibinfo  {publisher}
  {Springer},\ \bibinfo {year} {1991})\BibitemShut {NoStop}%
\bibitem [{\citenamefont {Percus}(2012)}]{percus2012combinatorial}%
  \BibitemOpen
  \bibfield  {author} {\bibinfo {author} {\bibfnamefont {J.~K.}\ \bibnamefont
  {Percus}},\ }\href@noop {} {\emph {\bibinfo {title} {Combinatorial
  methods}}},\ Vol.~\bibinfo {volume} {4}\ (\bibinfo  {publisher} {Springer
  Science \& Business Media},\ \bibinfo {year} {2012})\BibitemShut {NoStop}%
\bibitem [{\citenamefont {Garc{\'\i}a-Patr{\'o}n}\ \emph
  {et~al.}(2009)\citenamefont {Garc{\'\i}a-Patr{\'o}n}, \citenamefont
  {Pirandola}, \citenamefont {Lloyd},\ and\ \citenamefont
  {Shapiro}}]{garcia2009reverse}%
  \BibitemOpen
  \bibfield  {author} {\bibinfo {author} {\bibfnamefont {R.}~\bibnamefont
  {Garc{\'\i}a-Patr{\'o}n}}, \bibinfo {author} {\bibfnamefont {S.}~\bibnamefont
  {Pirandola}}, \bibinfo {author} {\bibfnamefont {S.}~\bibnamefont {Lloyd}},\
  and\ \bibinfo {author} {\bibfnamefont {J.~H.}\ \bibnamefont {Shapiro}},\
  }\href@noop {} {\bibfield  {journal} {\bibinfo  {journal} {Physical review
  letters}\ }\textbf {\bibinfo {volume} {102}},\ \bibinfo {pages} {210501}
  (\bibinfo {year} {2009})}\BibitemShut {NoStop}%
\bibitem [{\citenamefont {Reddy}\ \emph {et~al.}(2019)\citenamefont {Reddy},
  \citenamefont {Nerem}, \citenamefont {Lita}, \citenamefont {Nam},
  \citenamefont {Mirin},\ and\ \citenamefont {Verma}}]{reddy2019exceeding}%
  \BibitemOpen
  \bibfield  {author} {\bibinfo {author} {\bibfnamefont {D.~V.}\ \bibnamefont
  {Reddy}}, \bibinfo {author} {\bibfnamefont {R.~R.}\ \bibnamefont {Nerem}},
  \bibinfo {author} {\bibfnamefont {A.~E.}\ \bibnamefont {Lita}}, \bibinfo
  {author} {\bibfnamefont {S.~W.}\ \bibnamefont {Nam}}, \bibinfo {author}
  {\bibfnamefont {R.~P.}\ \bibnamefont {Mirin}},\ and\ \bibinfo {author}
  {\bibfnamefont {V.~B.}\ \bibnamefont {Verma}},\ }in\ \href@noop {} {\emph
  {\bibinfo {booktitle} {CLEO: QELS\_Fundamental Science}}}\ (\bibinfo
  {organization} {Optical Society of America},\ \bibinfo {year} {2019})\ pp.\
  \bibinfo {pages} {FF1A--3}\BibitemShut {NoStop}%
\bibitem [{\citenamefont {Lita}\ \emph {et~al.}(2008)\citenamefont {Lita},
  \citenamefont {Miller},\ and\ \citenamefont {Nam}}]{lita2008counting}%
  \BibitemOpen
  \bibfield  {author} {\bibinfo {author} {\bibfnamefont {A.~E.}\ \bibnamefont
  {Lita}}, \bibinfo {author} {\bibfnamefont {A.~J.}\ \bibnamefont {Miller}},\
  and\ \bibinfo {author} {\bibfnamefont {S.~W.}\ \bibnamefont {Nam}},\
  }\href@noop {} {\bibfield  {journal} {\bibinfo  {journal} {Optics express}\
  }\textbf {\bibinfo {volume} {16}},\ \bibinfo {pages} {3032} (\bibinfo {year}
  {2008})}\BibitemShut {NoStop}%
\bibitem [{\citenamefont {Marsili}\ \emph {et~al.}(2013)\citenamefont
  {Marsili}, \citenamefont {Verma}, \citenamefont {Stern}, \citenamefont
  {Harrington}, \citenamefont {Lita}, \citenamefont {Gerrits}, \citenamefont
  {Vayshenker}, \citenamefont {Baek}, \citenamefont {Shaw}, \citenamefont
  {Mirin} \emph {et~al.}}]{marsili2013detecting}%
  \BibitemOpen
  \bibfield  {author} {\bibinfo {author} {\bibfnamefont {F.}~\bibnamefont
  {Marsili}}, \bibinfo {author} {\bibfnamefont {V.~B.}\ \bibnamefont {Verma}},
  \bibinfo {author} {\bibfnamefont {J.~A.}\ \bibnamefont {Stern}}, \bibinfo
  {author} {\bibfnamefont {S.}~\bibnamefont {Harrington}}, \bibinfo {author}
  {\bibfnamefont {A.~E.}\ \bibnamefont {Lita}}, \bibinfo {author}
  {\bibfnamefont {T.}~\bibnamefont {Gerrits}}, \bibinfo {author} {\bibfnamefont
  {I.}~\bibnamefont {Vayshenker}}, \bibinfo {author} {\bibfnamefont
  {B.}~\bibnamefont {Baek}}, \bibinfo {author} {\bibfnamefont {M.~D.}\
  \bibnamefont {Shaw}}, \bibinfo {author} {\bibfnamefont {R.~P.}\ \bibnamefont
  {Mirin}}, \emph {et~al.},\ }\href@noop {} {\bibfield  {journal} {\bibinfo
  {journal} {Nature Photonics}\ }\textbf {\bibinfo {volume} {7}},\ \bibinfo
  {pages} {210} (\bibinfo {year} {2013})}\BibitemShut {NoStop}%
\bibitem [{\citenamefont {Miller}\ \emph {et~al.}(2003)\citenamefont {Miller},
  \citenamefont {Nam}, \citenamefont {Martinis},\ and\ \citenamefont
  {Sergienko}}]{miller2003demonstration}%
  \BibitemOpen
  \bibfield  {author} {\bibinfo {author} {\bibfnamefont {A.~J.}\ \bibnamefont
  {Miller}}, \bibinfo {author} {\bibfnamefont {S.~W.}\ \bibnamefont {Nam}},
  \bibinfo {author} {\bibfnamefont {J.~M.}\ \bibnamefont {Martinis}},\ and\
  \bibinfo {author} {\bibfnamefont {A.~V.}\ \bibnamefont {Sergienko}},\
  }\href@noop {} {\bibfield  {journal} {\bibinfo  {journal} {Applied Physics
  Letters}\ }\textbf {\bibinfo {volume} {83}},\ \bibinfo {pages} {791}
  (\bibinfo {year} {2003})}\BibitemShut {NoStop}%
\bibitem [{\citenamefont {Slussarenko}\ and\ \citenamefont
  {Pryde}(2019)}]{slussarenko2019photonic}%
  \BibitemOpen
  \bibfield  {author} {\bibinfo {author} {\bibfnamefont {S.}~\bibnamefont
  {Slussarenko}}\ and\ \bibinfo {author} {\bibfnamefont {G.~J.}\ \bibnamefont
  {Pryde}},\ }\href@noop {} {\bibfield  {journal} {\bibinfo  {journal} {Applied
  Physics Reviews}\ }\textbf {\bibinfo {volume} {6}},\ \bibinfo {pages}
  {041303} (\bibinfo {year} {2019})}\BibitemShut {NoStop}%
\bibitem [{\citenamefont {Su}\ \emph {et~al.}(2019)\citenamefont {Su},
  \citenamefont {Myers},\ and\ \citenamefont {Sabapathy}}]{su2019conversion}%
  \BibitemOpen
  \bibfield  {author} {\bibinfo {author} {\bibfnamefont {D.}~\bibnamefont
  {Su}}, \bibinfo {author} {\bibfnamefont {C.~R.}\ \bibnamefont {Myers}},\ and\
  \bibinfo {author} {\bibfnamefont {K.~K.}\ \bibnamefont {Sabapathy}},\
  }\href@noop {} {\bibfield  {journal} {\bibinfo  {journal} {Physical Review
  A}\ }\textbf {\bibinfo {volume} {100}},\ \bibinfo {pages} {052301} (\bibinfo
  {year} {2019})}\BibitemShut {NoStop}%
\end{thebibliography}%


\begin{thebibliography}{18}%
\makeatletter
\providecommand \@ifxundefined [1]{%
 \@ifx{#1\undefined}
}%
\providecommand \@ifnum [1]{%
 \ifnum #1\expandafter \@firstoftwo
 \else \expandafter \@secondoftwo
 \fi
}%
\providecommand \@ifx [1]{%
 \ifx #1\expandafter \@firstoftwo
 \else \expandafter \@secondoftwo
 \fi
}%
\providecommand \natexlab [1]{#1}%
\providecommand \enquote  [1]{``#1''}%
\providecommand \bibnamefont  [1]{#1}%
\providecommand \bibfnamefont [1]{#1}%
\providecommand \citenamefont [1]{#1}%
\providecommand \href@noop [0]{\@secondoftwo}%
\providecommand \href [0]{\begingroup \@sanitize@url \@href}%
\providecommand \@href[1]{\@@startlink{#1}\@@href}%
\providecommand \@@href[1]{\endgroup#1\@@endlink}%
\providecommand \@sanitize@url [0]{\catcode `\\12\catcode `\$12\catcode
  `\&12\catcode `\#12\catcode `\^12\catcode `\_12\catcode `\%12\relax}%
\providecommand \@@startlink[1]{}%
\providecommand \@@endlink[0]{}%
\providecommand \url  [0]{\begingroup\@sanitize@url \@url }%
\providecommand \@url [1]{\endgroup\@href {#1}{\urlprefix }}%
\providecommand \urlprefix  [0]{URL }%
\providecommand \Eprint [0]{\href }%
\providecommand \doibase [0]{https://doi.org/}%
\providecommand \selectlanguage [0]{\@gobble}%
\providecommand \bibinfo  [0]{\@secondoftwo}%
\providecommand \bibfield  [0]{\@secondoftwo}%
\providecommand \translation [1]{[#1]}%
\providecommand \BibitemOpen [0]{}%
\providecommand \bibitemStop [0]{}%
\providecommand \bibitemNoStop [0]{.\EOS\space}%
\providecommand \EOS [0]{\spacefactor3000\relax}%
\providecommand \BibitemShut  [1]{\csname bibitem#1\endcsname}%
\let\auto@bib@innerbib\@empty
\bibitem [{\citenamefont {Scheel}(2004)}]{scheel2004permanents}%
  \BibitemOpen
  \bibfield  {author} {\bibinfo {author} {\bibfnamefont {S.}~\bibnamefont
  {Scheel}},\ }\href@noop {} {\bibfield  {journal} {\bibinfo  {journal} {arXiv
  preprint quant-ph/0406127}\ } (\bibinfo {year} {2004})}\BibitemShut {NoStop}%
\bibitem [{\citenamefont {Gard}\ \emph {et~al.}(2015)\citenamefont {Gard},
  \citenamefont {Motes}, \citenamefont {Olson}, \citenamefont {Rohde},\ and\
  \citenamefont {Dowling}}]{gard2015introduction}%
  \BibitemOpen
  \bibfield  {author} {\bibinfo {author} {\bibfnamefont {B.~T.}\ \bibnamefont
  {Gard}}, \bibinfo {author} {\bibfnamefont {K.~R.}\ \bibnamefont {Motes}},
  \bibinfo {author} {\bibfnamefont {J.~P.}\ \bibnamefont {Olson}}, \bibinfo
  {author} {\bibfnamefont {P.~P.}\ \bibnamefont {Rohde}},\ and\ \bibinfo
  {author} {\bibfnamefont {J.~P.}\ \bibnamefont {Dowling}},\ }in\ \href@noop {}
  {\emph {\bibinfo {booktitle} {From atomic to mesoscale: The role of quantum
  coherence in systems of various complexities}}}\ (\bibinfo  {publisher}
  {World Scientific},\ \bibinfo {year} {2015})\ pp.\ \bibinfo {pages}
  {167--192}\BibitemShut {NoStop}%
\bibitem [{\citenamefont {Lund}\ \emph {et~al.}(2017)\citenamefont {Lund},
  \citenamefont {Bremner},\ and\ \citenamefont {Ralph}}]{lund2017quantum}%
  \BibitemOpen
  \bibfield  {author} {\bibinfo {author} {\bibfnamefont {A.~P.}\ \bibnamefont
  {Lund}}, \bibinfo {author} {\bibfnamefont {M.~J.}\ \bibnamefont {Bremner}},\
  and\ \bibinfo {author} {\bibfnamefont {T.~C.}\ \bibnamefont {Ralph}},\
  }\href@noop {} {\bibfield  {journal} {\bibinfo  {journal} {npj Quantum
  Information}\ }\textbf {\bibinfo {volume} {3}},\ \bibinfo {pages} {1}
  (\bibinfo {year} {2017})}\BibitemShut {NoStop}%
\bibitem [{\citenamefont {Brualdi}\ \emph {et~al.}(1991)\citenamefont
  {Brualdi}, \citenamefont {Ryser} \emph {et~al.}}]{brualdi1991combinatorial}%
  \BibitemOpen
  \bibfield  {author} {\bibinfo {author} {\bibfnamefont {R.~A.}\ \bibnamefont
  {Brualdi}}, \bibinfo {author} {\bibfnamefont {H.~J.}\ \bibnamefont {Ryser}},
  \emph {et~al.},\ }\href@noop {} {\emph {\bibinfo {title} {Combinatorial
  matrix theory}}},\ Vol.~\bibinfo {volume} {39}\ (\bibinfo  {publisher}
  {Springer},\ \bibinfo {year} {1991})\BibitemShut {NoStop}%
\bibitem [{\citenamefont {Percus}(2012)}]{percus2012combinatorial}%
  \BibitemOpen
  \bibfield  {author} {\bibinfo {author} {\bibfnamefont {J.~K.}\ \bibnamefont
  {Percus}},\ }\href@noop {} {\emph {\bibinfo {title} {Combinatorial
  methods}}},\ Vol.~\bibinfo {volume} {4}\ (\bibinfo  {publisher} {Springer
  Science \& Business Media},\ \bibinfo {year} {2012})\BibitemShut {NoStop}%
\bibitem [{\citenamefont {Su}\ \emph {et~al.}(2019)\citenamefont {Su},
  \citenamefont {Myers},\ and\ \citenamefont {Sabapathy}}]{su2019conversion}%
  \BibitemOpen
  \bibfield  {author} {\bibinfo {author} {\bibfnamefont {D.}~\bibnamefont
  {Su}}, \bibinfo {author} {\bibfnamefont {C.~R.}\ \bibnamefont {Myers}},\ and\
  \bibinfo {author} {\bibfnamefont {K.~K.}\ \bibnamefont {Sabapathy}},\
  }\href@noop {} {\bibfield  {journal} {\bibinfo  {journal} {Physical Review
  A}\ }\textbf {\bibinfo {volume} {100}},\ \bibinfo {pages} {052301} (\bibinfo
  {year} {2019})}\BibitemShut {NoStop}%
\bibitem [{\citenamefont {Winnel}\ \emph {et~al.}(2020)\citenamefont {Winnel},
  \citenamefont {Hosseinidehaj},\ and\ \citenamefont
  {Ralph}}]{winnel2020generalized}%
  \BibitemOpen
  \bibfield  {author} {\bibinfo {author} {\bibfnamefont {M.~S.}\ \bibnamefont
  {Winnel}}, \bibinfo {author} {\bibfnamefont {N.}~\bibnamefont
  {Hosseinidehaj}},\ and\ \bibinfo {author} {\bibfnamefont {T.~C.}\
  \bibnamefont {Ralph}},\ }\href@noop {} {\bibfield  {journal} {\bibinfo
  {journal} {Physical Review A}\ }\textbf {\bibinfo {volume} {102}},\ \bibinfo
  {pages} {063715} (\bibinfo {year} {2020})}\BibitemShut {NoStop}%
\bibitem [{\citenamefont {Ralph}\ and\ \citenamefont
  {Lund}(2009)}]{ralph2009nondeterministic}%
  \BibitemOpen
  \bibfield  {author} {\bibinfo {author} {\bibfnamefont {T.~C.}\ \bibnamefont
  {Ralph}}\ and\ \bibinfo {author} {\bibfnamefont {A.~P.}\ \bibnamefont
  {Lund}},\ }in\ \href@noop {} {\emph {\bibinfo {booktitle} {AIP Conference
  Proceedings}}},\ Vol.\ \bibinfo {volume} {1110}\ (\bibinfo {organization}
  {American Institute of Physics},\ \bibinfo {year} {2009})\ pp.\ \bibinfo
  {pages} {155--160}\BibitemShut {NoStop}%
\bibitem [{\citenamefont {Tserkis}\ \emph {et~al.}(2019)\citenamefont
  {Tserkis}, \citenamefont {Onoe},\ and\ \citenamefont
  {Ralph}}]{tserkis2019quantifying}%
  \BibitemOpen
  \bibfield  {author} {\bibinfo {author} {\bibfnamefont {S.}~\bibnamefont
  {Tserkis}}, \bibinfo {author} {\bibfnamefont {S.}~\bibnamefont {Onoe}},\ and\
  \bibinfo {author} {\bibfnamefont {T.~C.}\ \bibnamefont {Ralph}},\ }\href@noop
  {} {\bibfield  {journal} {\bibinfo  {journal} {Physical Review A}\ }\textbf
  {\bibinfo {volume} {99}},\ \bibinfo {pages} {052337} (\bibinfo {year}
  {2019})}\BibitemShut {NoStop}%
\bibitem [{\citenamefont {Garc{\'\i}a-Patr{\'o}n}\ \emph
  {et~al.}(2009)\citenamefont {Garc{\'\i}a-Patr{\'o}n}, \citenamefont
  {Pirandola}, \citenamefont {Lloyd},\ and\ \citenamefont
  {Shapiro}}]{garcia2009reverse}%
  \BibitemOpen
  \bibfield  {author} {\bibinfo {author} {\bibfnamefont {R.}~\bibnamefont
  {Garc{\'\i}a-Patr{\'o}n}}, \bibinfo {author} {\bibfnamefont {S.}~\bibnamefont
  {Pirandola}}, \bibinfo {author} {\bibfnamefont {S.}~\bibnamefont {Lloyd}},\
  and\ \bibinfo {author} {\bibfnamefont {J.~H.}\ \bibnamefont {Shapiro}},\
  }\href@noop {} {\bibfield  {journal} {\bibinfo  {journal} {Physical review
  letters}\ }\textbf {\bibinfo {volume} {102}},\ \bibinfo {pages} {210501}
  (\bibinfo {year} {2009})}\BibitemShut {NoStop}%
\bibitem [{\citenamefont {Winnel}\ \emph {et~al.}(2021)\citenamefont {Winnel},
  \citenamefont {Guanzon}, \citenamefont {Hosseinidehaj},\ and\ \citenamefont
  {Ralph}}]{winnel2021overcoming}%
  \BibitemOpen
  \bibfield  {author} {\bibinfo {author} {\bibfnamefont {M.~S.}\ \bibnamefont
  {Winnel}}, \bibinfo {author} {\bibfnamefont {J.~J.}\ \bibnamefont {Guanzon}},
  \bibinfo {author} {\bibfnamefont {N.}~\bibnamefont {Hosseinidehaj}},\ and\
  \bibinfo {author} {\bibfnamefont {T.~C.}\ \bibnamefont {Ralph}},\ }\href@noop
  {} {\bibfield  {journal} {\bibinfo  {journal} {arXiv preprint
  arXiv:2105.03586}\ } (\bibinfo {year} {2021})}\BibitemShut {NoStop}%
\bibitem [{\citenamefont {Reddy}\ \emph {et~al.}(2019)\citenamefont {Reddy},
  \citenamefont {Nerem}, \citenamefont {Lita}, \citenamefont {Nam},
  \citenamefont {Mirin},\ and\ \citenamefont {Verma}}]{reddy2019exceeding}%
  \BibitemOpen
  \bibfield  {author} {\bibinfo {author} {\bibfnamefont {D.~V.}\ \bibnamefont
  {Reddy}}, \bibinfo {author} {\bibfnamefont {R.~R.}\ \bibnamefont {Nerem}},
  \bibinfo {author} {\bibfnamefont {A.~E.}\ \bibnamefont {Lita}}, \bibinfo
  {author} {\bibfnamefont {S.~W.}\ \bibnamefont {Nam}}, \bibinfo {author}
  {\bibfnamefont {R.~P.}\ \bibnamefont {Mirin}},\ and\ \bibinfo {author}
  {\bibfnamefont {V.~B.}\ \bibnamefont {Verma}},\ }in\ \href@noop {} {\emph
  {\bibinfo {booktitle} {CLEO: QELS\_Fundamental Science}}}\ (\bibinfo
  {organization} {Optical Society of America},\ \bibinfo {year} {2019})\ pp.\
  \bibinfo {pages} {FF1A--3}\BibitemShut {NoStop}%
\bibitem [{\citenamefont {Lita}\ \emph {et~al.}(2008)\citenamefont {Lita},
  \citenamefont {Miller},\ and\ \citenamefont {Nam}}]{lita2008counting}%
  \BibitemOpen
  \bibfield  {author} {\bibinfo {author} {\bibfnamefont {A.~E.}\ \bibnamefont
  {Lita}}, \bibinfo {author} {\bibfnamefont {A.~J.}\ \bibnamefont {Miller}},\
  and\ \bibinfo {author} {\bibfnamefont {S.~W.}\ \bibnamefont {Nam}},\
  }\href@noop {} {\bibfield  {journal} {\bibinfo  {journal} {Optics express}\
  }\textbf {\bibinfo {volume} {16}},\ \bibinfo {pages} {3032} (\bibinfo {year}
  {2008})}\BibitemShut {NoStop}%
\bibitem [{\citenamefont {Marsili}\ \emph {et~al.}(2013)\citenamefont
  {Marsili}, \citenamefont {Verma}, \citenamefont {Stern}, \citenamefont
  {Harrington}, \citenamefont {Lita}, \citenamefont {Gerrits}, \citenamefont
  {Vayshenker}, \citenamefont {Baek}, \citenamefont {Shaw}, \citenamefont
  {Mirin} \emph {et~al.}}]{marsili2013detecting}%
  \BibitemOpen
  \bibfield  {author} {\bibinfo {author} {\bibfnamefont {F.}~\bibnamefont
  {Marsili}}, \bibinfo {author} {\bibfnamefont {V.~B.}\ \bibnamefont {Verma}},
  \bibinfo {author} {\bibfnamefont {J.~A.}\ \bibnamefont {Stern}}, \bibinfo
  {author} {\bibfnamefont {S.}~\bibnamefont {Harrington}}, \bibinfo {author}
  {\bibfnamefont {A.~E.}\ \bibnamefont {Lita}}, \bibinfo {author}
  {\bibfnamefont {T.}~\bibnamefont {Gerrits}}, \bibinfo {author} {\bibfnamefont
  {I.}~\bibnamefont {Vayshenker}}, \bibinfo {author} {\bibfnamefont
  {B.}~\bibnamefont {Baek}}, \bibinfo {author} {\bibfnamefont {M.~D.}\
  \bibnamefont {Shaw}}, \bibinfo {author} {\bibfnamefont {R.~P.}\ \bibnamefont
  {Mirin}}, \emph {et~al.},\ }\href@noop {} {\bibfield  {journal} {\bibinfo
  {journal} {Nature Photonics}\ }\textbf {\bibinfo {volume} {7}},\ \bibinfo
  {pages} {210} (\bibinfo {year} {2013})}\BibitemShut {NoStop}%
\bibitem [{\citenamefont {Miller}\ \emph {et~al.}(2003)\citenamefont {Miller},
  \citenamefont {Nam}, \citenamefont {Martinis},\ and\ \citenamefont
  {Sergienko}}]{miller2003demonstration}%
  \BibitemOpen
  \bibfield  {author} {\bibinfo {author} {\bibfnamefont {A.~J.}\ \bibnamefont
  {Miller}}, \bibinfo {author} {\bibfnamefont {S.~W.}\ \bibnamefont {Nam}},
  \bibinfo {author} {\bibfnamefont {J.~M.}\ \bibnamefont {Martinis}},\ and\
  \bibinfo {author} {\bibfnamefont {A.~V.}\ \bibnamefont {Sergienko}},\
  }\href@noop {} {\bibfield  {journal} {\bibinfo  {journal} {Applied Physics
  Letters}\ }\textbf {\bibinfo {volume} {83}},\ \bibinfo {pages} {791}
  (\bibinfo {year} {2003})}\BibitemShut {NoStop}%
\bibitem [{\citenamefont {Slussarenko}\ and\ \citenamefont
  {Pryde}(2019)}]{slussarenko2019photonic}%
  \BibitemOpen
  \bibfield  {author} {\bibinfo {author} {\bibfnamefont {S.}~\bibnamefont
  {Slussarenko}}\ and\ \bibinfo {author} {\bibfnamefont {G.~J.}\ \bibnamefont
  {Pryde}},\ }\href@noop {} {\bibfield  {journal} {\bibinfo  {journal} {Applied
  Physics Reviews}\ }\textbf {\bibinfo {volume} {6}},\ \bibinfo {pages}
  {041303} (\bibinfo {year} {2019})}\BibitemShut {NoStop}%
\bibitem [{\citenamefont {Xiang}\ \emph {et~al.}(2010)\citenamefont {Xiang},
  \citenamefont {Ralph}, \citenamefont {Lund}, \citenamefont {Walk},\ and\
  \citenamefont {Pryde}}]{xiang2010heralded}%
  \BibitemOpen
  \bibfield  {author} {\bibinfo {author} {\bibfnamefont {G.-Y.}\ \bibnamefont
  {Xiang}}, \bibinfo {author} {\bibfnamefont {T.~C.}\ \bibnamefont {Ralph}},
  \bibinfo {author} {\bibfnamefont {A.~P.}\ \bibnamefont {Lund}}, \bibinfo
  {author} {\bibfnamefont {N.}~\bibnamefont {Walk}},\ and\ \bibinfo {author}
  {\bibfnamefont {G.~J.}\ \bibnamefont {Pryde}},\ }\href@noop {} {\bibfield
  {journal} {\bibinfo  {journal} {Nature Photonics}\ }\textbf {\bibinfo
  {volume} {4}},\ \bibinfo {pages} {316} (\bibinfo {year} {2010})}\BibitemShut
  {NoStop}%
\bibitem [{\citenamefont {Andersen}\ and\ \citenamefont
  {Ralph}(2013)}]{andersen2013high}%
  \BibitemOpen
  \bibfield  {author} {\bibinfo {author} {\bibfnamefont {U.~L.}\ \bibnamefont
  {Andersen}}\ and\ \bibinfo {author} {\bibfnamefont {T.~C.}\ \bibnamefont
  {Ralph}},\ }\href@noop {} {\bibfield  {journal} {\bibinfo  {journal}
  {Physical review letters}\ }\textbf {\bibinfo {volume} {111}},\ \bibinfo
  {pages} {050504} (\bibinfo {year} {2013})}\BibitemShut {NoStop}%
\end{thebibliography}%

\end{document}


\title{Supplementary Material for: ``Ideal Quantum Tele-amplification up to a Selected Energy Cut-off using Linear Optics''}

\author{Joshua J. Guanzon}
\email{joshua.guanzon@uq.net.au}
\affiliation{Centre for Quantum Computation and Communication Technology, School of Mathematics and Physics, The University of Queensland, St Lucia, Queensland 4072, Australia}

\author{Matthew S. Winnel}
\affiliation{Centre for Quantum Computation and Communication Technology, School of Mathematics and Physics, The University of Queensland, St Lucia, Queensland 4072, Australia}

\author{Austin P. Lund}
\affiliation{Centre for Quantum Computation and Communication Technology, School of Mathematics and Physics, The University of Queensland, St Lucia, Queensland 4072, Australia}
\affiliation{Dahlem Center for Complex Quantum Systems, Freie Universit\"at Berlin, 14195 Berlin, Germany}

\author{Timothy C. Ralph}
\affiliation{Centre for Quantum Computation and Communication Technology, School of Mathematics and Physics, The University of Queensland, St Lucia, Queensland 4072, Australia}

\date{\today}

\maketitle

\section{Proof of n Quantum Scissors with Bunched Photons Resource} 

Here we will prove that the $n$ quantum scissor ($n$-QS), upon successful measurements, implements the following ideal NLA transformation
\begin{align}
    |\psi\rangle &\equiv \sum_{j=0}^{\infty} c_j|j\rangle \xrightarrow{n\text{-QS}} |g\psi_n\rangle = N \sum_{j=0}^{n} g^j c_j|j\rangle, \label{eq:QS}
\end{align}
with a chosen $g\in(0,\infty)$ gain, up to the $n$ Fock state. We will firstly prove this for the version that requires $n$ bunched photons (BP) as the resource input, as shown in Fig.~1(a).

The action of the gain BS, with $\tau = g^2/(g^2+1)\in[0,1]$ transmissivity, is described by the following transformation $a_1^\dagger \rightarrow \sqrt{1-\tau}a_2^\dagger -\sqrt{\tau} a_1^\dagger$; this just describes how the photons are scattered in the system. The BP resource state $|0\rangle|n\rangle$ transforms due to the gain BS $B_2(\tau)$ as 
\begin{align}
    B_2|0\rangle|n\rangle &= \frac{1}{\sqrt{n!}} B_2(a_1^\dagger)^n |0\rangle|0\rangle \nonumber \\ 
    &= \frac{1}{\sqrt{n!}} (\sqrt{1-\tau} a_2^\dagger-\sqrt{\tau} a_1^\dagger)^n|0\rangle|0\rangle \nonumber \\ 
    &= \frac{1}{\sqrt{n!}} \sum_{j=0}^n  \tau^{j/2} (1-\tau)^{(n-j)/2}(-1)^j \binom{n}{j} \nonumber \\
    &\quad \times(a_1^\dagger)^j (a_2^\dagger)^{n-j} |0\rangle|0\rangle \nonumber \\ 
    &= \frac{(1-\tau)^{n/2}}{\sqrt{n!}} \sum_{j=0}^n \left(\frac{\tau}{1-\tau}\right)^{j/2} (-1)^j \binom{n}{j}  \nonumber \\ 
    &\quad \times \sqrt{j!(n-j)!} |j\rangle|n-j\rangle \nonumber \\ 
    &= \frac{1}{(g^2+1)^{n/2}} \sum_{j=0}^n g^j(-1)^j \sqrt{\binom{n}{j}}  |j\rangle|n-j\rangle, \label{eq:n0}
\end{align}
where we used the binomial expansion with $\binom{n}{j} = \frac{n!}{j!(n-j)!}$, and substituting in our transmissivity setting of $g=\sqrt{\frac{\tau}{1-\tau}}$. We can use Eq.~\eqref{eq:n0} to show that the entire input state transforms due to the gain BS as  
\begin{align}
    (B_2\otimes \mathbb{I}_n) &|0\rangle|n\rangle|\psi\rangle \otimes^{n-1} |0\rangle \nonumber \\ 
    &= \sum_{k=0}^\infty c_k (B_2|0\rangle|n\rangle) |k\rangle \otimes^{n-1} |0\rangle \nonumber \\ 
    &= \frac{1}{(g^2+1)^{n/2}} \sum_{j=0}^n \sum_{k=0}^\infty g^j c_k (-1)^j \sqrt{\binom{n}{j}} \nonumber \\ 
    &\quad \times |j\rangle|n-j\rangle |k\rangle \otimes^{n-1} |0\rangle. \label{eq:n0all}
\end{align}

The next step is to funnel the last $n+1$ modes of Eq.~\eqref{eq:n0all} into a quantum Fourier transformation (QFT) $F_{n+1}$, and then detect $n$ single photons $\langle0| \otimes^n \langle 1|$. This results in the following output state  
\begin{align}
    |g\psi_n\rangle &= \frac{1}{(g^2+1)^{n/2}} \sum_{j=0}^n \sum_{k=0}^\infty g^j c_k (-1)^j \sqrt{\binom{n}{j}}  |j\rangle \nonumber \\ 
    &\quad \times \langle 0| \otimes^n \langle 1| F_{n+1} |n-j\rangle|k\rangle \otimes^{n-1} |0\rangle \nonumber \\ 
    &= \frac{1}{(g^2+1)^{n/2}} \sum_{j=0}^n  g^j c_j (-1)^j   \sqrt{\binom{n}{j}} |j\rangle \nonumber \\ 
    &\quad \times \langle 0| \otimes^n \langle 1| F_{n+1} |n-j\rangle|j\rangle \otimes^{n-1} |0\rangle \nonumber \\
    &\stackrel{?}{=} \frac{\sqrt{n!}}{(n+1)^{n/2}} \frac{1}{(g^2+1)^{n/2}} \sum_{j=0}^n g^j c_j |j\rangle, \label{eq:amp} 
\end{align}
where the second equality is due to the unitary and photon number conserving properties of $F_{n+1}$, which means the only non-zero probability amplitudes are the $k=j$ terms. To prove the $n$-QSBP transformation in Eq.~\eqref{eq:QS}, we need to prove the last equality in Eq.~\eqref{eq:amp}. Note that we define $\stackrel{?}{=}$ as the equality that will be proven in subsequent sections. In particular, we will prove that the probability amplitude 
\begin{align}
    \langle 0| \otimes^n \langle 1| F_{n+1} |n-j\rangle|j\rangle &\otimes^{n-1} |0\rangle \nonumber \\ 
    &\stackrel{?}{=} \frac{(-1)^j\sqrt{j!(n-j)!}}{(n+1)^{n/2}}, \label{eq:amp1}
\end{align}
in the next few sections.   
  
\subsection{Probability Amplitudes of \texorpdfstring{$F_{n+1}$}{QFT}}

We note that a linear optical network is a system composed of simple standard optical elements such as beam-splitters (BS) and phase shifts (PS). The action of any linear optical system with $m$ modes is encapsulated by an $m\times m$ unitary matrix $U_m$, which describes how the photons scatter in a linear manner $\vec{a}^\dagger \rightarrow U_m \vec{a}^\dagger$. A QFT with $m$ modes is defined as 
\begin{align}
    (F_m)_{j,k} \equiv \frac{\omega_m^{(j-1)(k-1)}}{\sqrt{m}},\quad \omega_m\equiv e^{-\frac{2i\pi}{m}}.
\end{align}
This means a QFT with $m=n+1$ modes, as needed for the $n$-QS operation, will have the following scattering matrix 
\begin{align}
    F_{n+1} = \frac{1}{\sqrt{n+1}}
        \overbrace{\begin{bmatrix}
            1 & 1 & \cdots & 1 \\ 
            1 & \omega_{n+1} & \cdots & \omega^n_{n+1} \\
            \vdots & \vdots & \ddots & \vdots  \\ 
            1 & \omega^n_{n+1} & \cdots & \omega^{2n}_{n+1} 
        \end{bmatrix} }^{n+1 \text{ columns}} 
        \left. \phantom{\begin{bmatrix}
            1 \\ 
            1 \\
            \vdots \\ 
            1 
        \end{bmatrix}\hspace{-2.3em}} \right\}{\scriptstyle n+1 \text{ rows}}. 
\end{align}
The probability amplitude in Eq.~\eqref{eq:amp1} is related to an $n\times n$ matrix $\Omega_{n,j}$, built from particular components of this $F_{n+1}$ matrix~\cite{scheel2004permanents}. More precisely, $\Omega_{n,j}$ is composed of $n-j$ duplicates of the first column and $j$ duplicates of the second column of $F_{n+1}$, but only the last $n$ rows, as follows 
\begin{align}
    \Omega_{n,j} = \overbrace{\left[ \begin{matrix}
            1 & \cdots & 1 \\
            \vdots & \vdots & \vdots \\ 
            1 & \cdots & 1   
        \end{matrix}\right. }^{n-j \text{ copies}} \overbrace{\left.\begin{matrix}
            \omega_{n+1} & \cdots & \omega_{n+1} \\
            \vdots & \vdots & \vdots \\ 
            \omega_{n+1}^n & \cdots & \omega_{n+1}^n  
        \end{matrix}\right]}^{j \text{ copies}} 
        \left. \phantom{\begin{bmatrix}
            1 \\ 
            \vdots \\ 
            1 
        \end{bmatrix}\hspace{-2.3em}} \right\} {\scriptstyle n \text{ rows}}. \label{eq:omeganj}
\end{align}
\tc{Based on references \cite{scheel2004permanents,gard2015introduction,lund2017quantum}, one can show that this} matrix is related to the probability amplitude as 
\begin{align}
\langle 0| &\otimes^n \langle 1| F_{n+1} |n-j\rangle|j\rangle \otimes^{n-1} |0\rangle  \nonumber \\ 
&= \frac{1}{\sqrt{j!(n-j)!}} \text{Per} 
    \left( \frac{\Omega_{n,j}}{\sqrt{n+1}} \right) \nonumber \\ 
&= \frac{1}{\sqrt{j!(n-j)!}} \frac{1}{(n+1)^{n/2}} \text{Per} 
    \left( \Omega_{n,j} \right), \label{eq:amp2}
\end{align}
where the extra factor $[j!(n-j)!]^{-1/2}$ accounts for the repeated columns~\cite{scheel2004permanents,gard2015introduction,lund2017quantum}. Note we used the matrix permanent function, defined as 
\begin{align}
    \text{Per}(U_m) \equiv \sum_{\sigma\in S_m} \prod^m_{j=1} (U_m)_{j,\sigma(j)}, 
\end{align}
in which the summation is over the symmetric group $S_m$ or all permutations of $\{1,\cdots,m\}$; the permanent is similar to the determinant, but without the sign changes. 

\tc{The first equality of Eq.~\eqref{eq:amp2} takes advantage of the known connection between linear optical networks, and the permanent of a matrix built from the scattering matrix elements~\cite{scheel2004permanents}. However, this connection rose in importance more recently due to Boson Sampling problems (as calculating the permanent is computational difficult). The particular relation in Eq.~\eqref{eq:amp2} is essentially the square root of Eq.~(7) in Ref.~\cite{lund2017quantum}, but for a particular scattering matrix $F_{n+1}$ (note that in comparison to Ref.~\cite{lund2017quantum} we are considering the time reversed or complex conjugate, hence the rows and columns are switched around).}

In the second equality of Eq.~\eqref{eq:amp2}, we used the properties of the permanent in which multiplying any single row or column of $U_m$ by a scalar $s$ changes $\text{Per}(U_m)$ to $s\text{Per}(U_m)$~\cite{brualdi1991combinatorial}. By comparing Eq.~\eqref{eq:amp1} and Eq.~\eqref{eq:amp2}, it's clear we must prove that 
\begin{align}
    \text{Per} 
    \left( \Omega_{n,j} \right) \stackrel{?}{=} (-1)^j j!(n-j)!, \label{eq:amp3}
\end{align}
to verify the probability amplitude. 

\subsection{Matrix Permanents of \texorpdfstring{$\Omega_{n,j}$}{Omega Matrix}}

Let us consider the simplest cases of $j$, and generalise from there. Consider $j=0$ first, then $\Omega_{n,0}$ is just a matrix with 1 for all it's elements $(\Omega_{n,0})_{p,q} = 1$. Hence   
\begin{align}
    \text{Per} \left( \Omega_{n,0} \right) &= \text{Per} \overbrace{\begin{bmatrix}
            1 & \cdots & 1 \\
            \vdots & \vdots & \vdots  \\ 
            1 & \cdots & 1 
        \end{bmatrix} }^{n \text{ columns}} 
        \left. \phantom{\begin{bmatrix}
            1 \\ 
            \vdots \\ 
            1 
        \end{bmatrix}\hspace{-2.2em}} \right\} {\scriptstyle n \text{ rows}} \nonumber \\ 
        &=    \sum_{\sigma\in S_n} \prod^n_{j=1} 1 = \sum_{\sigma\in S_n} 1 = n!, 
\end{align}
which follows because there are $n!$ permutations which make up $S_n$. This result clearly matches Eq.~\eqref{eq:amp3}. 

Next, consider $j=1$, then we want to evalute the permanent of 
\begin{align}
    \Omega_{n,1} = \overbrace{\left[ \begin{matrix}
            1 & \cdots & 1 \\
            \vdots & \vdots & \vdots \\ 
            1 & \cdots & 1   
        \end{matrix}\right. }^{n-1 \text{ copies}} \left.\begin{matrix}
            \omega_{n+1}  \\
            \vdots  \\ 
            \omega_{n+1}^n   
        \end{matrix}\right] \left. \phantom{\begin{bmatrix}
            1 \\ 
            \vdots \\ 
            1 
        \end{bmatrix}\hspace{-2.3em}} \right\}  {\scriptstyle n \text{ rows}}. 
\end{align}
Note that just like the Laplace cofactor expansion for determinants, we can expand a permanent over a particular row or column. We could expand over the last column with the $\omega_{n+1}$ phase factors, however for the sake of generalising to higher $j$ it's easier to think about expanding over the $1$ elements. In other words 
\begin{align}
    \text{Per} \left( \Omega_{n,1} \right) &= \left( \sum_{k=1}^n \omega_{n+1}^k \right) \text{Per} \overbrace{\begin{bmatrix}
            1 & \cdots & 1 \\
            \vdots & \vdots & \vdots  \\ 
            1 & \cdots & 1 
        \end{bmatrix} }^{n-1 \text{ columns}} 
        \left. \phantom{\begin{bmatrix}
            1 \\ 
            \vdots \\ 
            1 
        \end{bmatrix}\hspace{-2.2em}} \right\}  {\scriptstyle n-1 \text{ rows}} \nonumber \\ 
        &= \left( \sum_{k=1}^n \omega_{n+1}^k \right) (n-1)! \nonumber \\ &\stackrel{?}{=} (-1) (n-1)!.
\end{align}
Given that $\sum_{k=1}^n \omega_{n+1}^k \stackrel{?}{=} -1$ is true, this expression matches Eq.~\eqref{eq:amp3}. 

Next, consider $j=2$, then we want to evaluate the permanent of
\begin{align}
    \Omega_{n,2} = \overbrace{\left[ \begin{matrix}
            1 & \cdots & 1 \\
            \vdots & \vdots & \vdots \\ 
            1 & \cdots & 1   
        \end{matrix}\right. }^{n-2 \text{ copies}} \left.\begin{matrix}
            \omega_{n+1} & \omega_{n+1}  \\
            \vdots & \vdots \\ 
            \omega_{n+1}^n & \omega_{n+1}^n   
        \end{matrix}\right] \left. \phantom{\begin{bmatrix}
            1 \\ 
            \vdots \\ 
            1 
        \end{bmatrix}\hspace{-2.3em}} \right\}  {\scriptstyle n \text{ rows}}. 
\end{align}
By expanding the permanent over the \tc{columns with only $1$ elements}, it must be the case that 
\begin{align}
    \text{Per} \left( \Omega_{n,2} \right) &= \sum^n_{k_1=1} \sum^n_{k_2=k_1+1} \text{Per}
    \begin{bmatrix}
        \omega_{n+1}^{k_1} & \omega_{n+1}^{k_1} \\ 
        \omega_{n+1}^{k_2} & \omega_{n+1}^{k_2}
    \end{bmatrix}
    \nonumber \\ 
    &\quad \times \text{Per} \overbrace{\begin{bmatrix}
            1 & \cdots & 1 \\
            \vdots & \vdots & \vdots  \\ 
            1 & \cdots & 1 
        \end{bmatrix} }^{n-2 \text{ columns}} 
        \left. \phantom{\begin{bmatrix}
            1 \\ 
            \vdots \\ 
            1 
        \end{bmatrix}\hspace{-2.2em}} \right\}  {\scriptstyle n-2 \text{ rows}} \nonumber \\ 
    &= \sum^n_{k_1=1} \sum^n_{k_2=k_1+1} \omega_{n+1}^{k_1+k_2}\text{Per}
    \begin{bmatrix}
        1 & 1 \\ 
        1 & 1
    \end{bmatrix}
     (n-2)! \nonumber \\ 
    &= \sum^n_{k_1=1} \sum^n_{k_2=k_1+1} \omega_{n+1}^{k_1+k_2} 2! (n-2)! \nonumber \\
    &\stackrel{?}{=} (-1)^2 2! (n-2)!.
\end{align}
Given that $\left(\sum^n_{k_1=1} \sum^n_{k_2=k_1+1} \omega_{n+1}^{k_1+k_2}
    \right) \stackrel{?}{=} (-1)^2$ is true, this expression also matches Eq.~\eqref{eq:amp3}. 

It is clear by generalisation to arbitrary $j$, by expanding the permanent over the $1$ columns, it must be the case that 
\begin{align}
    \text{Per} &\left( \Omega_{n,j} \right) \nonumber \\ 
    &= \sum^n_{k_1=1} \cdots \sum^n_{k_j=k_{j-1}+1} \text{Per}
    \overbrace{\begin{bmatrix}
            \omega_{n+1}^{k_1} & \cdots & \omega_{n+1}^{k_1} \\ 
            \vdots & \vdots & \vdots \\
            \omega_{n+1}^{k_j} & \cdots & \omega_{n+1}^{k_j} \\
        \end{bmatrix} }^{j \text{ copies}} 
        \left. \phantom{\begin{bmatrix}
            1 \\ 
            \vdots \\ 
            1
        \end{bmatrix}\hspace{-2.2em}} \right\}  {\scriptstyle j \text{ rows}}
    \nonumber \\ 
    &\quad \times \text{Per} \overbrace{\begin{bmatrix}
            1 & \cdots & 1 \\
            \vdots & \vdots & \vdots  \\ 
            1 & \cdots & 1 
        \end{bmatrix} }^{n-j \text{ columns}} 
        \left. \phantom{\begin{bmatrix}
            1 \\ 
            \vdots \\ 
            1 
        \end{bmatrix}\hspace{-2.2em}} \right\}  {\scriptstyle n-j \text{ rows}} \nonumber \\ 
    &= \sum^n_{k_1=1} \cdots \sum^n_{k_j=k_{j-1}+1} \omega_{n+1}^{k_1+\cdots+k_j} \text{Per}
    \overbrace{\begin{bmatrix}
            1 & \cdots & 1 \\ 
            \vdots & \vdots & \vdots \\
            1 & \cdots & 1 \\
        \end{bmatrix} }^{j \text{ copies}} 
        \left. \phantom{\begin{bmatrix}
            1 \\ 
            \vdots \\ 
            1
        \end{bmatrix}\hspace{-2.2em}} \right\}  {\scriptstyle j \text{ rows}}
    \nonumber \\ 
    &\quad \times (n-j)! \nonumber \\ 
    &= \sum^n_{k_1=1} \cdots \sum^n_{k_j=k_{j-1}+1} \omega_{n+1}^{k_1+\cdots+k_j}j! (n-j)! \nonumber \\ 
    &\stackrel{?}{=} (-1)^j j! (n-j)!. \label{eq:amp4}
\end{align}
It is possible to get the same result by letting 
\begin{align}
    \Omega_{n,j} &= A_{n,j} + B_{n,j}, \\ 
     A_{n,j} &= \overbrace{\left[ \begin{matrix}
            1 & \cdots & 1 \\
            \vdots & \vdots & \vdots \\ 
            1 & \cdots & 1   
        \end{matrix}\right. }^{n-j \text{ copies}} \overbrace{\left.\begin{matrix}
            0 & \cdots & 0 \\
            \vdots & \vdots & \vdots \\ 
            0 & \cdots & 0  
        \end{matrix}\right]}^{j \text{ copies}} 
        \left. \phantom{\begin{bmatrix}
            1 \\ 
            \vdots \\ 
            1 
        \end{bmatrix}\hspace{-2.2em}} \right\} {\scriptstyle n \text{ rows}}, \\ 
    B_{n,j} &= \overbrace{\left[ \begin{matrix}
            0 & \cdots & 0 \\
            \vdots & \vdots & \vdots \\ 
            0 & \cdots & 0   
        \end{matrix}\right. }^{n-j \text{ copies}} \overbrace{\left.\begin{matrix}
            \omega_{n+1} & \cdots & \omega_{n+1} \\
            \vdots & \vdots & \vdots \\ 
            \omega_{n+1}^n & \cdots & \omega_{n+1}^n  
        \end{matrix}\right]}^{j \text{ copies}} 
        \left. \phantom{\begin{bmatrix}
            0 \\ 
            \vdots \\ 
            0 
        \end{bmatrix}\hspace{-2.2em}} \right\} {\scriptstyle n \text{ rows}}, 
\end{align}
and then resolving the permanent using following expansion rule 
\begin{align}
    \text{Per}(\Omega_{n,j}) &= \text{Per}(A_{n,j} + B_{n,j}) \nonumber \\ 
    &= \sum_{x,y} \text{Per}(A_{n,j})_{p \in x, q \in y} \text{Per}(B_{n,j})_{p \in \bar{x}, q \in \bar{y}},
\end{align}
in which we sum over $x$ and $y$ which are same sized subsets of $\{1,\cdots,n\}$ while $\bar{x}$ and $\bar{y}$ are the complementary subsets~\cite{percus2012combinatorial}. Due to the zeroes in $A_{n,j}$, only permanents of matrices up to $n-j$ sizes contribute to this summation. By comparing Eq.~\eqref{eq:amp3} with Eq.~\eqref{eq:amp4}, we have reduced down the problem of proving the $n$-QS operation to proving that
\begin{align}
    \sum^n_{k_1=1} \cdots \sum^n_{k_j=k_{j-1}+1} \omega_{n+1}^{k_1+\cdots+k_j}  &\stackrel{?}{=} (-1)^j. \label{eq:rou}
\end{align}
This is simply the summation of $(n+1)$th roots of unity. 

\subsection{Summation of Roots of Unity\texorpdfstring{ $\omega_{n+1}$}{}}

We will consider the simplest values of $j$, before generalising the pattern to all $j$ values. Consider $j=1$, we have the summation 
\begin{align}
    \sum_{k_1=1}^n \omega_{n+1}^{k_1} &= \frac{\omega_{n+1}(1-\omega_{n+1}^n)}{1-\omega_{n+1}} = \frac{\omega_{n+1}-1}{1-\omega_{n+1}} = -1, 
\end{align}
where we have used the finite geometric series and that $\omega_{n+1}^{n+1}= (e^{-\frac{2i\pi}{n+1}})^{n+1} = 1$. Note that more generally $\sum_{k_1=1}^n \omega_{n+1}^{q k_1} = -1, q \in \mathbb{Z}\tc{\backslash [(n+1)\mathbb{Z}]},$ is also true by the same computation. Next, consider $j=2$ 
\begin{align}
    \sum_{k_1=1}^n \sum_{k_2=k_1+1}^n \omega_{n+1}^{k_1+k_2} &= \sum_{k_1=1}^n \omega_{n+1}^{2k_1} \sum_{k_2=1}^{n-k_1} \omega_{n+1}^{k_2} \nonumber \\ 
    &= \sum_{k_1=1}^n \omega_{n+1}^{2k_1} \frac{\omega_{n+1}-\omega_{n+1}^{-k_1}}{1-\omega_{n+1}} \nonumber \\ 
    &= (-1)^2. 
\end{align}
Similarly, consider $j=3$
\begin{align}
    \sum_{k_1=1}^n &\sum_{k_2=k_1+1}^n \sum_{k_3=k_2+1}^n \omega_{n+1}^{k_1+k_2+k_3} \nonumber \\
    &= \sum_{k_1=1}^n  \omega_{n+1}^{3k_1} \sum_{k_2=1}^{n-k_1} \omega_{n+1}^{2k_2} \sum_{k_3=1}^{n-k_2} \omega_{n+1}^{k_3} \nonumber \\ 
    &= \sum_{k_1=1}^n  \omega_{n+1}^{3k_1} \sum_{k_2=1}^{n-k_1} \omega_{n+1}^{2k_2} \frac{\omega_{n+1}-\omega_{n+1}^{-k_2}}{1-\omega_{n+1}} \nonumber \\ 
    &= (-1)^3. 
\end{align}
This pattern will continue for $j\in\lbrace0,\cdots,n\rbrace$ such that 
\begin{align}
    \sum^n_{k_1=1} \cdots \sum^n_{k_j=k_{j-1}+1} \omega_{n+1}^{k_1+\cdots+k_j} = (-1)^j, \label{eq:rootsofunity}
\end{align}
as needed to be shown. 

In summary, we have shown the summation of roots of unity evaluates to Eq.~\eqref{eq:rou}. Hence we have proven the permanent of the $\Omega_{n,j}$ matrices are equivalent to Eq.~\eqref{eq:amp3}. It follows then that the probability amplitudes of the QFT scattering matrix given in Eq.~\eqref{eq:amp1} is true. Thus we have finally proven that the $n$-QSBP structure, for all arbitrary values of $n$, implements an NLA operation as given in Eq.~\eqref{eq:QS}.

\section{Proof of n Quantum Scissors with Single Photons Resource}

Here we prove the $n$-QS action given by Eq.~\eqref{eq:QS} can also be achieved using $n$ single photons (SP) as the resource, as shown in Fig.~1(b). We purposefully used the complex conjugate versions of the gain BS $B_2^\dagger(\tau)$ and the QFT $F_{n+1}^\dagger$, to make it clear that the $n$-QSSP is simply the temporal reverse of the $n$-QSBP in Fig.~1(a). This allows us to leverage the results from the previous section for this proof. 

We will use Eq.~\eqref{eq:n0} to show that 
\begin{align}
    \langle n | \langle 0| B_2^\dagger &= ( B_2 |n\rangle | 0\rangle )^\dagger \nonumber \\ 
    &= \left( \frac{1}{(g^2+1)^{n/2}} \sum_{j=0}^n g^j (-1)^j\sqrt{\binom{n}{j}} |j\rangle|n-j\rangle \right)^\dagger \nonumber \\ 
    &= \frac{1}{(g^2+1)^{n/2}} \sum_{j=0}^n g^j (-1)^j\sqrt{\binom{n}{j}} \langle j| \langle n-j|.
\end{align}
We will also use Eq.~\eqref{eq:amp} to show that
\begin{align}
    \langle n-j| \langle j | &\otimes^{n-1} \langle 0 | F_{n+1}^\dagger |0\rangle\otimes^n |1\rangle \nonumber \\ 
    &= [ \langle 0| \otimes^n \langle 1| F_{n+1} |n-j\rangle  |j\rangle \otimes^{n-1} |0\rangle ]^\dagger \nonumber \\ 
    &= \frac{(-1)^j\sqrt{j!(n-j)!}}{(n+1)^{n/2}} \nonumber \\ 
    &= \frac{\sqrt{n!}}{(n+1)^{n/2}}(-1)^j\sqrt{\binom{n}{j}^{-1}}, \\
    \Rightarrow \langle n-j|\otimes \mathbb{I}_1 &\otimes^{n-1} \langle 0 | F_{n+1}^\dagger |0\rangle\otimes^n |1\rangle \nonumber \\ 
    &= \frac{\sqrt{n!}}{(n+1)^{n/2}}(-1)^j\sqrt{\binom{n}{j}^{-1}} |j\rangle. \label{eq:nqsspF}
\end{align}

The input into the $n$-QSSP will transform as  
\begin{align}
    |g\psi_n\rangle &= (\langle n| \langle 0|B_2^\dagger) \otimes \mathbb{I}_1 \otimes^{n-1} \langle0|   F_{n+1}^\dagger  |\psi\rangle \otimes |0\rangle\otimes^n |1\rangle \nonumber \\ 
    &= \frac{1}{(g^2+1)^{n/2}} \sum_{k=0}^\infty \sum_{j=0}^n g^jc_k(-1)^j\sqrt{\binom{n}{j}}\nonumber \\ 
    &\quad \times \langle j | k \rangle \langle n-j| \otimes \mathbb{I}_1 \otimes^{n-1} \langle0|   F_{n+1}^\dagger |0\rangle \otimes^n |1\rangle \nonumber \\ 
    &=  \frac{1}{(g^2+1)^{n/2}} \sum_{j=0}^n g^j c_j(-1)^j\sqrt{\binom{n}{j}} \nonumber \\ 
    &\quad \times \langle n-j|\otimes \mathbb{I}_1 \otimes^{n-1} \langle 0| F_{n+1}^\dagger |0\rangle \otimes^n |1\rangle \nonumber \\ 
    &= \frac{\sqrt{n!}}{(n+1)^{n/2}} \frac{1}{(g^2+1)^{n/2}} \sum_{j=0}^n g^j c_j |j\rangle,
\end{align}
where we note by orthogonality of the Fock states $\langle k|j\rangle = \delta_{j,k}$. Hence we have proven that the $n$-QSSP implements the NLA operation in Eq.~\eqref{eq:QS}, with exactly the same factors as the $n$-QSBP.

\section{Probability of Success} \label{sec:probs}

The default probability of success $\mathbb{P}$ is calculated as  
\begin{align}
    \mathbb{P} &\equiv \langle g \psi_n | g\psi_n \rangle \nonumber \\ 
    &= \frac{n!}{(n+1)^n} \frac{1}{(g^2+1)^n} \sum_{j=0}^n\sum_{k=0}^n g^{j+k} c_j c_k^* \langle k|j\rangle \nonumber \\ 
    &= \frac{n!}{(n+1)^n} \frac{1}{(g^2+1)^n} \sum_{j=0}^n g^{2j} |c_j|^2. 
\end{align}
We will now detail two methods, one for each $n$-QS configuration, which can improve this success probability.

\subsection{Additional Measurements with \textit{n}-QSBP}

For the $n$-QSBP in Fig.~1(a), one could choose to accept the cases where the vacuum state $\langle 0|$ is instead measured in the $(m_0+1)$th output mode of the QFT, in which $m_0\in\lbrace0,\cdots,n\rbrace$. In other words, we can accept any detection of the form $\otimes^{m_0} \langle 1| \otimes \langle 0| \otimes^{n-m_0} \langle 1|$, where $m_0=0$ is the default measurement case we considered in our proof. The output state requires a simple phase correction $C_1(m_0)$, which depends on where the vacuum was measured. We will prove here that these extra measurements also performs the $n$-QS NLA in Eq.~\eqref{eq:QS}, given we apply a phase correction to the output state. 

Recall that to calculate the necessary probability amplitudes after the QFT, we needed the permanent of a particular matrix $\Omega_{n,j,m_0}$. In the default measurement case $\Omega_{n,j}\equiv\Omega_{n,j,0}$ in Eq.~\eqref{eq:omeganj} was made from copies of the first two columns of the QFT's scattering matrix $F_{n+1}$, ignoring the first row. If the vacuum measurement $\langle 0 |$ was instead detected in the $(m_0+1)$th output mode, then when constructing $\Omega_{n,j,m_0}$ we need to ignore the $(m_0+1)$th row of the $F_{n+1}$ matrix as follows 
\begin{align}
    \Omega_{n,j,m_0} &= \overbrace{\left[ \begin{matrix}
            1 & \cdots & 1 \\
            \vdots & \vdots & \vdots \\ 
            1 & \cdots & 1  \\ 
            1 & \cdots & 1 \\ 
            \vdots & \vdots & \vdots \\ 
            1 & \cdots & 1 
        \end{matrix}\right. }^{n-j \text{ copies}} \overbrace{\left.\begin{matrix}
            1 & \cdots & 1 \\
            \vdots & \vdots & \vdots \\ 
            \omega_{n+1}^{m_0-1} & \cdots & \omega_{n+1}^{m_0-1} \\ 
            \omega_{n+1}^{m_0+1} & \cdots & \omega_{n+1}^{m_0+1} \\
            \vdots & \vdots & \vdots \\ 
            \omega_{n+1}^n & \cdots & \omega_{n+1}^n 
        \end{matrix}\right]}^{j \text{ copies}} 
        \left. \phantom{\begin{bmatrix}
            1 \\
            \vdots \\ 
            1 \\ 
            1 \\ 
            \vdots \\ 
            1 
        \end{bmatrix}\hspace{-2.3em}} \right\} {\scriptstyle n \text{ rows}}. 
\end{align}
One useful property of the permanent is that $\text{Per}(U_m') = s\text{Per}(U_m)$, given $U_m'$ is $U_m$ except a column (or row) is multiplied by a scalar $s$. The permanent also has the property of being invariant under rearrangement of the columns or rows. Hence we can consider rearranging $\Omega_{n,j,m_0}$ as follows
\begin{align}
    \begin{bmatrix}
        1 \\ 
        \vdots \\
        \omega_{n+1}^{m_0-1} \\ 
        \omega_{n+1}^{m_0+1} \\
        \vdots \\ 
        \omega_{n+1}^n
    \end{bmatrix} = \omega_{n+1}^{m_0} \begin{bmatrix}
        \omega_{n+1}^{-m_0} \\ 
        \vdots \\
        \omega_{n+1}^{-1} \\ 
        \omega_{n+1} \\
        \vdots \\ 
        \omega_{n+1}^{n-m_0}
    \end{bmatrix} =&\  \omega_{n+1}^{m_0} \begin{bmatrix}
        \omega_{n+1}^{n+1-m_0} \\ 
        \vdots \\
        \omega_{n+1}^{n} \\ 
        \omega_{n+1} \\
        \vdots \\ 
        \omega_{n+1}^{n-m_0}
    \end{bmatrix} \nonumber \\ 
    \xrightarrow{\text{switch rows}}&\  \omega_{n+1}^{m_0} \begin{bmatrix}
        \omega_{n+1} \\
        \vdots \\ 
        \omega_{n+1}^{n}
    \end{bmatrix},
\end{align}
where we used the property $\omega_{n+1}^{n+1}=1$ on the top few elements, and switch the rows in the last step such that the powers are in consecutive order with no gaps. This means we can use the properties of the permanent to change $\text{Per} (\Omega_{n,j,m_0})$ to $\text{Per} (\Omega_{n,j})$ as follows 
\begin{align}
    \text{Per} \left( \Omega_{n,j,m_0} \right) &= \omega_{n+1}^{jm_0} \text{Per} \left( \Omega_{n,j} \right) \nonumber \\ 
    &= \omega_{n+1}^{jm_0} (-1)^j  j!(n-j)!, 
\end{align}
where we used Eq.~\eqref{eq:amp3}. This means the probability amplitude given we measured the vacuum state in the $(m_0+1)$th mode of the QFT is 
\begin{align}
    \otimes^{m_0} \langle 1| & \otimes \langle 0| \otimes^{n-m_0} \langle 1| F_{n+1} |n-j\rangle|j\rangle \otimes^{n-1} |0\rangle  \nonumber \\ 
    &= \frac{1}{\sqrt{j!(n-j)!}} \frac{1}{(n+1)^{n/2}} \text{Per} 
        \left( \Omega_{n,j,m_0} \right) \nonumber \\ 
    &= \omega_{n+1}^{jm_0} (-1)^j \frac{\sqrt{j!(n-j)!}}{(n+1)^{n/2}}, 
\end{align}
where we used Eq.~\eqref{eq:amp2}. Thus the $n$-QS operation gives 
\begin{align}
    &\frac{\sqrt{n!}}{(g^2+1)^{n/2}} \sum_{j=0}^n \frac{(-1)^j g^j c_j}{\sqrt{j!(n-j)!}} |j\rangle \nonumber \\ 
    &\quad \otimes^{m_0} \langle 1| \otimes \langle 0| \otimes^{n-m_0} \langle 1| F_{n+1} |n-j\rangle|j\rangle \otimes^{n-1} |0\rangle \nonumber \\
    &= \frac{\sqrt{n!}}{(n+1)^{n/2}} \frac{1}{(g^2+1)^{n/2}} \sum_{j=0}^n \omega_{n+1}^{jm_0} g^j c_j |j\rangle, 
\end{align}
where we used Eq.~\eqref{eq:amp1}. The phase correction $C_1(m_0)$ we must apply to the output should be 
\begin{align}
    C_1(m_0) = e^{\frac{2i\pi m_0}{n+1} a^\dagger_0 a_0}. 
\end{align}
This phase shift acts on a Fock state as follows 
\begin{align}
    C_1(m_0)|j\rangle &= e^{\frac{2i\pi}{n+1}j m_0}|j\rangle = \omega_{n+1}^{-jm_0}|j\rangle, 
\end{align}
which allows us to recover the output state $|g\psi_n\rangle$.  

We have just proven that we can measure the vacuum state in any of the $n+1$ output modes of the QFT, as long as we can phase correct the output state based on where the vacuum state was measured. These additional measurements provide an $n+1$ probability enhancement 
\begin{align}
    \mathbb{P}_{\text{BP}} &= (n+1)\mathbb{P} \nonumber \\ 
    &= \frac{(n+1)!}{(n+1)^n} \frac{1}{(g^2+1)^n} \sum_{j=0}^n g^{2j} |c_j|^2.
\end{align}
The numerator and denominator of the $(n+1)!(n+1)^{-n}$ factor are quite close to each other at low $n$ values. However, we can use the Stirling approximation bounds to gain a sense of it's scalability to higher $n$ values
\begin{align}
    \frac{\sqrt{2\pi}}{e}\frac{(n+1)^{3/2}}{e^{n}} \leq \frac{(n+1)!}{(n+1)^n} \leq \frac{(n+1)^{3/2}}{e^n}. 
\end{align}
We can see that the success probability will decrease exponentially as we increase the size of the $n$-QSBP, irrespective of what we do to mitigate the other factors. However, we will show in the next section that the $n$-QSSP with offline resource preparation can overcome this default order scaling, and hence is much more accessible at higher $n$ values.

\subsection{Resource Preparation with \textit{n}-QSSP} 

For the $n$-QS which uses $n$ single photons, it is possible to prepare the QFT $F_{n+1}$ operation and vacuum measurements $\otimes^{n-1} \langle 0|$ beforehand, as shown in the red dashed box in Fig.~1(b). The outputted two-mode state $|R_n\rangle$ from this operation can then be interpreted as the resource state itself. Here we determine what this resource state is, and prove that resource preparation helps to overcome the exponential scaling of the probability of success with respect to $n$. 

First consider the smallest non-trivial case with $n=2$. This is associated with the QFT scattering matrix 
\begin{align}
    F_3^\dagger = \frac{1}{\sqrt{3}}\begin{bmatrix}
        1 & 1 & 1 \\ 
        1 & \omega_3^{-1} & \omega_3^{-2} \\ 
        1 & \omega_3^{-2} & \omega_3^{-4} 
    \end{bmatrix}.
\end{align}
This means our resource state of two single photons $|011\rangle$ transforms due to $F_3^\dagger$ as 
\begin{align}
    a_2^\dagger a_3^\dagger &\rightarrow \frac{1}{3} ( a_1^\dagger + \omega_3^{-1} a_2^\dagger + \omega_3^{-2} a_3 ^\dagger ) ( a_1^\dagger + \omega_3^{-2} a_2^\dagger + \omega_3^{-4} a_3 ^\dagger ) \nonumber \\ 
    &= \frac{1}{3} [ (a_1^\dagger)^2 + (a_2^\dagger)^2 + (a_3^\dagger)^2 - a_1^\dagger a_2^\dagger - a_1^\dagger a_3^\dagger - a_2^\dagger a_3^\dagger ], \label{eq:a3term}
\end{align}
which we used the summation of roots of unity $\omega_3^{-1}+\omega_3^{-2} = -1$. We can then consider the vacuum state measurement in the last mode $\mathbb{I}_2 \otimes |0\rangle\langle 0|$, which results in  
\begin{align}
    \frac{1}{3} [ \sqrt{2}&(|200\rangle + |020\rangle + |002\rangle) - (|110\rangle + |101\rangle + |011\rangle)  ] \nonumber \\ 
    &\rightarrow |R_2\rangle = \frac{1}{3} [ \sqrt{2}|20\rangle - |11\rangle + \sqrt{2}|02\rangle ]. 
\end{align}
Notice to get this state we could have just ignored all output mode operators besides $a_1^\dagger$ and $a_2^\dagger$ in Eq.~\eqref{eq:a3term}, which we will do for larger $n$ sizes. The probability of preparing this state is 
\begin{align}
    P_2 = \langle R_2 | R_2 \rangle = 5/9.  
\end{align}
If we prepare this state beforehand, we enhance the default probability of success by $1/P_2=9/5$, which is smaller than probability enhancement due to the $n+1=3$ possible measurements for $2$-QSBP. However, for larger $n$-QSSP we expect $1/P_n$ to scale significantly more quickly than $n+1$. 

Next, consider the $3$-QSSP which is associated with the scattering QFT matrix of 
\begin{align}
    F_4^\dagger &= \frac{1}{2}
        \begin{bmatrix}
            1 & 1 & 1 & 1 \\ 
            1 & \omega^{-1}_4 & \omega^{-2}_4 & \omega^{-3}_4 \\
            1 & \omega^{-2}_4 & \omega^{-4}_4 & \omega^{-6}_4 \\ 
            1 & \omega^{-3}_4 & \omega^{-6}_4 & \omega^{-9}_4 
        \end{bmatrix}. 
\end{align}
We will see how the $|0111\rangle$ state transforms due to $F_4^\dagger$. Since $\mathbb{I}_2 \otimes |00\rangle \langle 00|$ will be detected, we may as well ignore all output creation operators besides those that act on the first two output modes. This corresponds to
\begin{align}
    a_2^\dagger a_3^\dagger a_4^\dagger &\rightarrow \frac{1}{8} ( a_1^\dagger + \omega_4^{-1} a_2^\dagger) ( a_1^\dagger + \omega_4^{-2} a_2^\dagger) ( a_1^\dagger + \omega_4^{-3} a_2^\dagger) \nonumber \\ 
    &= \frac{1}{8} [(a_1^\dagger)^3 - (a_1^\dagger)^2 a_2^\dagger + a_1^\dagger (a_2^\dagger)^2 - (a_2^\dagger)^3], 
\end{align}
which we once again used the summation of roots of unity $\omega^{-1}_4+\omega^{-2}_4+\omega^{-3}_4=-1$. This corresponds to an output state of
\begin{align}
    |R_3\rangle &= \frac{1}{8} [ \sqrt{3!} |30\rangle - \sqrt{2} |20\rangle + \sqrt{2} |02\rangle - \sqrt{3!} |03\rangle ],
\end{align}
where the probability of preparing this state is 
\begin{align}
    P_3 = \langle R_3 | R_3 \rangle = 1/4. 
\end{align}
This means the probability enhancement $1/P_3=4$ due to preparing the $|R_3\rangle$ resource state beforehand for the $3$-QSSP, is the same as the enhancement accounting for the $n+1=4$ possible measurement outcomes for the $3$-QSBP. Hence $n=3$ is when these two cases break-even, with resource prepared $n$-QSSP having a better overall scaling for larger $n$ values. 

We will now consider resource preparation for a completely arbitrary $n$-QSSP. The $n$ single photons state $|0\rangle\otimes^{n}|1\rangle$ is transformed due to the QFT $F_{n+1}^\dagger$ and vacuum measurements as
\begin{align}
    \prod_{j=2}^{n+1} a_j^\dagger &\rightarrow \prod_{j=2}^{n+1} \frac{1}{\sqrt{n+1}} (a_1^\dagger + \omega_{n+1}^{-j+1} a_2^\dagger ) \nonumber \\ 
    &= \frac{1}{(n+1)^{n/2}}\prod_{j=1}^{n} (a_1^\dagger + \omega_{n+1}^{-j} a_2^\dagger ).
\end{align}
Let us consider expanding over the product by considering the first few highest order $a_1^\dagger$ terms
\begin{align}
    &\frac{1}{(n+1)^{n/2}} \bigg[(a_1^\dagger)^n + \bigg( \sum_{j_1=1}^n \omega_{n+1}^{-j_1}  \bigg) (a_1^\dagger)^{n-1} a_2^\dagger \nonumber \\ 
    &\quad + \bigg( \sum_{j_1=1}^n \sum_{j_2=j_1+1}^n \omega_{n+1}^{-j_1-j_2}  \bigg) (a_1^\dagger)^{n-2} (a_2^\dagger)^2 + \cdots \bigg] \nonumber \\ 
    &= \frac{1}{(n+1)^{n/2}} \sum_{j=0}^n (-1)^j (a_1^\dagger)^{n-j} (a_2^\dagger)^j.
\end{align}
This once again includes summations of roots of unity, which we have solved for the general case in Eq.~\eqref{eq:rootsofunity}. Hence we have shown that the unnormalised prepared state is 
\begin{align}
    |R_n\rangle &= \frac{1}{(n+1)^{n/2}} \sum_{j=0}^n(-1)^j \sqrt{(n-j)!j!} |n-j\rangle| j\rangle \nonumber\\
    &= \frac{\sqrt{n!}}{(n+1)^{n/2}} \sum_{j=0}^n(-1)^j \sqrt{\binom{n}{j}^{-1}} |n-j\rangle| j\rangle. 
\end{align}
This result is consistent with our previous result in Eq.~\eqref{eq:nqsspF}. This means the probability to prepare this state is given by
\begin{align}
    P_n &= \langle R_n | R_n \rangle = \frac{1}{(n+1)^n} \sum_{j=0}^n(n-j)!j! \nonumber \\ 
    &= \frac{(n+1)!}{(n+1)^n} 2^{-(n+1)}[B_2(n+2,0)-i\pi] \nonumber \\ 
    &< \frac{(n+1)!}{(n+1)^n}.
\end{align}
This expression contains the beta function defined as $B_z(a,b) \equiv \int^z_0 t^{a-1}(1-t)^{b-1} dt$. We can show numerically that $B_2(n+2,0)-i\pi=\text{Re}[B_2(n+2,0)]<2^{n+1}$, which holds $\forall n >1$ as the scaling appears to be a tiny bit more slower than $2^{n+1}$. Hence we have shown that if we prepare the $|R_n\rangle$ resource beforehand, our $n$-QSSP actually has a success probability scaling of at least
\begin{align}
    \mathbb{P}_{\text{SP}} &= \frac{\mathbb{P}}{P_n} > \frac{1}{n+1}  \frac{1}{(g^2+1)^n} \sum_{j=0}^n g^{2j} |c_j|^2. 
\end{align}
This is clearly much better than the guaranteed exponential decrease scaling of the default configuration, even accounting for extra measurements in the $n$-QSBP (for cases greater than $n=3$). In other words, due to resource preparation, we showed that the success probability will not necessarily decrease exponentially as $n$ increases; it will depend on the input state being amplified $|\psi\rangle$ and the chosen gain $g$.

\tc{\subsection{Summary of Success Probability Improvements}}

In summary, both $n$-QS protocols in Fig.~1, without any modifications, have the same default success probability $\mathbb{P}(n,g,\{c_j\}_{j=0}^n)$ due to time symmetry; the $n$-QSBP is the time-reverse of the $n$-QSSP, and vice versa. We can improve this for the BP configuration in Fig.~1(a), since the QFT's symmetrical action on the $n+1$ modes means measuring the vacuum state on a different mode results in the same output state (up to a correctable phase). Thus, this QFT symmetry means the success probability of the BP configuration can be increased to $\mathbb{P}_{\text{BP}} = (n+1)\mathbb{P}$, since we can accept $n+1$ total measurements. In contrast, the SP configuration has two different groups of measurements (after the QFT and after the beam-splitter). We can allow the measurements after the QFT to happen first, thus producing a two-mode entangled state just after the dashed red box in Fig.~1(b). It doesn't matter how this two-mode entangled state resource is generated, we could potentially make this resource state using a better configuration (such as a Gaussian boson sampling type-device, as discussed in Ref.~\cite{su2019conversion}). If we assume we already have this two-mode entangled state resource (i.e. we prepared it offline), the success probability of the SP configuration improves by $\mathbb{P}_{\text{SP}} = \mathbb{P}/P_n$ since we remove the probability of generating this resource. 

\begin{figure}[htbp]
    \begin{center}
        \includegraphics[width=\linewidth]{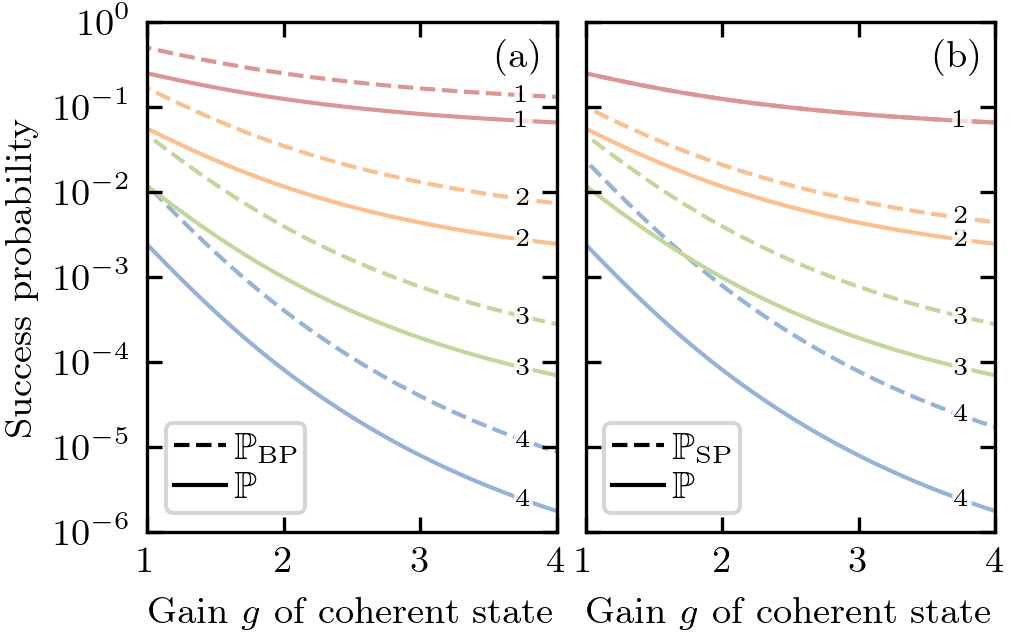}
        \caption{\label{fig:probvsg_BPvsSP} 
            \tc{It is possible to improve success probability (dashed lines) of our scheme, beyond the default configuration (solid lines). The small numbers at the end of each line refer to the $n$ size of the QS device. Here we consider amplifying a coherent state, with $\alpha=0.3$ amplitude, via (a) the $n$-QSBP, or (b) the $n$-QSSP. }} 
    \end{center}
\end{figure}

\tc{In Fig.~\ref{fig:probvsg_BPvsSP}, we give a concrete example of how the improvements explained in the previous subsections impact the success probability. In this graph, we consider amplifying a coherent state $|\alpha\rangle$ with our $n$-QS using (a) the bunched photons input configuration, or (b) the single photons input configuration. We can see by the solid lines that by default (i.e. without any improvements) the two configurations have the same probability of success $\mathbb{P}$. The $n+1$ total measurements allowed in the BP configuration improves the success probability by $\mathbb{P}_{\text{BP}} = (n+1)\mathbb{P}$. The resource preparation in the SP configuration has no impact on $n=1$ (since there is nothing to prepare), however does improve the situation for $n>1$ sizes by $\mathbb{P}_{\text{SP}} = \mathbb{P}/P_n$. Notice that Fig.~\ref{fig:probvsg_BPvsSP} confirms our previous analysis that  
\begin{align}
    \mathbb{P}_{\text{BP}} &> \mathbb{P}_{\text{SP}},\quad n \in \{1,2\}, \nonumber \\ 
    \mathbb{P}_{\text{BP}} &= \mathbb{P}_{\text{SP}},\quad n = 3, \nonumber \\ 
    \mathbb{P}_{\text{BP}} &< \mathbb{P}_{\text{SP}},\quad n \geq 4.
\end{align}
We know that the last inequality is true for all $n\geq4$ because $(n+1)$ is only a linear improvement to BP, while $1/P_n$ is an exponential improvement to SP. Note this doesn't take into account experimental imperfections; see the last section for this analysis. }

\section{Fidelity} 

\tc{\subsection{General Input States}}

We may quantify the quality of an NLA operation by its fidelity $F$, which is the overlap between its output state against a perfect NLA operation's output state. The normalised output from the $n$-QS protocol is 
\begin{align}
    |g\widetilde{\psi}_n\rangle \equiv \frac{|g\psi_n\rangle}{\sqrt{\mathbb{P}}}  = \frac{ \sum_{j=0}^n g^j c_j |j\rangle }{\sqrt{\sum_{j=0}^n g^{2j} |c_j|^2}},
\end{align}
while the normalised output due to a perfect NLA operation $|g\psi\rangle \equiv g^{a^\dagger a}|\psi\rangle = \sum_{j=0}^\infty g^j c_j |j\rangle$ is  
\begin{align}
    |g\widetilde{\psi}\rangle \equiv \frac{|g\psi\rangle}{\sqrt{\langle g\psi|g\psi\rangle}} &= \frac{\sum_{j=0}^\infty g^j c_j |j\rangle }{\sqrt{\sum_{j=0}^\infty g^{2j} |c_j|^2}}.
\end{align}
This means the fidelity of the $n$-QS is calculated by the overlap between these two states as follows
\begin{align}
    F &\equiv | \langle g \widetilde{\psi}|g \widetilde{\psi}_n \rangle |^2 \\ 
    &= \frac{\left(\sum_{j=0}^n\sum_{k=0}^\infty g^{j+k} c_j c_k^* \langle k|j\rangle\right)^2}{\left(\sum_{j=0}^\infty g^{2j} |c_j|^2\right)\left(\sum_{j=0}^n g^{2j} |c_j|^2\right)} \\
    &=\frac{ \sum_{j=0}^n g^{2j} |c_j|^2 }{\sum_{j=0}^\infty g^{2j} |c_j|^2}. \label{eq:Ftruncation}
\end{align}
Clearly $\lim_{n\rightarrow\infty}F = 1$ because $\lim_{n\rightarrow\infty}|g\widetilde{\psi}_n\rangle = |g\widetilde{\psi}\rangle$. More precisely, one can see from Eq.~\eqref{eq:Ftruncation} that any infidelity $1-F>0$ is purely due to the effect of the truncation. This is in contrast with other linear-optical NLA protocols which also introduce a distortion effect on the Fock basis coefficients.

\tc{\subsection{Specific Input States}}

\tc{Let us now consider specific values of $c_j$ for the input state $|\psi\rangle \equiv \sum_{j=0}^\infty c_j |j\rangle$. If the input state is of the form $|\psi_{n_\text{max}}\rangle = \sum_{j=0}^{n_\text{max}} c_j |j\rangle$ (i.e. it has a hard cut-off in the Fock basis coefficients where $c_j = 0, \forall j > n_\text{max}$), then an $n_\text{max}$-QS device is enough to amplify this state with perfect fidelity $F=1$. As far as we know, this is not possible using any other linear-optical NLA protocols for general $n_\text{max}$, assuming finite resources. This is because we do not need to rely on splitting the input state into many modes, amplifying each mode separately, and then recombining all the modes together (which is an imperfect process in general).}  

\tc{It is also interesting to consider continuous variable states which have no hard cut-off in the Fock basis. In this paper, we consider coherent states defined as 
\begin{align}
    |\alpha\rangle \equiv e^{-|\alpha|^2/2} \sum_{j=0}^\infty \frac{1}{\sqrt{j!}} \alpha^j |j\rangle,  
\end{align}
which has an amplitude parameter $\alpha\in\mathbb{C}$. This state's photon numbers have a Poisson distribution, with an $\langle a^\dagger a\rangle = |\alpha|^2$ average and $\text{Var}( a^\dagger a) = |\alpha|^2$ variance. We should expect our $n$-QS device to work with good fidelity if the $n$ size is a few standard deviation or so above the average photon number $n>|\alpha|^2+2|\alpha|$. We note that a perfect amplification/de-amplification process outputs $g^{a^\dagger a}|\alpha\rangle \rightarrow |g\alpha\rangle$, where $g\alpha$ is the resultant amplitude (this can be seen by acting $g^{a^\dagger a}$ to the above coherent state definition).}

\tc{We also consider the single-mode squeezed vacuum (SMSV) states as inputs, which are defined as 
\begin{align}
    |s\rangle \equiv (1-s^2)^{1/4} \sum_{j=0}^\infty \frac{\sqrt{(2j)!}}{2^j j!} s^j |2j\rangle, 
\end{align}
given a squeezing parameter $s\in[0,1]$. These states have interesting Fock state structure, where only even photon numbers are possible. This means it is best to just use even sized $n$-QS devices (in contrast to the competing protocol), as shown in the main paper. We note that a perfect amplification/de-amplification process outputs $g^{a^\dagger a}|s\rangle \rightarrow |g^2s\rangle$, where we can treat $g^2s$ as the resultant amount of squeezing. If we consider $g^2_\text{max}s=1$, this corresponds to an infinitely squeezed SMSV state which is spread out over all Fock components. However the output state of any finite sized NLA device, for example from our $n$-QS device
\begin{align}
    |g^2s_n\rangle \propto \sum_{j=0}^{\lfloor n/2 \rfloor} \frac{\sqrt{(2j)!}}{2^j j!} (g^2s)^j |2j\rangle, 
\end{align}
is restricted to a finite Fock basis $[0,n]$. Hence the overlap of $|g^2s_n\rangle$ and $|g^2s\rangle$ will tend towards zero (i.e. $F\rightarrow 0$) as $g^2s\rightarrow1$. This effect is seen clearly in Fig.~2(c) in the main paper, where in that case $s\approx 0.29$ hence $g_\text{max}\approx1.9$ is the maximum amount of gain possible for any finite sized amplifier.}

\section{Comparison with Other NLA Protocols} 

\begin{figure*}[htbp]
    \begin{center}
        \includegraphics[width=\linewidth]{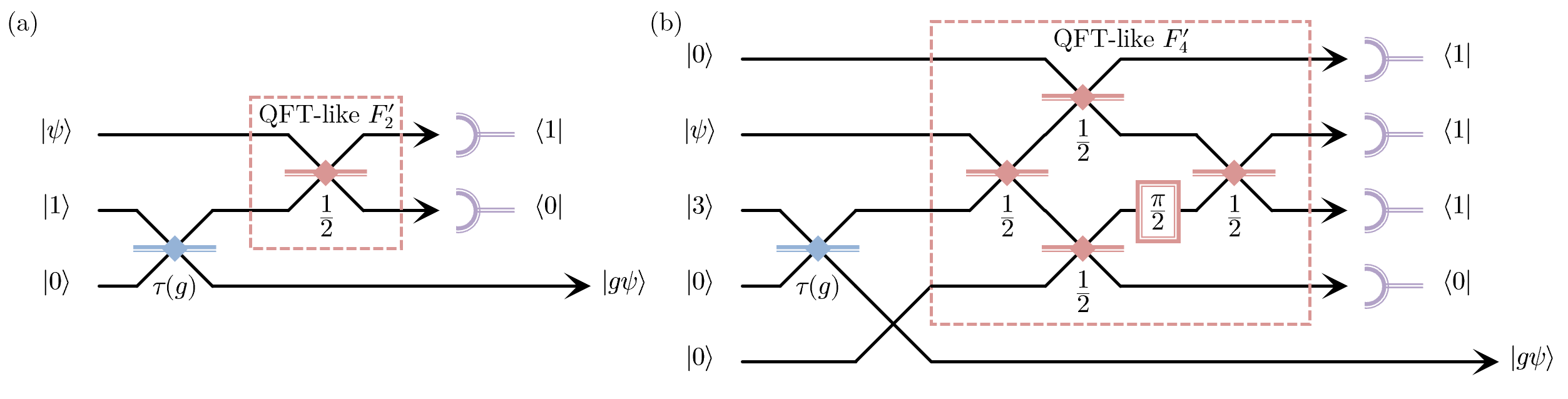}
        \caption{\label{fig:13QSold} 
            Schematic of the (a) $1$-QS and (b) $3$-QSBP, as described in previous literature~\cite{winnel2020generalized}. The photon scattering associated with the red dashed boxes, after the gain BS in blue, is functionally equivalent to QFTs.}
    \end{center}
\end{figure*} 

\subsection{1-Photon Quantum Scissors in Parallel} 

The $n$-X10 protocol is a scalable protocol consisting of $n$ copies of $1$-QS in parallel~\cite{ralph2009nondeterministic}. The $n$-X10 protocol has the output state 
\begin{align}
    |g\phi_n\rangle = \frac{n!}{(g^2+1)^{n/2}} \sum^n_{j=0} \frac{1}{(n-j)!n^j} g^j c_j |j\rangle, 
\end{align}
which means the success probability will be
\begin{align}
    \mathbb{P}_\text{X10} &\equiv \langle g\phi_n | g\phi_n \rangle \\
    &= \frac{(n!)^2}{(g^2+1)^n} \sum_{j=0}^n \frac{1}{[(n-j)!n^j]^2} g^{2j} |c_j|^2. 
\end{align}
Hence the normalised output state will be
\begin{align}
    |g\widetilde{\phi}_n\rangle \equiv \frac{|g\phi_n\rangle}{\sqrt{\mathbb{P}_\text{X10}}} = \frac{\sum^n_{j=0} \frac{g^j c_j |j\rangle}{(n-j)!n^j}}{\sqrt{\sum_{j=0}^n \frac{g^{2j} |c_j|^2}{[(n-j)!n^j]^2}}},
\end{align}
and the fidelity of an $n$-X10 device is 
\begin{align}
    F_\text{X10} &\equiv | \langle g \widetilde{\psi}|g \widetilde{\phi}_n \rangle |^2 \\ 
    &= \frac{\sum^n_{j=0} \frac{g^{2j} |c_j|^2 }{(n-j)!n^j}}{\sqrt{\sum_{j=0}^n \sum_{k=0}^\infty\frac{g^{2(j+k)} |c_j|^2|c_k|^2}{[(n-j)!n^j]^2}}}.
\end{align}
In the limit of large sizes then $\lim_{n\rightarrow\infty}F_\text{X10}=1$. 

\subsection{Scattering of Existing n-Photons Quantum Scissors are Functionally QFTs}

Here we point out that the $n$-QS in past literature, which exist only for $n\in\{1,3,7\}$ with bunched photons, have components which scatter photons in the same manner as our protocol using QFTs~\cite{winnel2020generalized}. 

Let us firstly consider the $1$-QS. It is clear from Fig.~\ref{fig:13QSold}(a) that the scattering matrix in the red dashed box is given by
\begin{align}
    F'_2 = B_2(\tfrac{1}{2}) = \frac{1}{\sqrt{2}}\begin{bmatrix}
        1 & 1 \\ 
        -1 & 1
    \end{bmatrix}. 
\end{align}
We note that this is clearly similar to our $1$-QS which uses a two mode QFT
\begin{align}
    F_2 &= \begin{bmatrix}
        1 & 0 \\ 
        0 & -1
    \end{bmatrix} \frac{1}{\sqrt{2}}\begin{bmatrix}
        1 & 1 \\ 
        -1 & 1
    \end{bmatrix} = C_1^{(2)}(\pi) F_2',  
\end{align}
except it has a trivial $\pi$ phase shift. Hence we have shown that our protocol and the existing protocol for $1$-QS has functionally the same photon scattering properties. 

Similarly, we can consider the previous literature on the $3$-QSBP~\cite{winnel2020generalized}. The red dashed box in Fig.~\ref{fig:13QSold}(b) has a scattering matrix given by 
\begin{align}
    F'_4 &= B_2^{(2,3)}(\tfrac{1}{2})C_1^{(3)}(\tfrac{\pi}{2}) B_2^{(1,2)}(\tfrac{1}{2}) B_2^{(3,4)}(\tfrac{1}{2}) B_2^{(2,3)}(\tfrac{1}{2}) \nonumber \\ 
    &= \frac{1}{2} \begin{bmatrix}
        \sqrt{2} & 1 & 1 & 0 \\ 
        -1 & e^{-i\pi/4} & e^{i\pi/4} & e^{i\pi/2} \\ 
        1 & e^{-i3\pi/4} & e^{i 3\pi/4} & e^{i\pi/2} \\ 
        0 & 1 & -1 & \sqrt{2} 
    \end{bmatrix}. 
\end{align}
Note only the two middle columns are important for the scattering dynamics, since the inputs in this case are injected into the two middle modes. This can be compared to the scattering matrix of the four mode QFT 
\begin{align}
    F_4 &= \frac{1}{2} \begin{bmatrix}
        1 & 1 & 1 & 1 \\ 
        1 & e^{-i\pi/2} & -1 & e^{i\pi/2} \\ 
        1 & -1 & 1 & -1 \\ 
        1 & e^{i\pi/2} & -1 & e^{-i\pi/2} 
    \end{bmatrix} \nonumber \\ 
    &= \begin{bmatrix}
        1 & 0 & 0 & 0 \\ 
        0 & 0 & e^{3i\pi/4} & 0 \\ 
        0 & 0 & 0 & 1 \\ 
        0 & e^{i\pi/4} & 0 & 0
    \end{bmatrix} \nonumber \\ 
    &\quad \times  
    \frac{1}{2} \begin{bmatrix}
        1 & 1 & 1 & 1 \\ 
        e^{-i\pi/4} & e^{i\pi/4} & -e^{-i\pi/4} & e^{-i3\pi/4} \\ 
        e^{-i3\pi/4} & e^{i3\pi/4} & -e^{-i3\pi/4} & e^{-i\pi/4} \\ 
        1 & -1 & 1 & -1 
    \end{bmatrix}. 
\end{align}
We can see that after some trivial permutations and phase factors, the first two columns of $F_4$ are functionally equivalent to the two middle columns of $F_4'$. Hence the photon scattering dynamics associated with the $3$-QSBP in the past literature is shown to be equivalent to our protocol.

\section{Loss Analysis}

\subsection{Analytical Expression for the Output State}

Suppose we transmit an arbitrary state $\rho$ through a lossy channel with some non-unity transmissivity $\eta\in[0,1]$. We can determine how $\rho$ is transformed by this lossy channel using the following quantum operation
\begin{align}
    \rho \rightarrow \sum_{l=0}^\infty A_l \rho A_l^\dagger,\quad A_l = \sqrt{\frac{(1-\eta)^l}{l!}} \eta^{\frac{a^\dagger a}{2}} a^l,   
\end{align}
in which this set of Kraus operators satisfy $\sum_{l=0}^\infty A_l^\dagger A_l = \mathbb{I}$. The annihilation operator $a^l$ means we can naturally interpret the summation variable $l$ as the amount of lost photons. Full transmissivity $\eta=1$ means only the $l=0$ term is non-zero where $A_0=\mathbb{I}$, such that the output state will be the same as the input state. In contrast, no transmissivity $\eta=0$ means $A_l=|0\rangle\langle l|$, in other words the output state will just be vacuum. 

\begin{figure}[htbp]
    \begin{center}
        \includegraphics[width=\linewidth]{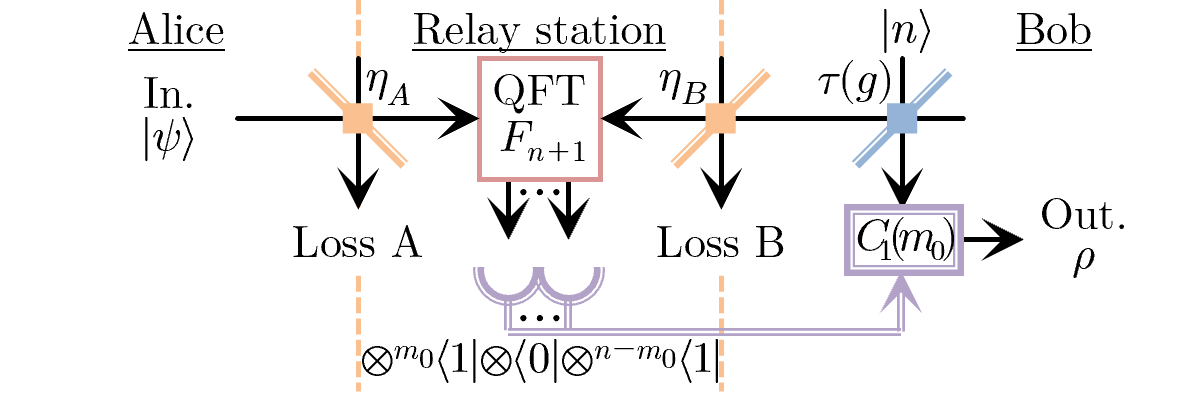}
        \caption{\label{fig:lossnq} 
            We consider a lossy $n$-QS protocol with an arbitrary state input $|\psi\rangle$, where Alice's channel has $\eta_A$ transmissivity and Bob's channel has $\eta_B$ transmissivity.} 
    \end{center}
\end{figure}

We can now consider the situation where the $n$-QS has two lossy channels, as depicted in Fig.~\ref{fig:lossnq}. We will label the transmissivity of the channel on Alice's side as $\eta_A\in[0,1]$, and on Bob's side as $\eta_B\in[0,1]$. The relay station is allowed to measure the zero state in any mode $m_0$ (as described previously in Sec.~\ref{sec:probs}), however for simplicity we will just consider $m_0=0$ with $P_{n+1} = |0\rangle\langle0|\otimes^n |1\rangle\langle1|$ detection events. We may then calculate the output state of this lossy $n$-QS as follows 
\begin{widetext}
    \begin{align}
        \rho &= \sum_{l_A,l_B}^\infty \text{Tr}_{n+1} \left(P_{n+1} F_{n+1} A_{l_B} B_2 |0\rangle\langle0| \otimes |n\rangle\langle n| B_2^\dagger A_{l_B}^\dagger \otimes A_{l_A} |\psi\rangle\langle\psi| A_{l_A}^\dagger \otimes^{n-1}|0\rangle\langle0| F_{n+1}^\dagger P_{n+1}^\dagger \right) \nonumber \\ 
        &= \sum_{l_A,l_B}^\infty \sum_{k_1,k_2}^\infty c_{k_1} c_{k_2}^* \langle0|\otimes^n\langle1| F_{n+1} A_{l_B} B_2 |0\rangle\langle0| \otimes |n\rangle\langle n| B_2^\dagger A_{l_B}^\dagger \otimes A_{l_A} |k_1\rangle\langle k_2| A_{l_A}^\dagger \otimes^{n-1}|0\rangle\langle0| F_{n+1}^\dagger |0\rangle\otimes^n|1\rangle \nonumber \\ 
        &= \frac{n!}{(n+1)^n} \frac{1}{(g^2+1)^n} \sum_{l_A=0}^\infty \sum_{l_B=0}^n \sum_{j_1=0}^{n-l_B} \sum_{j_2=0}^{n-l_B} c_{j_1+l_A+l_B} c^*_{j_2+l_A+l_B} (g\sqrt{\eta_A/\eta_B})^{j_1+j_2} \frac{(1-\eta_A)^{l_A}(1-\eta_B)^{l_B} \eta_A^{l_B} \eta_B^{n-l_B}}{l_A!l_B!} \nonumber \\
        &\quad \times  \sqrt{\frac{(j_1+l_A+l_B)!(j_2+l_A+l_B)!}{j_1!j_2!}} |j_1\rangle\langle j_2|. \label{eq:lossout}
\end{align}
\end{widetext}
Note that $\text{Tr}_{n+1}$ means we are tracing over the $n+1$ output modes of the QFT measurement. The resultant output expression is quite complicated, however one can see the effect of loss on the gain by considering the no loss terms $l_A=l_B=0$ as follows 
\begin{align}
        \rho_0 &= \frac{n!}{(n+1)^n} \frac{1}{(g^2+1)^n} \sum_{j_1=0}^{n} \sum_{j_2=0}^{n} c_{j_1} c^*_{j_2} (g\sqrt{\eta_A/\eta_B})^{j_1+j_2} \nonumber \\
        &\quad \times |j_1\rangle\langle j_2|. 
\end{align}
Hence we can take $g\sqrt{\eta_A/\eta_B}$ to be the effective gain. Interestingly, if we can manufacture a scenario where Alice and Bob experience the same loss $\eta_A=\eta_B$, then the losses cancel out such that the effective gain is simply $g$; this idea will be explored more in a later section. 

A typical quantum procedure Alice may want to perform is the distribution of one part of an entangled state to Bob. Here we will consider Alice transmitting one mode of a two-mode squeezed vacuum (TMSV) state $|\chi\rangle=\sqrt{1-\chi^2} \sum_{k=0}^\infty \chi^k |k\rangle |k\rangle$, where $\chi\in[0,1]$ is the squeezing parameter, through a lossy $n$-QS as shown in Fig.~3(b). We may use Eq.~\eqref{eq:lossout} to determine the output state for this situation
\begin{widetext}
    \begin{align}
        \rho_{AB} &= \frac{n!}{(n+1)^n} \frac{1-\chi^2}{(g^2+1)^n} \sum_{l_A=0}^\infty \sum_{l_B=0}^n \sum_{j_1=0}^{n-l_B} \sum_{j_2=0}^{n-l_B} (g\chi \sqrt{\eta_A/\eta_B})^{j_1+j_2} \frac{\chi^{2(l_A+l_B)} (1-\eta_A)^{l_A}(1-\eta_B)^{l_B} \eta_A^{l_B} \eta_B^{n-l_B}}{l_A!l_B!} \nonumber \\
        &\quad \times  \sqrt{\frac{(j_1+l_A+l_B)!(j_2+l_A+l_B)!}{j_1!j_2!}} |j_1+l_A+l_B\rangle\langle j_2+l_A+l_B|\otimes |j_1\rangle\langle j_2|. \label{eq:rhoabloss}
\end{align}
\end{widetext}
Similar to before, consider just the no loss terms where $l_A=l_B=0$ as follows
\begin{align}
        \rho_{AB,0} &= \frac{n!}{(n+1)^n} \frac{1}{(g^2+1)^n} \sum_{j_1=0}^{n} \sum_{j_2=0}^{n} (g\chi\sqrt{\eta_A/\eta_B})^{j_1+j_2} \nonumber \\
        &\quad \times |j_1\rangle\langle j_2| \otimes |j_1\rangle\langle j_2|. \label{eq:rhoab0}
\end{align}
Hence the coefficient $g\chi \sqrt{\eta_A/\eta_B}$ can be thought of as an overall squeezing factor up to the $n$ Fock basis. In fact, if the gain beam-splitter is set to $g=g_\text{max}$ such that $g_\text{max}\chi \sqrt{\eta_A/\eta_B}\approx1$ is true, then the logarithmic negativity (LN), an upper bound to distillable entanglement, is roughly maximised as illustrated in Fig.~5(a). This is simply because Eq.~\eqref{eq:rhoab0} becomes a state that is uniformly distributed and hence maximally entangled up to the $n$ Fock state. 

The success probability for this lossy $n$-QS protocol with $|\chi\rangle$ input can be calculated as
\begin{align}
    \mathbb{P}_\text{loss} &= \text{Tr}(\rho_{AB}) \nonumber \\ 
    &= \frac{n!}{(n+1)^n} \frac{1-\chi^2}{(g^2+1)^n} \sum_{l_B=0}^n \sum_{j=0}^{n-l_B} (g\chi \sqrt{\eta_A/\eta_B})^{2j} \nonumber \\
    &\quad \times \frac{\chi^{2l_B}(1-\eta_B)^{l_B} \eta_A^{l_B} \eta_B^{n-l_B}}{(1-\chi^2+\chi^2\eta_A)^{1+j+l_B}}  \frac{(j+l_B)!}{j!l_B!}.
\end{align}
Note that as the size $n$ of our protocol increases, we expect the output to become more similar to a CV-type state. This is illustrated clearly by comparing the Gaussian LN to the full LN as in Fig.~5(a), where the $g$ range in which they are equivalent increases with increasing $n$. The covariance matrix of $\rho_{AB}$ encapsulates the CV properties of the state, and is calculated using the following expectation values 
\begin{align}
    \langle a_A^\dagger a_A\rangle &= \text{Tr}(a_A^\dagger a_A \frac{\rho_{AB}}{\mathbb{P}_\text{loss}}),\\ 
    \langle a_B^\dagger a_B\rangle &= \text{Tr}(a_B^\dagger a_B \frac{\rho_{AB}}{\mathbb{P}_\text{loss}}), \\ 
    \langle a_A^\dagger a_B^\dagger \rangle = \langle a_A a_B \rangle &= \text{Tr}(a_A a_B \frac{\rho_{AB}}{\mathbb{P}_\text{loss}}).
\end{align}
We then determine the covariance matrix which will have the following structure
\begin{align}
    \Gamma_{AB} &= \begin{bmatrix}
        a & 0 & c & 0 \\ 
        0 & a & 0 & -c \\ 
        c & 0 & b & 0 \\ 
        0 & -c & 0 & b
    \end{bmatrix}, \label{eq:covmat} \\ 
    a &= 1 + 2\langle a^\dagger_A a_A \rangle, \\ 
    b &= 1 + 2\langle a^\dagger_B a_B \rangle, \\ 
    c &= \langle a_A a_B \rangle + \langle a_A^\dagger a_B^\dagger \rangle. 
\end{align}
From this matrix, one can now easily calculate various entanglement metrics, like the LN mentioned previously.

\subsection{Entanglement Distribution and Distillation} \label{sec:entdist}

Here we consider how well our $n$-QS protocol can distribute and distill entanglement in a situation with loss. The previous section derived analytical expressions for the output from an TMSV input; in this section we will set squeezing to $\chi=0.25$ for our plots. This situation is depicted in Fig.~3(b), where the $n$-QS has lossy channels with $\eta_A$ and $\eta_B$ transmissivities. In fact, one can consider the $n$-QS to be distributed over a single lossy channel with $\eta\equiv\eta_A\eta_B$ overall transmissivity. Based on experimental data on fibres, we set the transmissivity to decrease exponentially with distance travelled as $\eta=10^{-0.02d}$, where $d$ is the distance between Alice and Bob in km. We consider placing the QFT measurement component either at (i) the middle $(\eta_A=\sqrt{\eta},\eta_B=\sqrt{\eta})$, or (ii) the end $(\eta_A=\eta,\eta_B=1)$ of the channel. Scenario (i) is where the QFT measurement is treated as a quantum relay station in the middle of the channel, in which the distance between Alice and the station $d/2$ is equal to the distance between the station and Bob $d/2$. In contrast, scenario (ii) is simply where the input state experiences loss for the total distance $d$, and then is amplified back up by the NLA $n$-QS at the end. 

\begin{figure}[htbp]
    \begin{center}
        \includegraphics[width=\linewidth]{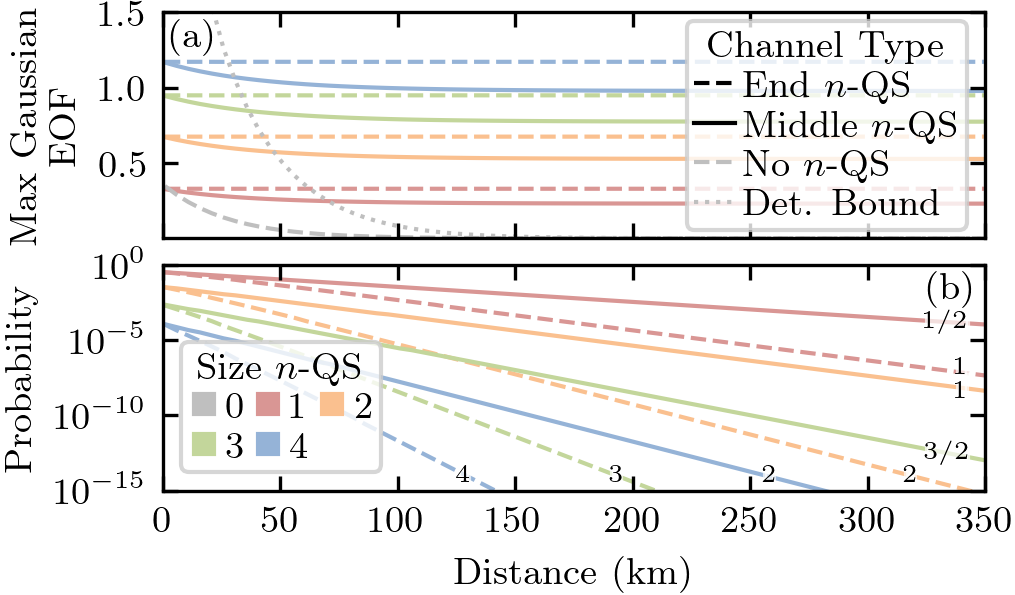}
        \caption{\label{fig:maxgeof} 
            We consider the entanglement of a two-mode squeezed vacuum (TMSV) state with $\chi=0.25$ squeezing, transmitted through a lossy channel with distance $d$. The maximum Gaussian entanglement of formation (EOF) that can be achieved by our $n$-QS protocol at various sizes $n$ is given in (a). We also consider placing the QFT measurement at the end (dashed lines) or in the middle (solid lines) of the channel. This can be contrast with no $n$-QS (dashed gray line), and the deterministic bound (dotted gray line). The associated success probability is given in (b), where the small numbers refers to the long distance scaling of the line with respect to transmissivity (e.g. the success probability given by the solid red line has $\eta^{1/2}$ scaling).} 
    \end{center}
\end{figure}

As described in the main text, the entanglement of formation (EOF) is a measure of entanglement, with Gaussian EOF being particularly relevant for CV type entanglement. We used our calculated covariance matrix in Eq.~\eqref{eq:covmat}, and the numerical estimation technique given in Ref.~\cite{tserkis2019quantifying}, to determine the Gaussian EOF for various gain $g$ settings. We then found the most optimal $g$ gain which corresponds to maximum entanglement, as summarised in Fig.~\ref{fig:maxgeof}(a). We can clearly see that both the end and the middle configurations tend towards a non-zero value as the distance $d$ (and loss) increases. Hence both configurations can be called loss-tolerant in that it is in \tc{principle} possible to distill some amount of entanglement irrespective of the distance or amount of loss. 

\tc{The reason why the entanglement in the end configuration stays the same is because we can completely counteract loss by increasing $g\in[0,\infty]$ gain (recall that $g$ can be chosen by simply modifying one beam-splitter's transmissivity). Consider simplifying the output state in Eq.~\eqref{eq:rhoabloss} for the end configuration case $\eta_A=\eta$ and $\eta_B=1$, which gives  
\begin{align}
        \rho_{AB} &\approx \frac{n!}{(n+1)^n} \frac{1-\chi^2}{(g^2+1)^n} \sum_{l=0}^\infty \sum_{j_1=0}^n \sum_{j_2=0}^n (g\chi \sqrt{\eta})^{j_1+j_2} \chi^{2l} \nonumber \\
        &\quad \times  \sqrt{ \frac{(j_1+l)!(j_2+l)!}{(l!)^2 j_1!j_2!} } |j_1+l\rangle\langle j_2+l|\otimes |j_1\rangle\langle j_2|.
\end{align}
Note that we simplified this expression by assuming $\eta\ll 1$, which is valid for most of Fig.~\ref{fig:maxgeof} considering the transmissivity $\eta$ of the channel goes to zero exponentially as distance increases. In this expression, one can see that by setting $g\propto 1/\sqrt{\eta}$ we can completely counteract the effect of loss on the output state. Hence the EOF, which is determined from this output state, will also remain the same irrespective of distance.}

\tc{The EOF lines in Fig.~\ref{fig:maxgeof}(a)} may be compared to the gray dashed line, which is without the $n$-QS protocol (i.e. a pure loss channel). This can also be compared to the deterministic bound given by the gray dotted line, which is the best that can be done without a probabilistic protocol; this is associated with sending an infinitely squeezed $\chi=1$ TMSV state through a pure loss channel. Due to loss-tolerance, all $n$-QS will eventually beat the deterministic bound, with larger $n$ sizes overcoming the bound at shorter distances because the output is an entangled state with more Fock basis terms. 

\begin{figure}[htbp]
    \begin{center}
        \includegraphics[width=\linewidth]{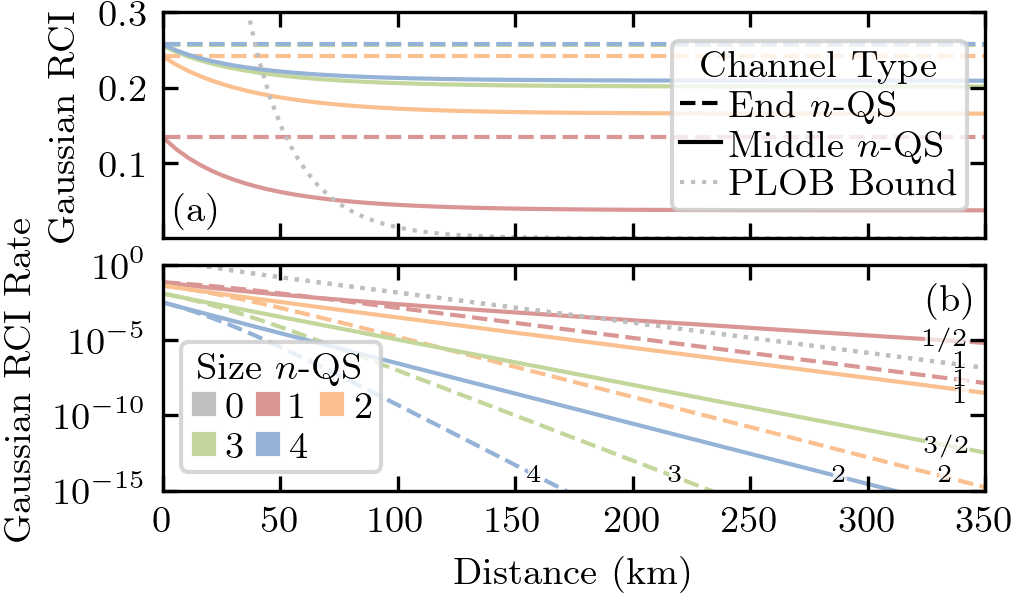}
        \caption{\label{fig:GRCIRate} 
            Similar to Fig.~\ref{fig:maxgeof}, except we consider information transferred. The Gaussian reverse coherent information (RCI) achieved by our $n$-QS protocol at various sizes $n$ is given in (a). Note that this is not with maximised RCI, but with $g=1/\sqrt{\eta}$ for the end configurations and $g=1$ for the middle configurations. The associated information rate is given in (b), where the small numbers refers to the long distance scaling. This can be contrast with the repeaterless PLOB bound (dotted gray line).}
    \end{center}
\end{figure}

It is experimentally preferable if we could generate the correct output at a fast rate. In this regard, we show in Fig.~\ref{fig:GRCIRate}(b) the effect of distance on our protocol's probability of success. The small numbers $x$ at the ends of each line refer to the long distance scaling of success probability $\mathbb{P}_\text{loss}\propto\eta^x=10^{-0.02dx}$. We can see that the $n$-QS success probability scales with transmissivity $\mathbb{P}_\text{loss}\propto\eta^n$ by placing the QFT measurement at the end, however this is improved to $\mathbb{P}_\text{loss}\propto\eta^{n/2}$ by placing the QFT measurement at the middle. Therefore, it is clearly advantageous to have the QFT measurement as a relay station, as we can potentially have orders of magnitude improvement in success probability with some loss in the amount of distillable entanglement. 

We also consider the reverse coherent information (RCI), which is a lower bound on the distillable entanglement in bits per use~\cite{garcia2009reverse}; in particular, we calculate the Gaussian RCI which only deals with the covariance matrix~\cite{winnel2020generalized}. Unlike previously with the EOF, we have not set the gain beam-splitter to maximise the Gaussian RCI. Instead, we simply set $g=1/\sqrt{\eta}$ for the end configuration and $g=1$ for the middle configuration; these gain scaling settings are sufficient to achieve loss tolerance as shown in Fig.~\ref{fig:GRCIRate}(a). In this graph we can clearly see the effect of the truncation on the entanglement at $d=0$, since $g=1$ for both cases. Recall that the input state $|\chi\rangle$ has the coefficients $\chi^k$, and since $\chi=0.25$ each additional term will contribute significantly less to the entanglement in this parameter regime; this is in contrast to Fig.~\ref{fig:maxgeof}(a). 

We can also calculate the information rate by multiplying RCI by the success probability as shown in Fig.~\ref{fig:GRCIRate}(b). This can then be compared to the PLOB bound, the bound that can be achieved without any repeaters. Naturally, the smaller order scissors will have better success probability, but pays a penalty in terms of the quality of the output state since it can't keep higher order terms. Note that quantum repeaters can improve the rate-distance scaling, which can be extremely useful for higher-order scissors. In other words, by dividing the lossy channel into additional shorter segments, a chain of the aforementioned single-node n-QS protocols (in series) can independently distil entanglement of the shorter links to improve the rate-distance scaling, as was shown for the $1$-QS in Ref.~\cite{winnel2021overcoming}.

\tc{\section{Experimental Imperfections Analysis}}

\tc{In this section, we will analyse the overall effect of various experimental imperfections to our quantum scissor devices. One imperfection is inefficient detectors, which we will model by applying loss channels with $\tau_d = 0.7$ transmissivity just before each detection. We will also include noise by simulating dark-count error detection events, which is modelled by mixing each detector mode with thermal states of particular energy to achieve a $C_d=10^{-8}$ single-photon dark-count probability. These parameters are readily achievable using either superconducting nanowires or transition-edge sensors (TES), which have both been shown to have $\tau_d>0.95$ detector efficiencies, with TES in particular also having very low $C_d<10^{-10}$ dark-count probabilities~\cite{reddy2019exceeding,lita2008counting,marsili2013detecting,miller2003demonstration}. }

\begin{figure}[t]
    \begin{center}
        \includegraphics[width=\linewidth]{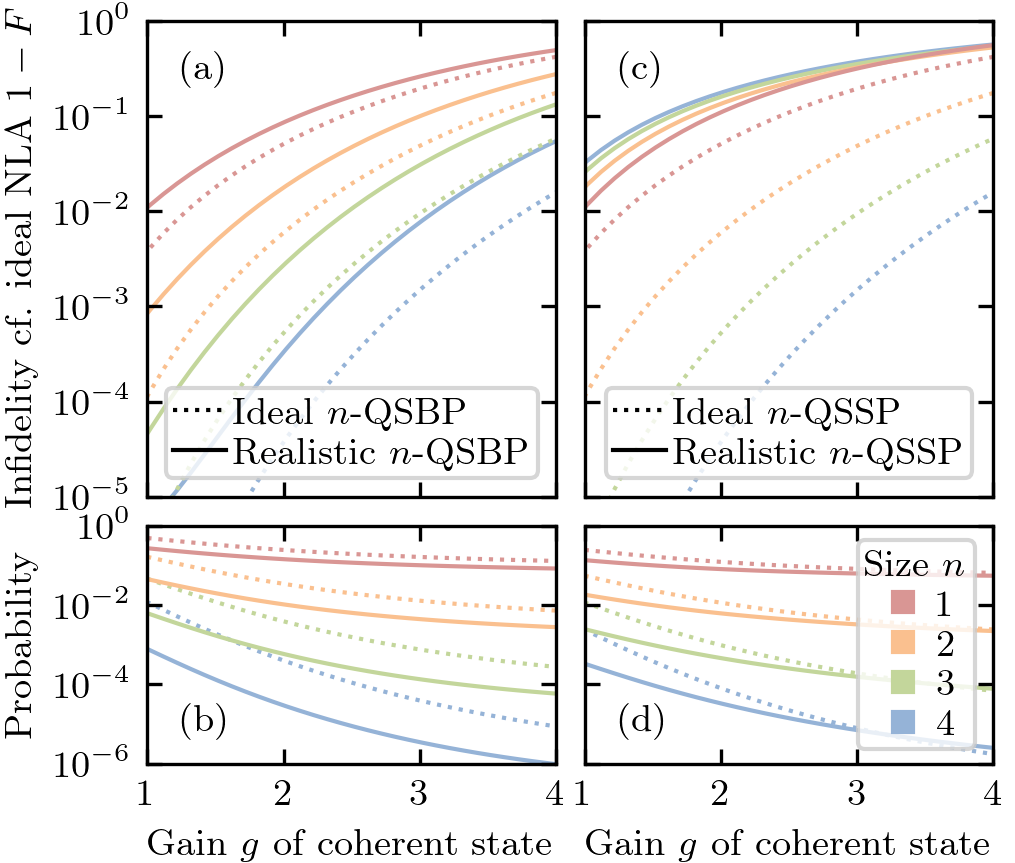}
        \caption{\label{fig:fidvsg_loss_BPvsSP} 
            \tc{How our $n$-QS amplification protocol is affected by experimental imperfections (solid lines), in contrast to the ideal case (dotted lines). The realistic solid lines include imperfect resources ($\tau_r=0.7$ transmissivity), and inefficient detectors ($\tau_d=0.7$ efficiency) with dark-counts ($10^{-8}$ probability). The left side is the bunched photons resource configuration given in Fig.~1(a), while the right side is the single photons resource configuration \textit{without} resource preparation given in Fig.~1(b). The input is a coherent state input with $\alpha=0.3$ amplitude. (a) and (c) shows the infidelity $1-F$ relative to a perfect NLA output state, while (b) and (d) shows the probability of success.}}
    \end{center}
\end{figure}

\tc{We will also consider imperfect resource states, which in the ideal case should be $|n\rangle$ for $n$-QSBP or $\otimes^n|1\rangle$ for $n$-QSSP. These resources can be generated by spontaneous parametric down-conversion and heralded detections, hence we will simply model this imperfect process by applying loss channels with $\tau_r = \tau_d = 0.7$ transmissivity to the initially ideal resource states. This translates to a realistic overall transmissivity $\tau_r\tau_d$, from resource to detection, of less than 50\%. Note that photon generation and detection technologies have improved enormously in recent decades~\cite{slussarenko2019photonic}, with further improvements being likely to continue into the future. }

\tc{In the subsequent sections, we will numerically regenerate realistic versions of the primary graphs in this paper, by taking into account all of the above experimental imperfections. We will contrast this with the ideal case (i.e. without any imperfections), as well as realistic versions of the competing protocols when appropriate. }

\tc{\subsection{Amplifier}}

\begin{figure}[t]
    \begin{center}
        \includegraphics[width=0.543\linewidth]{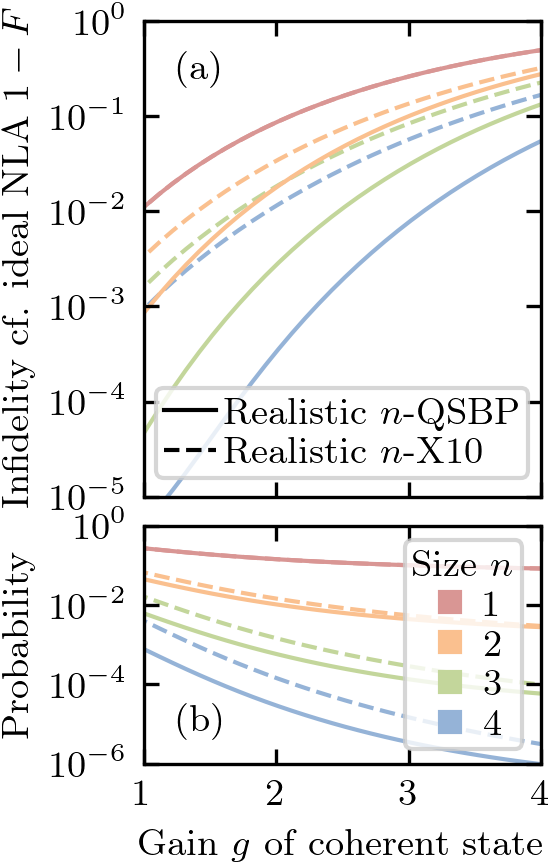}
        \caption{\label{fig:fidvsg_loss_nQSBPvsnX10} 
            \tc{A comparison of our $n$-QSBP NLA protocol against the $n$-X10 NLA protocol~\cite{xiang2010heralded}, given the same realistic conditions as outlined in Fig.~\ref{fig:fidvsg_loss_BPvsSP} caption. This graph has no significant differences in contrast to ideal conditions in Fig.~2(a) and (b), hence the benefit of our protocol holds even in imperfect experimental conditions.}}
    \end{center}
\end{figure}

\tc{We can see the effect of imperfections on the $n$-QS NLA protocol in Fig.~\ref{fig:fidvsg_loss_BPvsSP}. The left hand side of this graph shows that the $n$-QSBP configuration (given schematically by the left side of Fig.~1) still achieves exponential fidelity scaling with increased protocol size $n$, similar to the ideal case. Imperfections simply vertically shift each curve, and it is clearly still useful to increase the size of this protocol $n$. In contrast, the right hand side of Fig.~\ref{fig:fidvsg_loss_BPvsSP} shows that experimental imperfections for the $n$-QSSP configuration (given schematically by the right side of Fig.~1) causes a collapse in fidelity scaling, as one increases protocol size $n$.}

\tc{One primary reason for this stark contrast between the two configurations is the multiple vacuum measurements used for the $n$-QSSP case; one can imagine easily losing a photon just before a particular detector, which vacuum measurements will misidentify. In contrast, the $n$-QSBP splits up the light as equally as possible via the $n+1$ mode QFT, which minimises the possibility of bunched photons at the output and hence an error single-photon detection; also losing a single-photon before the detector just means the protocol will fail (i.e. decreases the probability with limited affect on fidelity). Furthermore, it is known that $n$-QSBP is resource loss-resistant, as lost resource photons results in the device acting similar to a lower-sized $k$-QSBP, where $k<n$, as has been discussed for coherent state inputs in the $3$-QSBP in Ref.~\cite{winnel2020generalized}. This effect can be seen in Fig.~\ref{fig:fidvsg_loss_BPvsSP}(a) and (b) as the realistic QSBP fidelity curves are quite similar to lower-sized ideal QSBP fidelity curves.}

\begin{figure}[t]
    \begin{center}
        \includegraphics[width=\linewidth]{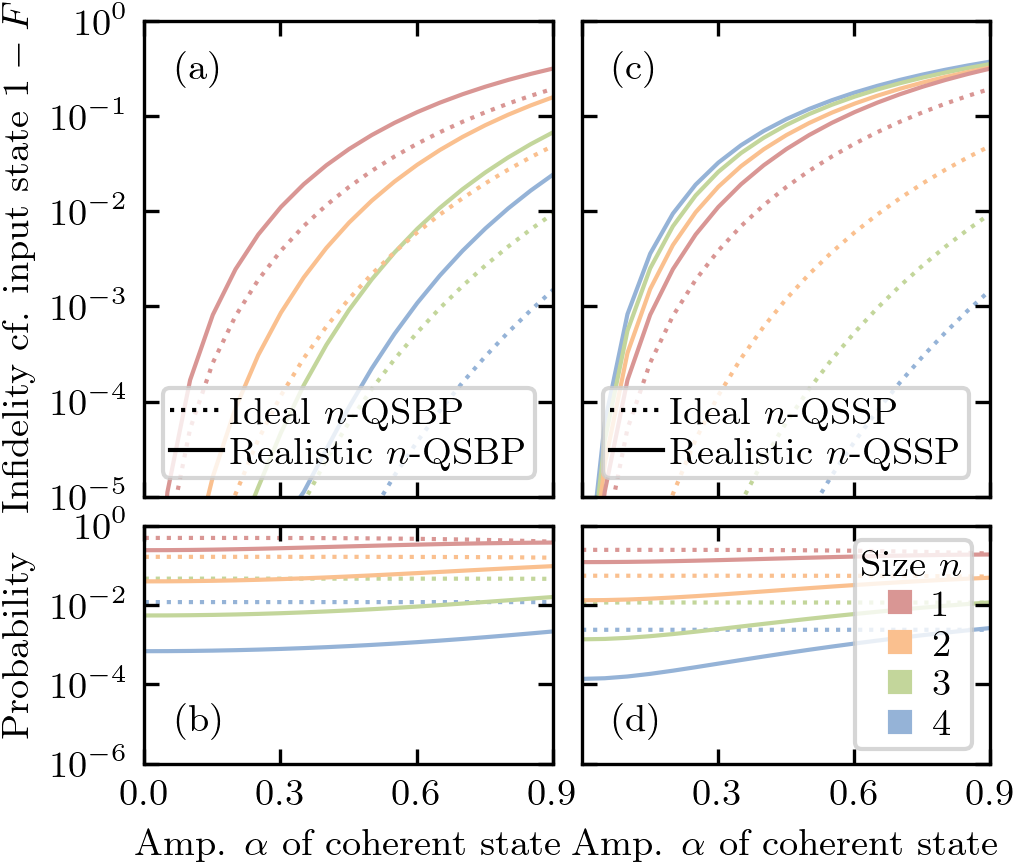}
        \caption{\label{fig:fidvsamp_loss_BPvsSP} 
            \tc{How our $n$-QS teleportation protocol is affected by experimental imperfections (solid lines), in contrast to the ideal case (dotted lines). The same realistic conditions as outlined in Fig.~\ref{fig:fidvsg_loss_BPvsSP} caption was used. We consider teleporting an $\alpha$ amplitude coherent state. (a) and (c) is the infidelity $1-F$ of the teleported output state relative to the input state, while (b) and (d) is the protocol's probability of success.}}
    \end{center}
\end{figure}

\tc{We note that these results do not completely discount the $n$-QSSP configuration, because we can prepare beforehand the two mode resource state $|R_n\rangle$, given by the red dashed box in Fig.~1(b), as previously discussed in Sec.~\ref{sec:probs}. One can consider replacing this red box with a better device which also generates the required $|R_n\rangle$ state (for example, using a Gaussian boson sampling type-device, as discussed in Ref.~\cite{su2019conversion}). Assuming one has the $|R_n\rangle$ resource state, the $n$-QSSP device is effectively just three modes for all $n$, which may present less challenging experimental issues than an $n$-QSBP device, especially for larger $n$ sizes. Considering all these factors will require more detailed investigations beyond the scope of this paper.}

\tc{In Fig.~\ref{fig:fidvsg_loss_nQSBPvsnX10} we apply the same experimental imperfection considerations to both our $n$-QSBP protocol and the competing $n$-X10 NLA protocol. One can see that this graph is very similar to the ideal situation given in Fig.~2; imperfections also vertically shifts the $n$-X10 protocol. This is not surprising, as $n$-X10 also requires $n$ single-photon measurements and roughly the same amount of physical resources as $n$-QSBP; hence one would intuitively expect realistic imperfections to affect the two systems in a similar manner. Thus the ultimate outcome of our paper doesn't change by introducing experimental imperfections; our protocol is shown to provide substantial benefits over the alternative NLA protocol, which we have shown here can be achieved in practice via an imperfect linear-optical apparatus.}

\tc{ \subsection{Teleporter} }

\begin{figure}[t]
    \begin{center}
        \includegraphics[width=0.543\linewidth]{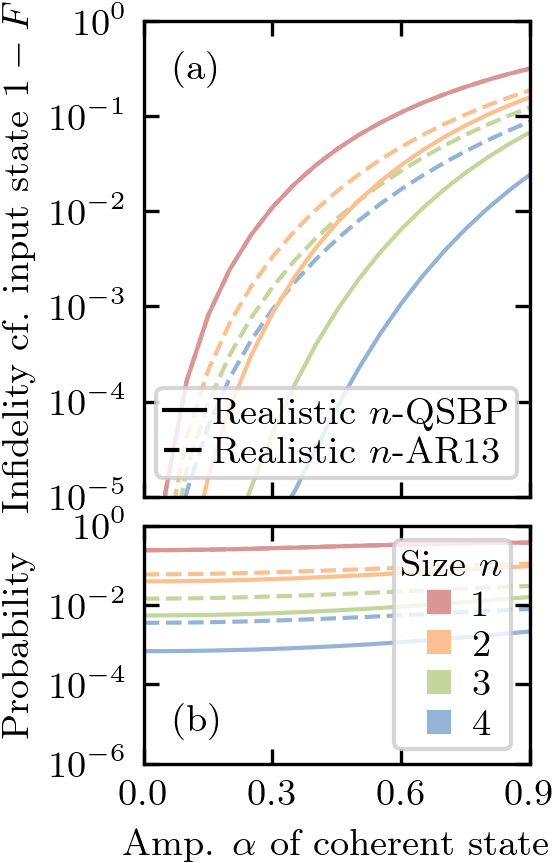}
        \caption{\label{fig:fidvsamp_loss_nQSBPvsnAR13} 
            \tc{A comparison of our $n$-QSBP teleportation protocol against the $n$-AR13 teleportation protocol~\cite{andersen2013high}, given the same realistic conditions as outlined in Fig.~\ref{fig:fidvsg_loss_BPvsSP} caption. Our protocol still has significant benefits in imperfect conditions; the ideal version of this graph is given in Fig.~4(a) and (b).}}
    \end{center}
\end{figure}

\tc{Here we consider the effect of imperfections on the $n$-QS for teleportation purposes (i.e. no gain $g=1$). This is summarised in Fig.~\ref{fig:fidvsamp_loss_BPvsSP}, which shows effectively the same results as the amplification situation. These results further confirms the analysis contrasting the two different $n$-QS configurations in the previous section.}

\tc{In Fig.~\ref{fig:fidvsamp_loss_nQSBPvsnAR13} we apply the same experimental imperfection considerations to both our $n$-QSBP protocol and the competing $n$-AR13 protocol. This result is ultimately the same as the ideal case given in Fig.~4. For example, in this regime we are considering, one can see that the $3$-QSBP has both better fidelity and success probability than the $4$-AR13, and requires less experimental resources.}

\tc{Note that we had to reduce the size $n$ range, compared to the ideal case given in Fig.~4, because we employed a different method of simulating the state through the device (a necessary change to model these additional experimental imperfections). Unfortunately, the computational time of this method scales exponentially with size $n$, which prevents us from exploring larger sized $n$ values in a timely manner. However, the trends in these graphs provide good evidence that our protocol can realistically beat previously established alternatives in the literature even for larger $n$ values.}

\tc{\subsection{Relay}}

\begin{figure}[t]
    \begin{center}
        \includegraphics[width=\linewidth]{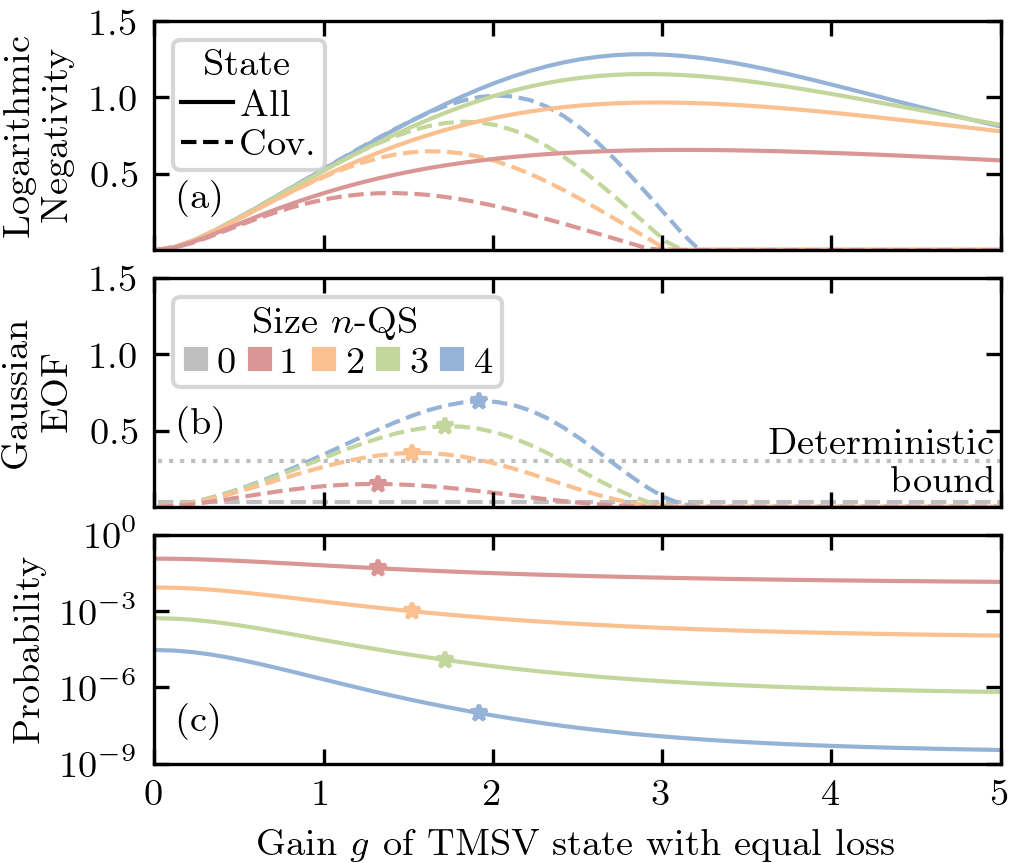}
        \caption{\label{fig:entvsg_loss} 
            \tc{The amount of entanglement which can be recovered, using an equally $\eta_A=\eta_B=\sqrt{\eta}$ distributed $n$-QSBP as a quantum relay (i.e. measurement is in the middle of the channel), as per Fig.~3(b). The same realistic conditions as outlined in Fig.~\ref{fig:fidvsg_loss_BPvsSP} caption was used. We consider a $\chi=0.25$ amplitude two-mode squeezed vacuum (TMSV) state into a lossy channel with $\eta=0.05$ total transmission. The entanglement is measured using (a) log negativity and (b) Gaussian entanglement of formation (EOF). The solid line considers all correlations (i.e. the entire state), while the dashed lines only considers second moments (i.e. the covariance matrix). (c) is the protocol's probability of success. The ideal version of this graph is given in Fig.~5.}}
    \end{center}
\end{figure}

\begin{figure}[t]
    \begin{center}
        \includegraphics[width=\linewidth]{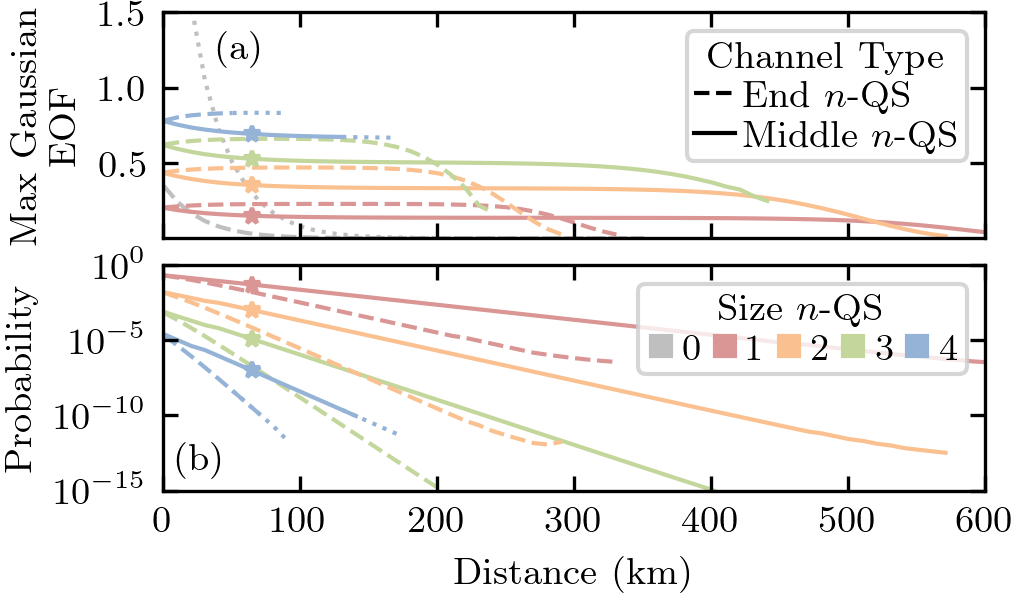}
        \caption{\label{fig:maxgeof_loss} 
            \tc{Similar to Fig.~\ref{fig:maxgeof}, however with realistic conditions as outlined in Fig.~\ref{fig:fidvsg_loss_BPvsSP} caption. Here we clearly indicate the connection with Fig.~\ref{fig:entvsg_loss} via star markers, which is at distance $d=-50\text{log}_{10}(\eta)\approx65\text{ km}$.}}
    \end{center}
\end{figure}

\tc{We consider the effect of experimental imperfections on the $n$-QSBP as a quantum relay, as given schematically in Fig.~3(b). Fig.~\ref{fig:entvsg_loss} shows that one can still realistically distill quantum entanglement through a lossy channel, with $\eta=0.05$ transmissivity. There are no issues with scaling to larger $n$ sizes in this regime. By contrasting these graphs to the ideal case given in Fig.~5, imperfections simply dampens the amount of entanglement. Clearly, the realistic version of the $2$-QSBP can still beat the deterministic bound in this parameter regime.}

\FloatBarrier 

\tc{Finally, in Fig.~\ref{fig:maxgeof_loss} we replicate Fig.~\ref{fig:maxgeof} with experimental imperfections. One can see that imperfections causes a vertical shift in the maximum amount of entanglement; this is predominately due to loss in the photon resource $\tau_r$ and photon detectors $\tau_d$. In addition, Fig.~\ref{fig:maxgeof_loss} also has an extra waterfall feature where there is now a maximum distance for which entanglement can be distributed. This effect is due to the dark-count rates, since at those distances the errors due to the dark-counts dominates over the correct outcomes. This reinforces our previous analysis in Sec.~\ref{sec:entdist}, which shows that having the QFT detection in the middle of the channel is more loss-resistant; this is clearly the case for this realistic scenario, where we can effectively double the maximum range in which entanglement can be distributed.}

\tc{Note in Fig.~\ref{fig:maxgeof_loss}, we are unable to observe the entirety of the $n=4$ lines with satisfactory accuracy, hence the maximum distance is not clear for these cases. This is because of computational precision errors associated with small numbers, as well as computational difficulty in simulating states with very large Hilbert spaces. However, these partial $n=4$ lines have been included as the curves can still tell us how much imperfections affects the maximum amount of entanglement, in contrast to the ideal case given in Fig.~\ref{fig:maxgeof}.}

\bibliography{supp}